\documentclass[eqsecnum,showkeys,secnumarabic,showpacs,twocolumn,hyperref,pra]{revtex4}
\usepackage{amsfonts}
\usepackage{amsmath}
\usepackage{amssymb}
\usepackage{graphicx}
\usepackage{hyperref}

\begin{document}

\preprint{}
\title{Extended Ensemble Theory, Spontaneous Symmetry Breaking, and Phase
Transitions}
\author{Ming-wen Xiao}
\affiliation{Department of Physics, Nanjing University, Nanjing 210093, China}

\begin{abstract}
In this paper, as a personal review, we suppose a possible extension of
Gibbs ensemble theory so that it can provide a reasonable description to
phase transitions and spontaneous symmetry breaking. The extension is
founded on three hypotheses, and can be regarded as a microscopic edition of
the Landau phenomenological theory of phase transitions. Within its
framework, the stable state of a system is determined by the evolution of
order parameter with temperature according to such a principle that the
entropy of the system will reach its minimum in this state. The evolution of
order parameter can cause change in representation of the system
Hamiltonian; different phases will realize different representations,
respectively; a phase transition amounts to a representation transformation.
Physically, it turns out that phase transitions originate from the automatic
interference among matter waves as temperature is cooled down. Typical
quantum many-body systems are studied with this extended ensemble theory. We
regain the Bardeen-Cooper-Schrieffer solution for the weak-coupling
superconductivity, and prove that it is stable. We find that
negative-temperature and laser phases arise from the same mechanism as phase
transitions, and that they are instable. For the ideal Bose gas, we
demonstrate that it will produce Bose-Einstein condensation (BEC) in the
thermodynamic limit, which confirms exactly Einstein's deep physical
insight. In contrast, there is no BEC either within the phonon gas in a
black body or within the ideal photon gas in a solid body. We prove that it
is not admissible to quantize Dirac field by using Bose-Einstein statistics.
We show that a structural phase transition belongs physically to the BEC
happening in configuration space, and that a double-well anharmonic system
will undergo a structural phase transition at a finite temperature. For the $%
O(N)$-symmetric vector model, we demonstrate that it will yield spontaneous
symmetry breaking and produce Goldstone bosons; and if it is coupled with a
gauge field, the gauge field will obtain a mass (Higgs mechanism). Also, we
show that an interacting Bose gas is stable only if the interaction is
repulsive. For the weak interaction case, we find that the BEC is a
\textquotedblleft $\lambda $\textquotedblright -transition and its
transition temperature can be lowered by the repulsive interaction. In
connection with liquid $^{4}\mathrm{He}$, it is found that the specific heat
at constant pressure $C_{P}$ will show a $T^{3}$ law at low temperatures,
which is in agreement with the experiment. If the system is further cooled
down, the theory predicts that $C_{P}$ will vanish linearly as $T\rightarrow
0$, which is anticipating experimental verifications.
\end{abstract}

\received{October 25, 2005}
\revised{\today}
\startpage{1}
\endpage{}
\maketitle
\tableofcontents

\section{Introduction}

\subsection{The Problem of Gibbs Ensemble Theory \label{PGET}}

Since Gibbs \cite{Gibbs} established the ensemble theory of statistical
mechanics, there has arisen the problem as to whether it has the ability to
describe phase transitions \cite{Yang1}.

Let us investigate the problem through the famous example, i.e., the
conventional low-$T_{c}$ superconductivity, which can be described by the
so-called Bardeen-Cooper-Schrieffer (BCS) Hamiltonian \cite{BCS1,BCS2,BCS3},
\begin{align}
H(c)& =\sum_{\mathbf{k}}\varepsilon (\mathbf{k})(c_{\mathbf{k}\uparrow
}^{\dag }c_{\mathbf{k}\uparrow }+c_{-\mathbf{k}\downarrow }^{\dag }c_{-%
\mathbf{k}\downarrow })  \notag \\
& -g\sum_{\mathbf{k},\mathbf{k}^{\prime }}c_{\mathbf{k}^{\prime }\uparrow
}^{\dag }c_{-\mathbf{k}^{\prime }\downarrow }^{\dag }c_{-\mathbf{k}%
\downarrow }c_{\mathbf{k}\uparrow },  \label{BCS}
\end{align}%
where $\varepsilon (\mathbf{k})$ denotes the energy relative to Fermi level,
$g>0$ the coupling strength, $c_{\mathbf{k}\uparrow }$ and $c_{\mathbf{k}%
\downarrow }$ ($c_{\mathbf{k}\uparrow }^{\dag }$ and $c_{\mathbf{k}%
\downarrow }^{\dag }$) the destruction (creation) operators for the
electrons with up and down spins, respectively. Here, the notation $H(c)$
expresses that the Hamiltonian is a function of $c_{\mathbf{k}\uparrow }$, $%
c_{\mathbf{k}\downarrow }$, $c_{\mathbf{k}\uparrow }^{\dag }$, and $c_{%
\mathbf{k}\downarrow }^{\dag }$.

For BCS superconductivity, what is the most important physically is that the
Hamiltonian $H(c)$ will remain invariant under the transformation,%
\begin{equation}
G(\vartheta ,c)H(c)G^{\dagger }(\vartheta ,c)=H(c),
\end{equation}%
where
\begin{equation}
G(\vartheta ,c)=e^{-i\vartheta \sum_{\mathbf{k}}(c_{\mathbf{k}\uparrow
}^{\dag }c_{\mathbf{k}\uparrow }+c_{\mathbf{k}\downarrow }^{\dag }c_{\mathbf{%
k}\downarrow })},\text{ }\vartheta \in \lbrack 0,2\pi ).  \label{Gauge}
\end{equation}%
This invariance is known as gauge symmetry. As a direct consequence of this
symmetry, one has%
\begin{equation}
\left\langle c_{-\mathbf{k}\downarrow }c_{\mathbf{k}\uparrow }\right\rangle =%
\mathrm{Tr}\big(c_{-\mathbf{k}\downarrow }c_{\mathbf{k}\uparrow }\rho (H(c))%
\big)=0,  \label{Zero}
\end{equation}%
where $\langle \cdots \rangle $ denotes the Gibbs ensemble average with
respect to the Hamiltonian $H(c)$, and
\begin{equation}
\rho (H)=\frac{e^{-\beta H}}{\mathrm{Tr}\!\left( e^{-\beta H}\right) }
\end{equation}%
is the statistical density operator where $\beta =1/\left( k_{B}T\right) $
with $k_{B}$ and $T$ being the Boltzmann constant and temperature,
respectively. To prove Eq. (\ref{Zero}), it is sufficient to heed that
\begin{equation}
G\left( \frac{\pi }{2},c\right) c_{-\mathbf{k}\downarrow }c_{\mathbf{k}%
\uparrow }G^{\dag }\left( \frac{\pi }{2},c\right) =-c_{-\mathbf{k}\downarrow
}c_{\mathbf{k}\uparrow }.  \label{Negative}
\end{equation}

Eq. (\ref{Zero}) indicates that the electron-pair amplitude $\left\langle
c_{-\mathbf{k}\downarrow }c_{\mathbf{k}\uparrow }\right\rangle $ can never
become nonzero according to Gibbs ensemble theory, in other words, there can
not happen superconductivity for the BCS Hamiltonian of Eq. (\ref{BCS})
within Gibbs ensemble theory no matter how low the temperature is. This
proves definitely that Gibbs ensemble theory has no ability to describe
phase transitions.

Even so, it is still hoped to do, as least as possible, some modification to
Gibbs theory so that it could describe phase transitions. Up to now, there
are two kinds of modification, they both arise from the theoretical studies
on the so-called Ising model \cite{Ising}, and become the popular and
predominant believes about phase transitions in the current world of
statistical physics.

The first is to introduce the thermodynamic limit in the end of calculation
\cite{Onsager,Kaufmann,Yang2,Yang3,Georgii,GEmch,Emch,Bratteli}, that is,%
\begin{equation}
\left\langle c_{-\mathbf{k}\downarrow }c_{\mathbf{k}\uparrow }\right\rangle
=\lim_{V\rightarrow +\infty }\mathrm{Tr}\big(c_{-\mathbf{k}\downarrow }c_{%
\mathbf{k}\uparrow }\rho (H(c))\big),  \label{Zero1}
\end{equation}%
where the $\lim_{V\rightarrow +\infty }$ means to take the thermodynamic
limit ($V\rightarrow +\infty $, $N\rightarrow +\infty $, and $N/V=$
constant) after performing the trace. Because%
\begin{equation}
\mathrm{Tr}\big(c_{-\mathbf{k}\downarrow }c_{\mathbf{k}\uparrow }\rho (H(c))%
\big)=0,
\end{equation}%
irrespective of the magnitudes of $V$, $N$, and $N/V$ ($H(c)$ is invariant
under gauge transformation, and that is true independently of $V$, $N$, and $%
N/V$, namely, the gauge symmetry has nothing to do with $V$, $N$, and $N/V$%
.), one has%
\begin{equation}
\lim_{V\rightarrow +\infty }\mathrm{Tr}\big(c_{-\mathbf{k}\downarrow }c_{%
\mathbf{k}\uparrow }\rho (H(c))\big)=0.  \label{Zero2}
\end{equation}%
Therefore, this kind of modification can not make Gibbs ensemble theory have
the ability to describe phase transitions. Indeed, the thermodynamic limit
is important for calculating the statistical average of observable, but
itself alone is insufficient to act as the physical criterion for phase
transitions.

The second is to introduce both the thermodynamic limit and an auxiliary
external field \cite{Ising,Kramers,Yang4,Huang},
\begin{widetext}
\begin{equation}
\left\langle c_{-\mathbf{k}\downarrow }c_{\mathbf{k}\uparrow }\right\rangle
=\lim_{\phi \rightarrow 0}\lim_{V\rightarrow +\infty }\mathrm{Tr\!}\left(
c_{-\mathbf{k}\downarrow }c_{\mathbf{k}\uparrow }\rho \Big(H(c)+\sum_{%
\mathbf{k}}(\phi c_{-\mathbf{k}\downarrow }c_{\mathbf{k}\uparrow }+\phi
^{\dag }c_{\mathbf{k}\uparrow }^{\dag }c_{-\mathbf{k}\downarrow }^{\dag })%
\Big)\right) ,  \label{Pair}
\end{equation}%
\end{widetext}
where, as pointed out by Huang \cite{Huang}, the limit $\phi \rightarrow 0$
must be taken after the thermodynamic limit $V\rightarrow +\infty $. The
purpose of introducing an auxiliary external field $\phi $ is to break the
gauge symmetry of the system, as can be easily seen from the term,%
\begin{equation}
\sum_{\mathbf{k}}(\phi c_{-\mathbf{k}\downarrow }c_{\mathbf{k}\uparrow
}+\phi ^{\dag }c_{\mathbf{k}\uparrow }^{\dag }c_{-\mathbf{k}\downarrow
}^{\dag }).  \label{Broken}
\end{equation}%
Since the gauge symmetry has been broken with regard to the entire
Hamiltonian,
\begin{equation}
H(c)+\sum_{\mathbf{k}}(\phi c_{-\mathbf{k}\downarrow }c_{\mathbf{k}\uparrow
}+\phi ^{\dag }c_{\mathbf{k}\uparrow }^{\dag }c_{-\mathbf{k}\downarrow
}^{\dag }),  \label{ExField}
\end{equation}%
in so far as the auxiliary external field $\phi $ is infinitesimal but
nonzero, the trace and the limit $V\rightarrow +\infty $ of Eq.(\ref{Pair})
become nonzero. Further, if the limit $\phi \rightarrow 0$ does not tend to
zero, the electron-pair amplitude $\left\langle c_{-\mathbf{k}\downarrow }c_{%
\mathbf{k}\uparrow }\right\rangle $ will get nonzero. One thus concludes
that there appear Cooper pairs in the system, and that the system has gone
into the superconducting phase. However, this conclusion can not hold in
physics, for the modification suffers the serious problem: The limit
procedure employed by Eq.(\ref{Pair}) is physically equivalent to the
scenario that an infinitesimal auxiliary external field is first added to
the system so as to induce the symmetry of the system to break down, and
taken off from the system finally. But, in the actual situation, the
inducement of such an external field is unnecessary, the symmetry itself
breaks down spontaneously and does not need any help of external force.
Anyhow, this kind of modification contains unphysical operations, and thus
can not be accepted in principle.

Here, it is also significant and worth while to give a brief discussion to
the two-dimensional Ising model because it can be solved exactly within the
two modifications,%
\begin{equation}
H_{\mathrm{I}}=-J\sum_{\langle ij\rangle}s_{i}s_{j},
\end{equation}
where $s_{i}=\pm1$ denotes the spin variable on site $i$, the symbol $%
\langle ij\rangle$ means that the sites $i$ and $j$ \ are the nearest
neighbors, and $J>0$ is the exchange coupling between a nearest-neighbor
pair of spins.

As is well known, Onsager \cite{Onsager} proved rigorously that the specific
heat of the Ising Hamiltonian $H_{\mathrm{I}}$ is singular at $T=T_{c}>0$
within the first modification of Gibbs ensemble theory. It hints that the
system might undergo a paramagnetic-ferromagnetic phase transition at $T_{c}$
in the thermodynamic limit. However, just as Eqs. (\ref{Zero1}--\ref{Zero2}%
), the magnetization $m$,
\begin{equation}
m=\lim_{N\rightarrow \infty }\mathrm{Tr}\big(s_{i}\rho (H_{\mathrm{I}})\big)%
=0,  \label{mIsing}
\end{equation}%
is always equal to zero due to the fact that the parity symmetry of $H_{%
\mathrm{I}}$ with respect to $s_{i}=1$ and $s_{i}=-1$ has nothing to do with
the magnitude of $N$ where $N$ represents the total number of sites. Eq. (%
\ref{mIsing}) proves definitely that there is no paramagnetic-ferromagnetic
phase transition at any temperature within the first modification of Gibbs
ensemble theory even if the specific heat is singular.

The above proof also shows that it can not be solved within the framework of
the first modification whether or not the onset of the singularity at $%
T=T_{c}$ manifests a phase transition. In order to justify that the
phenomenon occurring at $T=T_{c}$ is a phase transition, Yang \cite{Yang4}
went beyond the first modification and turned to the second one, which was,
in fact, suggested originally by Ising himself \cite{Ising} in studying the
one-dimensional Ising model. Wonderfully, Yang succeeded in proving
rigorously that%
\begin{eqnarray}
m &=&\lim_{B\rightarrow 0}\lim_{N\rightarrow \infty }\mathrm{Tr}\Big(%
s_{i}\rho \big(H_{\mathrm{I}}-B\sum_{i}s_{i}\big)\Big)  \notag \\
&=&\left\{
\begin{array}{ll}
\text{zero,} & T\geq T_{c} \\
\text{nonzero, } & T<T_{c},%
\end{array}%
\right.  \label{mnzero}
\end{eqnarray}%
where $B$ stands for an external magnetic field. Mathematically, it seems
reasonable to say from Eq. (\ref{mnzero}) that a paramagnetic-ferromagnetic
phase transition will occur at $T=T_{c}$. Nevertheless, as pointed out
above, Eq. (\ref{mnzero}) imports unphysical operations from the
introduction of the external field $B$, and this leads to the result that
the phase transition will not occur automatically but has to be driven and
decided from outside the system, which directly contradicts the basic
experimental fact that any phase transition occurs itself spontaneously.
Therefore, the second modification, or rather Ising's criterion, can not be
used as a physical criterion to justify whether there exists a
paramagnetic-ferromagnetic phase transition at $T=T_{c}$.

It should be stressed again that it is the two rigorous works of Refs. \cite%
{Onsager} and \cite{Yang4} that establish the two kinds of modification to
the Gibbs ensemble theory.

On all accounts, the two kinds of modification must both be discarded, and
it is necessary to modify or extend Gibbs ensemble theory anew. That is just
the main purpose of this paper.

Recently, Gibbs ensemble theory has been extended by C. Tsallis and his
followers with the conception of nonextensive entropy \cite%
{Tsallis1,Tsallis2,Tsallis3}, which is being under controversy. In this
paper, as a personal review, we would like to suppose another possible
extension of the Gibbs ensemble theory so that it can provide a reasonable
description to phase transitions and spontaneous symmetry breaking (SSB).

\subsection{Landau Phenomenological Theory}

Also, there is a macroscopic theory for phase transitions, i.e., Landau
phenomenological theory \cite{Landau}.

Phenomenological as it is, Landau theory succeeds in providing a unified
picture for all the second-kind phase transitions, e.g., superconductivity,
superfluidity, magnetism, and structural phase transitions. Physically,
Landau picture consists of two basic notions: order parameter and
variational principle. Order parameter describes the degree of order of the
system, it is zero in the disordered phase, and nonzero in the ordered one.
Variational principle yields the equation of motion of order parameter and
controls the evolution of order parameter with temperature, a system must
arrive at the minimum of Helmholtz or Gibbs free energy in its stable state.
According to this picture, a system will evolve with the evolution of its
order parameters, and produce phase transitions spontaneously, i.e., without
any help or drive from outside the system. In short, Order parameter and
variational principle are two characteristics of Landau theory.

On the other hand, Landau pointed out that a phase transition of the second
kind reflects physically the change in symmetry of the system: the
disordered phase has a higher symmetry than the ordered one. He found that
this change in symmetry can be described through the representations of the
symmetry group, a phase transition corresponds to a representation
transformation, different phases will realize different representations,
respectively. Accordingly, Landau established the relation between the order
parameter and the representation. Using this relation, it can be easily
determined from the variational principle which representation will be
realized at a certain temperature. Representation transformation and
spontaneous symmetry breaking are the other two characteristics of Landau
theory.

To sum up, a system will select and realize automatically different
representations at different temperatures according to the variational
principle of order parameter. That is the mechanism for phase transitions
discovered by Landau theory, it crystallizes Landau's ideas about phase
transitions: order parameter, variational principle, representation
transformation, and spontaneous symmetry breaking. Theoretically, this
mechanism gives a reasonable answer to the problem why a phase transition or
SSB can happen at a certain temperature. In applications, it also agrees
quite well with various phase transitions, its effectiveness and
universality are well known to condensed matter physicists. In a word,
Landau's ideas on phase transitions are successful and of great value in
physics!

Regrettably, it is impossible to deduce Landau theory from Gibbs ensemble
theory because there is no variational principle of order parameter within
the latter. This demonstrates again the deficiency of Gibbs ensemble theory.
In fact, this deficiency was pointed out by Born and Fucks \cite{Born} early
in 1938. In 1937, Mayer \cite{Mayer} published his famous paper on
gas-liquid transition, which is based directly on Gibbs ensemble theory.
This work was seriously doubted and questioned by Born and Fucks\ \cite{Born}%
: \textquotedblleft How can the gas molecules `know' when they have to
coagulate to form a liquid and solid?\textquotedblright\ Obviously, owing to
lack of Landau mechanism, it is hard for Gibbs ensemble theory to answer the
question posed by Born and Fucks.

Gibbs ensemble theory should be extended to incorporate the Landau's ideas
so as to describe phase transitions and answer the question posed by Born
and Fucks. That is another intention of this paper.

\section{The Extended Ensemble Theory \label{EET}}

In order to describe phase transitions and spontaneous symmetry breaking, we
shall, as a personal review, extend Gibbs ensemble theory with three
hypotheses, which can be stated, taking the BCS superconductivity as an
instance, as follows.

(1). The system Hamiltonian is represented by $H^{\prime}(\phi,c)$,
\begin{equation}
H^{\prime}(\phi,c)=e^{iD(\phi,c)}H(c)e^{-iD(\phi,c)},  \label{HPrime}
\end{equation}
where
\begin{equation}
D(\phi,c)=\sum_{\mathbf{k}}(\phi_{\mathbf{k}}c_{-\mathbf{k}\downarrow }c_{%
\mathbf{k}\uparrow}+\phi_{\mathbf{k}}^{\dag}c_{\mathbf{k}\uparrow}^{\dag
}c_{-\mathbf{k}\downarrow}^{\dag}).  \label{Dfactor}
\end{equation}
Here $\phi$ is an internal field of the system, it is also the order
parameter for BCS superconductivity. Evidently, $H^{\prime}(\phi,c)$ is not
invariant under the gauge transformation of $G(\vartheta,c)$ as long as $%
\phi\neq0$, namely, the gauge symmetry will be broken for $%
H^{\prime}(\phi,c) $ if $\phi\neq0$. From now on, we shall call $D(\phi,c)$
the phase-transition operator, for sake of convenience.

(2). The statistical average of an observable $F(c)$ is defined as
\begin{equation}
\mathcal{F}(\phi,\beta)=\langle F(c)\rangle=\mathrm{Tr}\big(F(c)\rho
(H^{\prime}(\phi,c))\big).  \label{Average}
\end{equation}
We remark that the average is now a function of both temperature and order
parameter, which is expressed explicitly by the two arguments of $\mathcal{F}%
(\phi,\beta)$.

(3). The order parameter $\phi $ is determined by the minimum of the entropy
of the system,
\begin{eqnarray}
\delta S &=&0,  \label{Variation} \\
\Delta S &\geq &0,  \label{Increment}
\end{eqnarray}%
where $S$ denotes the entropy,
\begin{eqnarray}
S(\phi ,\beta ) &=&\langle -\ln \!\left( \rho (H(c))\right) \rangle  \notag
\\
&=&-\mathrm{Tr}\big(\!\ln \!\left( \rho (H(c))\right) \rho (H^{\prime }(\phi
,c))\big).  \label{Entropy}
\end{eqnarray}%
As a function of the order parameter, the entropy controls the evolution of
state of the system with temperature.

The first hypothesis means that the system Hamiltonian can take different
representations at different temperatures, e.g., it can take the symmetric
representation ($\phi=0$) at a high temperature, and the asymmetric
representation ($\phi\neq0$) at a low temperature. Which representation it
will take is determined by the third hypothesis. After the representation is
so determined, the statistical average can be calculated with respect to
this representation, as stated in the second hypothesis.

Obviously, the extended theory will reduce to the original one if $\phi =0$,
that is to say, the original theory holds only for the normal phase of the
system. The broken-symmetry phase will be described by the extended theory.

Within the framework of the extended ensemble theory, a system will realize
different representations of the same system Hamiltonian at different
temperatures according to the principle of least entropy with respect to
order parameter, this mechanism for phase transitions is, in spirit, the
same as that given by Landau phenomenological theory; hence, it can be said
that Landau mechanism has been incorporated into the extended ensemble
theory. Also, because the order parameter is an internal field of the system
itself, symmetry breaking will occur in a completely spontaneous way rather
than forced by an external field, there is no unphysical operation within
the extended theory. In a word, the extended ensemble theory is
conceptionally in accordance with Landau's ideas on phase transitions: order
parameter, variational principle, representation transformation, and
spontaneous symmetry breaking.

In Sec. \ref{POPT}, we shall show further that phase transitions originate
physically from the wave nature of matter.

Now, let us establish the relationship between the extended ensemble theory
and thermodynamics, it can be implemented as follows.

First, the internal energy $U$ can be obtained from Eq. (\ref{Average}),
replacing $F(c)$ by the Hamiltonian $H(c)$. It is a function of temperature $%
T$ and volume $V$, that is,
\begin{equation}
U=U(T,V).
\end{equation}
From the internal energy $U(T,V)$, the specific heat at constant volume, $%
C_{V}$, can be calculated through%
\begin{equation}
C_{V}=C_{V}(T,V)=\left( \frac{\partial U}{\partial T}\right) _{V}.
\end{equation}

Then, we can obtain the thermodynamical entropy $S_{th}$,%
\begin{equation}
S_{th}=S_{th}(T,V)=\int_{0}^{T}C_{V}(T^{\prime},V)\frac{\mathrm{d}T^{\prime}%
}{T^{\prime}}.
\end{equation}
It is worth paying attention to the significant difference between the
thermodynamical entropy $S_{th}$ and the statistical entropy $S$ defining in
Eq. (\ref{Entropy}): $S$ measures the degree of order of the system, whereas
$S_{th}$ measures the amount of heat absorbed or rejected by the system,
i.e., $dQ=TdS_{th}$; besides, $S$ contains the basic information of the
system, it controls the evolution of state of the system, as a consequence, $%
S_{th}$ derives itself from $S$.

At last, we arrive at%
\begin{equation}
F=F(T,V)=U(T,V)-TS_{th}(T,V),
\end{equation}%
where $F$ is the Helmholtz free energy of the system, with $T$ and $V$ as
its natural variables. As a thermodynamic potential of the system, $F$ can
generate other thermodynamic quantities. By such a way, the extended
ensemble theory connects with thermodynamics.

In the end of this section, let us explain why we choose the BCS model
rather than the Ising model as the starting instance to develop the theory
for phase transitions and spontaneous symmetry breaking.

As is well known, Ising model $H_{\mathrm{I}}$ is a classical discrete
Hamiltonian. Simple as it is mathematically, it is, however, ill defined
from the point of view of physics. On one hand, a classical Hamiltonian, $%
H=H(q,p)$ where $q$ and $p$ denote the generalized coordinates and momenta
of the system, must be continuous\ in phase space, that is to say, $H(q,p)$
is a continuous function of $q$ and $p$; there does not exist any classical
system which corresponds to a discrete Hamiltonian; therefore, $H_{\mathrm{I}%
}$ does not correspond to a classical system. On the other hand, the
discrete quantity $s_{i}=\pm 1$ is not a quantum operator but a pure integer
variable, therefore, $H_{\mathrm{I}}$ does not correspond to a quantum
system, either. In a word, Ising model can not correspond to any physical
system; its solution, whether rigorous or not, has no real physical
relevance \cite{Yang1}, which was, in fact, realized and pointed out as the
"Ising disease" by people early in 1940's and 1950's \cite{Yang1}.

To understand that point further, let us first suppose that Ising model
could represent a real physical system, and then analyze the statistical
properties of this system in detail. First, if the system is not placed in
an external magnetic field, then, according to the Onsager's solution \cite%
{Ising,Huang}, its specific heat will diverge logarithmically at $T=T_{c}>0$%
; however, as indicated by Eq. (\ref{mIsing}), there appears no spontaneous
magnetization bellow $T_{c}$, namely, the state of the system will remain
nonmagnetic at any temperature, whether $T>T_{c}$ or $T<T_{c}$. There is no
paramagnetic-ferromagnetic phase transition in the Ising system if it is not
placed in an external magnetic field, though its specific heat is singular.
That is the first characteristic behavior of the Ising system. Secondly, if
the system is placed in an external magnetic field, then, according to the
Yang's solution \cite{Yang4}, there will appear an induced magnetization
bellow $T_{c}$. That is to say, the Ising system will transform from a
paramagnetic state at $T>T_{c}$ into a ferromagnetic state at $T<T_{c}$ when
it is placed in an external magnetic field, here, the transformation is an
induced transformation other than a spontaneous phase transition. That is
the second characteristic behavior of the Ising system. In sum, an Ising
system will exhibit two distinct behaviors, respectively, depending on
whether the system is placed in an external magnetic field or not. It will
exhibit the first characteristic behavior in the case without the
application of any external magnetic field and the second one in the case
with the application of an external magnetic field. The two behaviors should
be checked experimentally with respect to the same physical system. However,
such a physical system that can simultaneously show both the two
characteristic behaviors of the Ising model has never been observed and
reported experimentally. In other words, there does not exist the supposed
Ising system in nature at all! That is not surprising, it just reflects the
ill definition of the Ising model. Here, it should be pointed out that the
solutions of Onsager and Yang are not identical but essentially different
from each other, whether in the sense of mathematics or in the sense of
physics: In the limit $B\rightarrow 0$, the Yang's solution can not reduce
to the Onsager's solution, as can be easily seen by comparing Eq. (\ref%
{mIsing}) and Eq. (\ref{mnzero}); they represent the statistical properties
of the Ising system with and without the interaction of an external magnetic
field, respectively.

Needless to say, an Ising system is completely distinguishable from a real
physical system of phase transition. For the latter, the transition happens
itself spontaneously, needing no inducement of a corresponding external
field. Evidently, this nature of phase transitions is neither the same as
the first characteristic behavior of the Ising system, i.e., a singular
specific heat but no phase transition, nor the same as the second
characteristic behavior of the Ising system, i.e., a transformation induced
but not spontaneous. There are clear and definite discrepancies between the
statistical behavior of a real physical system of phase transition and those
of the Ising system. Any phase transition can not be described by Ising
model. Of course, if one disregards physically the ill definition of Ising
model, the distinction between the solutions of Onsager and Yang, and the
discrepancies between the statistical behavior of a real physical system of
phase transition and those of the Ising system, he can simulate some phase
transitions and critical phenomena quite well with Ising model \cite%
{Kadanoff1,Kadanoff2,Kadanoff3,Wilson1,Wilson2,Wilson3,Wilson4,Wilson5,Fisher}%
.

Theoretically, Ising model is ill defined; experimentally, Ising system does
not exist in nature at all. That is the reason why we do not choose it as
the starting instance to develop the theory of phase transitions and SSB. As
for the BCS model \cite{BCS1,BCS2,BCS3}, quite the contrary, it is of high
value in theory because it contributes a new quantum concept, i.e., Cooper
pairs, which is profound and rather hard to understand theoretically but yet
verified exactly by the quantized flux experiments. Actually, it is the BCS
theory that began our microscopic understanding of phase transitions.

Although Ising's criterion for phase transitions is problematic in physics,
it is much suggestive in mathematics, indeed. To some degree, it can be said
that the present modification is just equivalent to substituting the
external field proposed by Ising with an internal field determined by the
system itself, as can be seen from expanding the right-hand side of Eq. (\ref%
{HPrime}) to the linear term of $\phi $ and then comparing it with Eq. (\ref%
{ExField}).

\section{Application to the Ideal Fermi Gas \label{IFG}}

Let us first apply the extended ensemble theory to the ideal Fermi gas. It
is the simplest case, and we shall see that this case is exactly solvable.

For the ideal Fermi gas, there is no interaction ($g=0$), Eq. (\ref{BCS})
reduces to
\begin{equation}
H(c)=\sum_{\mathbf{k}}\varepsilon (\mathbf{k})(c_{\mathbf{k}\uparrow
}^{\dagger }c_{\mathbf{k}\uparrow }+c_{-\mathbf{k}\downarrow }^{\dag }c_{-%
\mathbf{k}\downarrow }),
\end{equation}%
where $\varepsilon (\mathbf{k})=\hbar ^{2}\mathbf{k}^{2}/\left( 2m\right)
-\mu $ with $m$ and $\mu $ being the mass of the fermions and the chemical
potential of the system, respectively.

To facilitate the calculation of the entropy, we reformulated Eq. (\ref%
{Entropy}) as
\begin{equation}
S(\phi ,\beta )=-\mathrm{Tr}\!\left( \ln \!\Big(\rho \big(H(e^{-iD(\phi
,c)}ce^{iD(\phi ,c)})\big)\Big)\rho (H(c))\right) .  \label{Sphi}
\end{equation}%
By use of Eq. (\ref{Dfactor}), we find
\begin{widetext}
\begin{subequations}
\label{Ctrans}
\begin{eqnarray}
e^{-iD(\phi ,c)}c_{\mathbf{k}\uparrow }e^{iD(\phi ,c)} &=&\cos \left( \theta
_{\mathbf{k}}\right) c_{\mathbf{k}\uparrow }+i\sin \left( \theta _{\mathbf{k}%
}\right) e^{-i\phi _{\mathbf{k}}}c_{-\mathbf{k}\downarrow }^{\dagger } \\
e^{-iD(\phi ,c)}c_{-\mathbf{k}\downarrow }e^{iD(\phi ,c)} &=&\cos \left(
\theta _{\mathbf{k}}\right) c_{-\mathbf{k}\downarrow }-i\sin \left( \theta _{%
\mathbf{k}}\right) e^{-i\phi _{\mathbf{k}}}c_{\mathbf{k}\uparrow }^{\dagger
},
\end{eqnarray}%
where $\theta _{\mathbf{k}}=|\phi _{\mathbf{k}}|$ and $\varphi _{\mathbf{k}%
}=\arg (\phi _{\mathbf{k}})$. Substituting them into Eq. (\ref{Sphi}), we
have
\end{subequations}
\begin{equation}
S(\phi ,\beta )=\ln \!\big(\mathrm{Tr}(e^{-\beta H(c)})\big)+2\beta \sum_{%
\mathbf{k}}\varepsilon (\mathbf{k})\left[ \cos ^{2}(\theta _{\mathbf{k}%
})f(\varepsilon (\mathbf{k}))+\sin ^{2}(\theta _{\mathbf{k}})f(-\varepsilon (%
\mathbf{k}))\right] ,  \label{Sphi1}
\end{equation}%
\end{widetext}
where%
\begin{equation}
f(\varepsilon )=\frac{1}{e^{\beta \varepsilon }+1}
\end{equation}%
is the Fermi distribution function.

From Eqs. (\ref{Sphi1}) and (\ref{Variation}), it follows that
\begin{equation}
\sin(2\theta_{\mathbf{k}})=0,  \label{IFGOrder}
\end{equation}
which is the equation of order parameter. Obviously, this equation has the
solutions,%
\begin{equation}
\theta_{\mathbf{k}}=0,\text{ }\pi/2\text{ },\pi,\text{ }3\pi/2.
\label{Theta}
\end{equation}

In addition, one can easily deduce from Eq. (\ref{Sphi1}) that%
\begin{equation}
\Delta S=\sum_{\mathbf{k}}\Delta s_{\mathbf{k}},
\end{equation}%
where
\begin{widetext}%
\begin{equation}
\Delta s_{\mathbf{k}}=\left\{
\begin{array}{rlll}
\beta \varepsilon (\mathbf{k})\tanh \!\left( \frac{\beta \varepsilon (%
\mathbf{k})}{2}\right) \left[ 1-\cos \left( 2\delta \theta _{\mathbf{k}%
}\right) \right] & \geq & 0,\text{ } & \theta _{\mathbf{k}}=0,\text{ or }\pi
\\
-\beta \varepsilon (\mathbf{k})\tanh \!\left( \frac{\beta \varepsilon (%
\mathbf{k})}{2}\right) \left[ 1-\cos \left( 2\delta \theta _{\mathbf{k}%
}\right) \right] & \leq & 0,\text{ } & \theta _{\mathbf{k}}=\pi /2,\text{ or
}3\pi /2,%
\end{array}%
\right.
\end{equation}%
with $\delta \theta _{\mathbf{k}}$ being the deviation of $\theta _{\mathbf{k%
}}$ from the corresponding solution.

For the solution that all $\theta _{\mathbf{k}}=0$ or $\pi $, we have $%
\Delta S\geq 0$, which meets the requirement of Eq. (\ref{Increment}),
thereby, this solution corresponds to a stable phase. Observe that the order
parameter $\phi =0$ when all $\theta _{\mathbf{k}}=0$, one recognizes
immediately that this phase is just the normal phase of the system, which
can also be seen from the following equation,
\begin{equation}
\mathcal{F}(\phi ,\beta )=\mathrm{Tr}\big(F(c)\rho (H^{\prime }(\phi ,c))%
\big)=\mathrm{Tr}\big(F(c)\rho (H(c))\big),\text{ }\forall \theta _{\mathbf{k%
}}=0,\text{ or }\pi .
\end{equation}%
We are familiar with the properties of this normal phase from the standard
books on quantum statistical mechanics. What it is worth stressing here is
that the normal phase is proved to be stable at any temperature within the
framework of the extended ensemble theory. Of course, that is reasonable,
and just what we expect.

As to the other solutions, they also have physical senses though they
correspond to instable phases because the entropies for them are not
minimal. Let us first discuss the phase with all $\theta _{\mathbf{k}}=\pi
/2 $ or $3\pi /2$. We find
\begin{equation}
\mathcal{F}(\phi ,\beta )=\mathrm{Tr}\big(F(c)\rho (H^{\prime }(\phi ,c))%
\big)=\mathrm{Tr}\!\left( F(c)\frac{e^{-\beta ^{\prime }H(c)}}{\mathrm{Tr}%
\!\left( e^{-\beta ^{\prime }H(c)}\right) }\right) ,\text{ }\forall \theta _{%
\mathbf{k}}=\frac{\pi }{2},\text{ or }\frac{3\pi }{2},
\end{equation}%
\end{widetext}
where $\beta ^{\prime }=1/(k_{B}T^{\prime })$ with $T^{\prime }=-T<0$. That
is to say, the temperature of the system goes negative. This shows that
negative temperatures root from the same microscopic mechanism as that for
phase transitions, so we call this instable phase the negative-temperature
phase. In comparison, the normal phase is said to be the
positive-temperature phase ($T>0$). This negative-temperature phase can not
be realized physically because its internal energy is unbounded above.

Apart from the negative-temperature phase, all the other instable phases are
partially negative-temperature phases, which means that the particles of the
system are distributed partially among the negative-temperature
(single-particle) states, and partially among the positive-temperature
(single-particle) states,
\begin{equation}
\langle c_{\mathbf{k}\sigma }^{\dag }c_{\mathbf{k}\sigma }\rangle =\left\{
\begin{array}{ll}
\frac{1}{e^{\beta \varepsilon (\mathbf{k})}+1},\text{ } & \theta _{\mathbf{k}%
}=0,\text{ or }\pi \\
\frac{1}{e^{\beta ^{\prime }\varepsilon (\mathbf{k})}+1},\text{ } & \theta _{%
\mathbf{k}}=\pi /2,\text{ or }3\pi /2.%
\end{array}%
\right.
\end{equation}%
That is to say, some of the states take $0$ or $\pi $, the others take $\pi
/2$ or $3\pi /2$. Each $\mathbf{k}$ with $\theta _{\mathbf{k}}=0$ or $\pi $
represents a positive-temperature state, and each $\mathbf{k}$ with $\theta
_{\mathbf{k}}=\pi /2$ or $3\pi /2$ represents a negative-temperature state.
Because the partially negative-temperature phase is an instable phase, it
can emit photons when particles transfer from a negative-temperature state
into a positive-temperature one. Therefore, a partially negative-temperature
system can constitute a laser.

To make that point more specified, let us consider a semiconductor that is
described by the following Hamiltonian,
\begin{equation}
H(c)=\sum_{n,\mathbf{k}}\varepsilon _{n}(\mathbf{k})(c_{n,\mathbf{k}\uparrow
}^{\dag }c_{n,\mathbf{k}\uparrow }+c_{n,-\mathbf{k}\downarrow }^{\dag }c_{n,-%
\mathbf{k}\downarrow }),
\end{equation}%
where $n$ denotes the energy-band index, $\varepsilon _{n}(\mathbf{k})$ the
energy of the electrons in the $n$th band relative to the Fermi level of the
system, and $c_{n,\mathbf{k}\sigma }$ the annihilation operator of the
electrons with momentum $\mathbf{k}$ and spin $\sigma $ in the $n$th band.
Replacing the $D(\phi ,c)$ of Eq. (\ref{Dfactor}) with
\begin{equation}
D(\phi ,c)=\sum_{n,\mathbf{k}}(\phi _{n,\mathbf{k}}c_{n,-\mathbf{k}%
\downarrow }c_{n,\mathbf{k}\uparrow }+\phi _{n,\mathbf{k}}^{\dag }c_{n,%
\mathbf{k}\uparrow }^{\dag }c_{n,-\mathbf{k}\downarrow }^{\dag }),
\end{equation}%
and following the same procedure as for the ideal Fermi gas, one can easily
get
\begin{equation}
\langle c_{n,\mathbf{k}\sigma }^{\dag }c_{n,\mathbf{k}\sigma }\rangle
=\left\{
\begin{array}{ll}
\frac{1}{e^{\beta \varepsilon _{n}(\mathbf{k})}+1},\text{ } & \theta _{n,%
\mathbf{k}}=0,\text{ or }\pi \\
\frac{1}{e^{\beta ^{\prime }\varepsilon _{n}(\mathbf{k})}+1},\text{ } &
\theta _{n,\mathbf{k}}=\pi /2,\text{ or }3\pi /2,%
\end{array}%
\right.
\end{equation}%
where $\theta _{n,\mathbf{k}}=|\phi _{n,\mathbf{k}}|$.

Suppose that there are, for example, two electrons per site. If the system
is at zero temperature and stays in the normal phase where all $\theta _{n,%
\mathbf{k}}=0$ or $\pi $, the valence band, i.e., the $0$th band, is fully
filled, the conduction band, i.e. the $1$st band, and all the other higher
bands are empty. This is the well-known picture for semiconductors (or
insulators), which is depicted in Fig. 1.
\begin{figure}[htbp]
\includegraphics[scale=0.35,angle=-90]{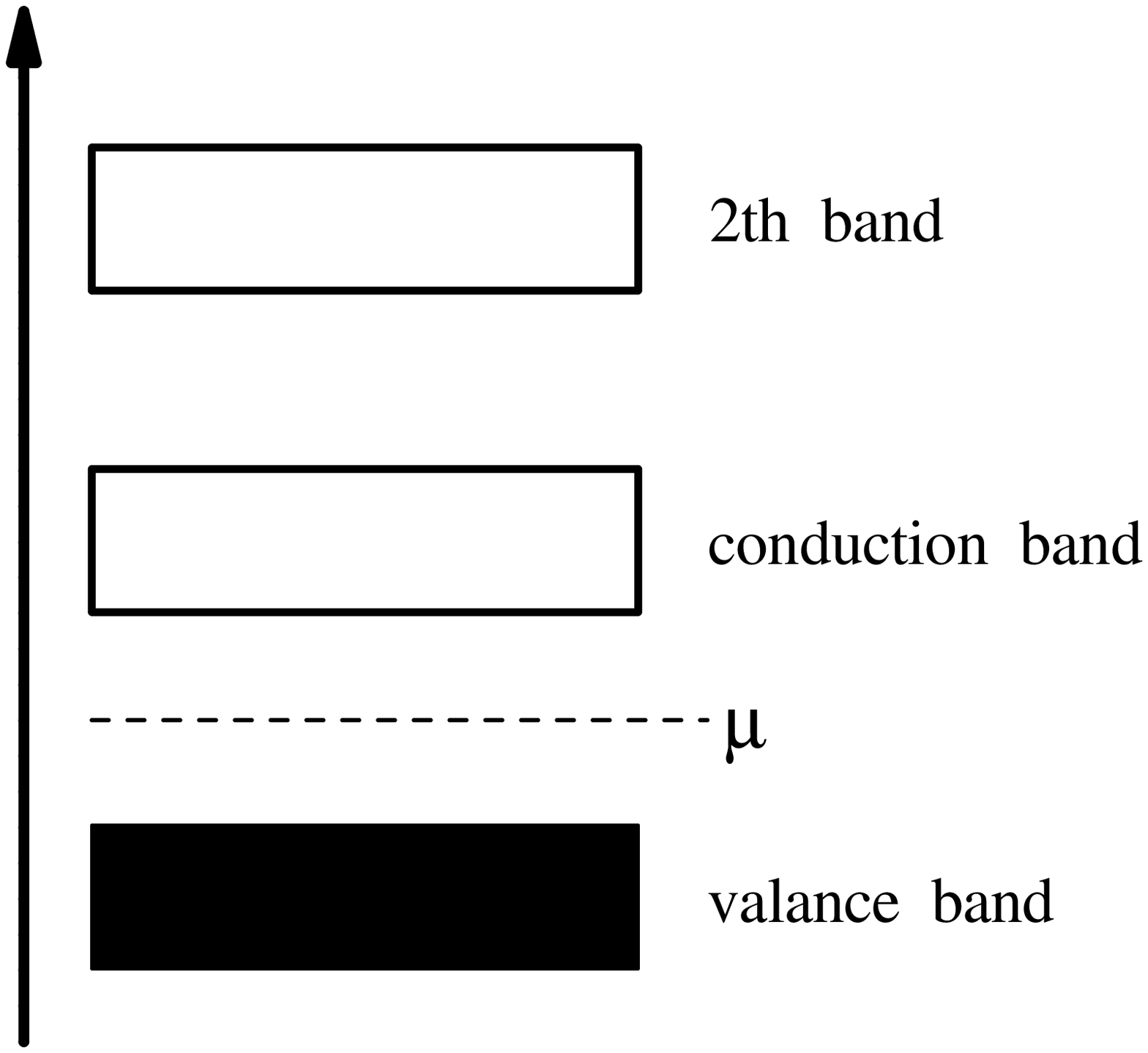}
\caption
{Energy bands for the normal (positive-temperature) phase of a semiconductor
where $\mu$ denotes the chemical potential.}
\end{figure}%

Now, let us consider such a partially negative-temperature phase,
\begin{equation}
\theta _{n,\mathbf{k}}=\left\{
\begin{array}{ll}
\pi /2,\text{ or }3\pi /2,\text{ } & n=1 \\
0,\text{ or }\pi ,\text{ } & n\neq 1.%
\end{array}%
\right.
\end{equation}%
If this phase is at zero temperature, the conduction band is fully filled,
all the other bands are empty. It implies that the population is inverted
between the valence and conduction bands, which is depicted in Fig. 2.
\begin{figure}[htbp]
\includegraphics[scale=0.35,angle=-90]{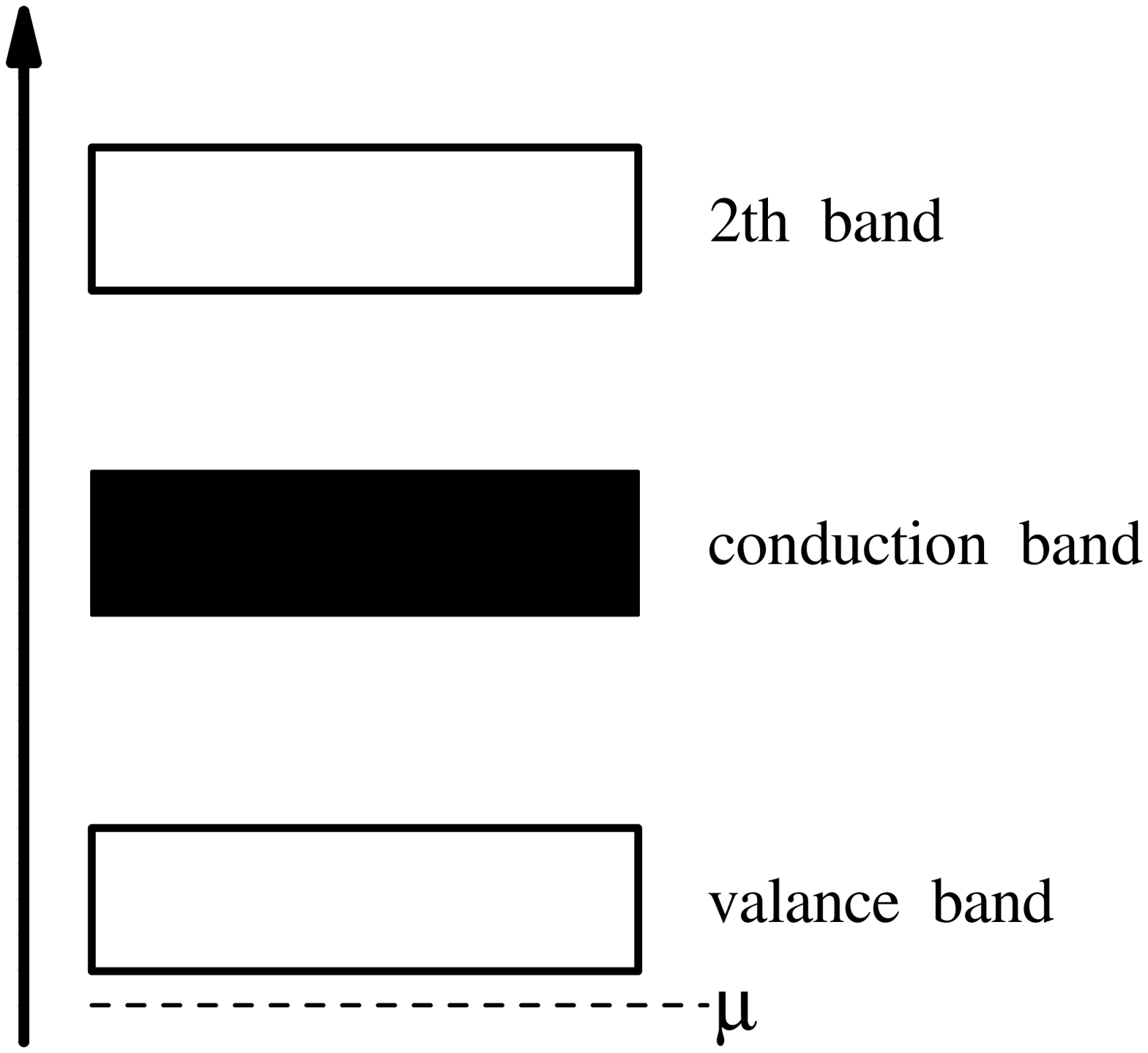}
\caption
{Energy bands for a partially negative-temperature phase of a semiconductor
where $\mu
$ denotes the chemical potential. Here, the population is inverted between the
valance and conduction bands.}
\end{figure}
As mentioned above, such a population-inversion phase is instable, it will
finally turn into the stable phase, i.e., the normal phase, with the
electrons in the conduction band hopping into the valence band and
simultaneously releasing energies to the surroundings. It will make a laser
if the energies are released by means of emitting photons. That is somewhat
an ideal semiconductor laser, it is a \textquotedblleft
two-level\textquotedblright\ system, a more realistic laser can be achieved
in the so-called \textquotedblleft three-level\textquotedblright\ system,%
\begin{equation}
\theta _{n,\mathbf{k}}=\left\{
\begin{array}{ll}
0,\text{ or }\pi ,\text{ } & n=0 \\
\pi /2,\text{ or }3\pi /2,\text{ } & \exists \mathbf{k}\text{ for }n=2 \\
0,\text{ or }\pi ,\text{ } & \text{otherwise,}%
\end{array}%
\right.
\end{equation}%
where the $0$th and $2$nd bands are partially filled, the $1$st band is
empty, which is depicted in Fig. 3. Here, the population is not inverted
between the $0$th and $2$nd bands, i.e., the population of the $2$nd band is
less than that of the $0$th band, but it is inverted between the $1$st and $%
2 $nd bands. This kind of \textquotedblleft three-level\textquotedblright\
semiconductor laser has been realized in experiments \cite{Laud}. For this
reason, we call the partially negative-temperature phase the laser phase.
Obviously, a laser phase cannot be described by the original Gibbs ensemble
theory \cite{Gibbs}.
\begin{figure}[htbp]
\includegraphics[scale=0.35,angle=-90]{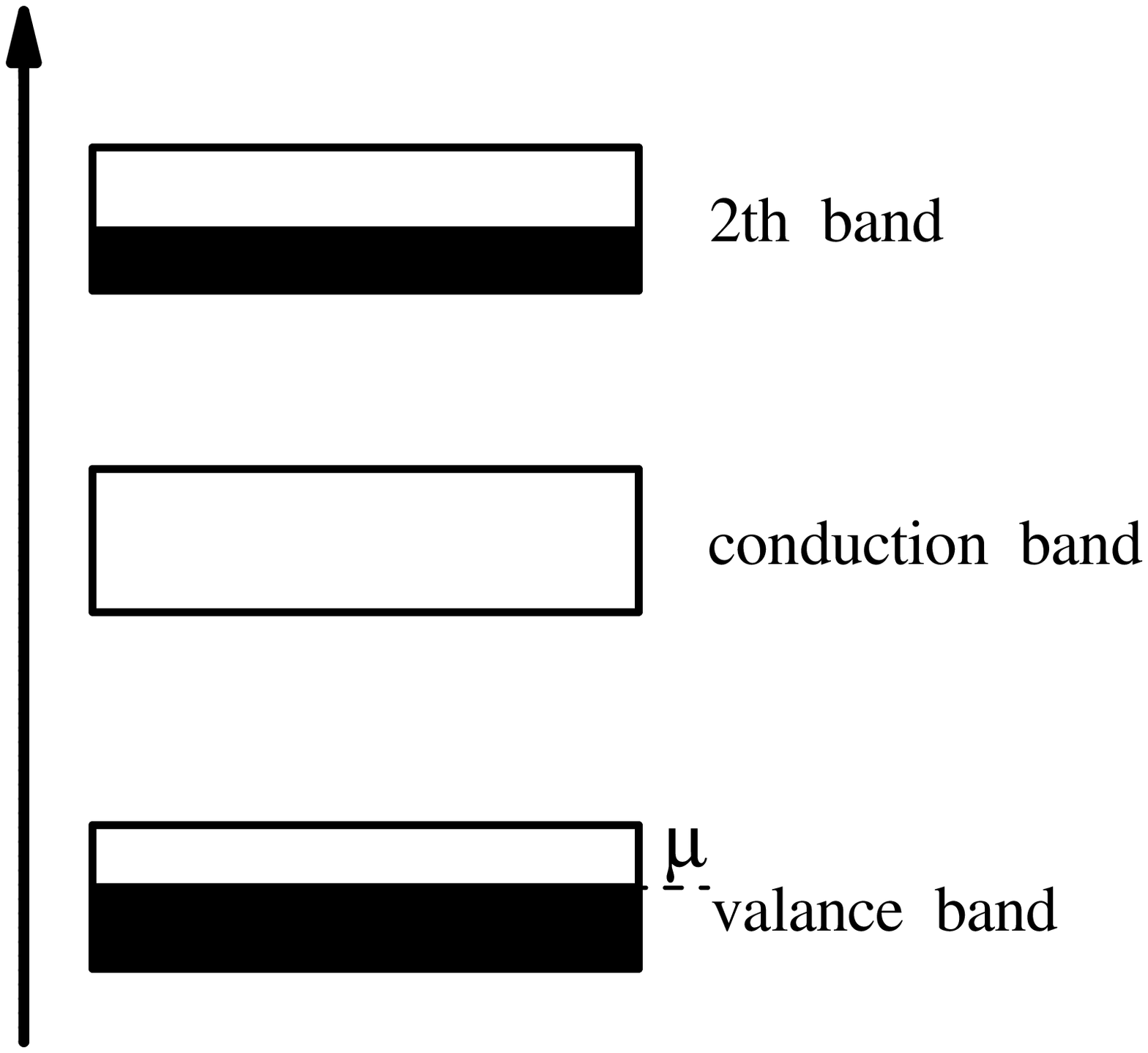}
\caption{Energy bands for a laser phase of a semiconductor where $\mu
$ denotes the
chemical potential. Here, both the valance and 2th bands are partially filled, the
population is inverted between the 2th and conduction bands.}
\end{figure}%

Traditionally, negative temperatures are something obscure, and hard to
understand why they are higher or hotter than positive temperatures \cite%
{Purcell,Ramsey,Pathria}. From the viewpoint of the extended ensemble
theory, the reason is that both the negative-temperature and laser phases
are instable, they will trend, by giving up parts of their internal energy
to the surroundings, towards the stable phase, viz., the
positive-temperature phase. That is the microscopic interpretation about
negative temperatures.

Finally, we note that the ideal Fermi gas cannot produce superconductivity
because there does not exist any Cooper pair in any case of the solutions of
Eq. (\ref{Theta}),
\begin{equation}
\left\langle c_{-\mathbf{k}\downarrow }c_{\mathbf{k}\uparrow }\right\rangle =%
\mathrm{Tr}\big(c_{-\mathbf{k}\downarrow }c_{\mathbf{k}\uparrow }\rho
(H^{\prime }(\phi ,c))\big)=0,
\end{equation}%
this result is just as expected, and right in accordance with the BCS theory
\cite{BCS1,BCS2,BCS3}.

In summary, it is proved, within the framework of the extended ensemble
theory, that the ideal Fermi gas can not produce superconductivity, its
normal phase is stable at any temperature. Besides, the extended ensemble
theory gives a microscopic interpretation to the negative-temperature and
laser phases: they originate from the same microscopic mechanism as that for
phase transitions, and most importantly they are instable.

\section{Application to BCS Superconductivity}

\subsection{The BCS\ Mean-Field Theory}

To study BCS superconductivity with the extended ensemble theory, one needs
first to solve Eq. (\ref{Variation}) to obtain the order parameter, and then
to prove that it can satisfy the requirement of Eq. (\ref{Increment}).
Unfortunately, that is rather difficult because it is impossible to
calculate $S(\phi ,\beta )$ rigorously when $H(c)$ contains an interaction.
As a result, we have to seek approximations. Before doing so, it is
worthwhile to give a brief survey to the self-consistent mean-field theory
due to BCS \cite{BCS2,BCS3}, which can be summarized as follows,
\begin{widetext}
\begin{gather}
H(c)\Longrightarrow H_{\mathrm{MF}}(c)=\sum_{\mathbf{k}}\varepsilon (\mathbf{%
k})(c_{\mathbf{k}\uparrow }^{\dag }c_{\mathbf{k}\uparrow }+c_{-\mathbf{k}%
\downarrow }^{\dag }c_{-\mathbf{k}\downarrow })-\Delta \sum_{\mathbf{k}}(c_{-%
\mathbf{k}\downarrow }c_{\mathbf{k}\uparrow }+c_{\mathbf{k}\uparrow }^{\dag
}c_{-\mathbf{k}\downarrow }^{\dag })+\frac{1}{g}\Delta ^{2},  \label{BCSMF1}
\\
\left\langle F(c)\right\rangle =\mathrm{Tr}\big(F(c)\rho \left( H(c)\right) %
\big)\Longrightarrow \left\langle F(c)\right\rangle =\mathrm{Tr}\big(%
F(c)\rho \left( H_{\mathrm{MF}}(c)\right) \big),  \label{BCSMF2}
\end{gather}%
where%
\begin{equation}
\Delta =g\sum_{\mathbf{k}}\left\langle c_{-\mathbf{k}\downarrow }c_{\mathbf{k%
}\uparrow }\right\rangle =g\sum_{\mathbf{k}}\mathrm{Tr}\big(c_{-\mathbf{k}%
\downarrow }c_{\mathbf{k}\uparrow }\rho \left( H_{\mathrm{MF}}(c)\right) %
\big).  \label{BCSMF3}
\end{equation}%
\end{widetext}
Eq. (\ref{BCSMF1}) and Eq. (\ref{BCSMF3}) constitute a pair of
self-consistent equations to determine both the mean-field Hamiltonian $H_{%
\mathrm{MF}}(c)$ and the energy gap $\Delta $. After $H_{\mathrm{MF}}(c)$ is
so determined, the average of observable $F(c)$ can be evaluated according
to Eq. (\ref{BCSMF2}).

Obviously, if $\Delta \neq 0$, the mean-field Hamiltonian $H_{\mathrm{MF}%
}(c) $ is not invariant under the gauge transformation of Eq. (\ref{Gauge}),
i.e., the gauge symmetry is broken with respect to $H_{\mathrm{MF}}(c)$. It
should be noted that, as shown in Eqs. (\ref{BCSMF2}) and (\ref{BCSMF3}), $%
H_{\mathrm{MF}}(c)$ must be used in place of $H(c)$ when calculating
statistical averages. Otherwise, $\Delta $ becomes zero, there is no Cooper
pair and superconductivity, as has been demonstrated in Eq. (\ref{Zero}).
That is the key point of the BCS mean-field theory. This kind of mean-field
theory is also widely used in studying other phase transitions and SSB,
Bogoliubov theory of a weakly interacting Bose gas being a famous example
\cite{Bogoliubov,Zagrebnov}. Nevertheless, as pointed out by Emch in Sec.
1.1.f of Ref. \cite{Emch}, it contains an inescapable paradox. That is, the
spectrum of the system Hamiltonian $H_{\mathrm{MF}}(c)$ depends on
temperature, which contradicts the fact that the spectrum of an operator is
an invariant property of this operator itself and should not depend on
temperature. Now, for the extended ensemble theory, the normal and
superconducting phases correspond, as stated in Sec. \ref{EET}, to the
symmetric and asymmetric representations of the same system Hamiltonian,
respectively, it excludes such paradox absolutely.

Although the approximation of Eqs. (\ref{BCSMF1}--\ref{BCSMF3}) contains the
paradox, its results are astonishingly in good agreement with the
experiments. Therefore, we need to search a formally identical but
essentially different approximation within the extended ensemble theory so
as to interpret the superconductivity anew.

In language of Green's function (GF), the above approximation can be
translated into the following formalism,
\begin{gather}
\langle \langle c_{\mathbf{k}\uparrow }|c_{\mathbf{k}\uparrow }^{\dag
}\rangle \rangle _{\omega }^{\mathrm{MF}}=\frac{\omega +\varepsilon (\mathbf{%
k})}{\omega ^{2}-\left[ \varepsilon ^{2}(\mathbf{k})+\Delta ^{2}\right] },
\label{GMF1} \\
\langle \langle c_{\mathbf{k}\uparrow }|c_{-\mathbf{k}\downarrow }\rangle
\rangle _{\omega }^{\mathrm{MF}}=\frac{\Delta }{\omega ^{2}-\left[
\varepsilon ^{2}(\mathbf{k})+\Delta ^{2}\right] },  \label{GMF2} \\
\Delta =-g\sum_{\mathbf{k}}\int_{-\infty }^{+\infty }\frac{\mathrm{d}\omega
}{\pi }f(\omega )\mathrm{Im}\langle \langle c_{\mathbf{k}\uparrow }|c_{-%
\mathbf{k}\downarrow }\rangle \rangle _{\omega }^{\mathrm{MF}},  \label{GMF3}
\end{gather}%
where $\langle \langle A|B\rangle \rangle _{\omega }^{\mathrm{MF}}$ denotes
the retarded Green's function defined with respect to $H_{\mathrm{MF}}(c)$.
Now, Eqs. (\ref{GMF2}) and (\ref{GMF3}) constitute a pair of self-consistent
equations to determine the two complex-valued functions, $\langle \langle c_{%
\mathbf{k}\uparrow }|c_{-\mathbf{k}\downarrow }\rangle \rangle _{\omega }^{%
\mathrm{MF}}$ and $\Delta $, in contrast to Eqs. (\ref{BCSMF1}) and (\ref%
{BCSMF3}) where an operator is involved. Therefore, Eqs. (\ref{GMF1}--\ref%
{GMF3}) are formally adaptable to transplanting into the extended ensemble
theory. As is well known, BCS superconductivity can be explained in terms of
these Green's functions \cite{BCS2,BCS3}. We shall reestablish them within
the framework of the extended ensemble theory. Of course, they must be
redefined, with respect to $H^{\prime }(\phi ,c)$ instead of $H_{\mathrm{MF}%
}(c)$.

\subsection{Interpretation of BCS Superconductivity}

Now, we devote ourselves to reestablishing a formalism similar to Eqs. (\ref%
{GMF1}--\ref{GMF3}), and interpreting the superconductivity with the
extended ensemble theory.

Following Eq. (\ref{Sphi}), the entropy of a BCS superconductor can be
written as%
\begin{align}
S(\phi ,\beta )& =\ln \!\Big(\mathrm{Tr}\big(e^{-\beta H(c)}\big)\Big)
\notag \\
& +\beta \sum_{\mathbf{k}}\varepsilon (\mathbf{k})(\overline{d_{\mathbf{k}%
\uparrow }^{\dag }d_{\mathbf{k}\uparrow }}+\overline{d_{-\mathbf{k}%
\downarrow }^{\dag }d_{-\mathbf{k}\downarrow }})  \notag \\
& -\beta g\sum_{\mathbf{k}}\sum_{\mathbf{k}^{\prime }}\overline{d_{\mathbf{k}%
^{\prime }\uparrow }^{\dag }d_{-\mathbf{k}^{\prime }\downarrow }^{\dag }d_{-%
\mathbf{k}\downarrow }d_{\mathbf{k}\uparrow }}\,,  \label{BCSEtr}
\end{align}%
where
\begin{subequations}
\label{dOpr}
\begin{eqnarray}
d_{\mathbf{k\uparrow }} &=&e^{-iD(\phi ,c)}c_{\mathbf{k\uparrow }}e^{iD(\phi
,c)} \\
d_{-\mathbf{k\downarrow }} &=&e^{-iD(\phi ,c)}c_{-\mathbf{k\downarrow }%
}e^{iD(\phi ,c)},
\end{eqnarray}%
and
\end{subequations}
\begin{equation}
\overline{A(c)}\equiv \mathrm{Tr}\big(A(c)\rho \left( H(c)\right) \big).
\end{equation}%
As the usual mean-field approximation, we decouple the last term on the
right-hand side of Eq. (\ref{BCSEtr}) as follows,
\begin{equation}
\overline{d_{\mathbf{k}^{\prime }\uparrow }^{\dag }d_{-\mathbf{k}^{\prime
}\downarrow }^{\dag }\times d_{-\mathbf{k}\downarrow }d_{\mathbf{k}\uparrow }%
}=\overline{d_{\mathbf{k}^{\prime }\uparrow }^{\dag }d_{-\mathbf{k}^{\prime
}\downarrow }^{\dag }}\times \overline{d_{-\mathbf{k}\downarrow }d_{\mathbf{k%
}\uparrow }}\text{\thinspace }.  \label{BCSFact}
\end{equation}%
It results in
\begin{widetext}%
\begin{align}
S(\phi ,\beta )& =\ln \!\Big(\mathrm{Tr}\big(e^{-\beta H(c)}\big)\Big)+\beta
\sum_{\mathbf{k}}\varepsilon (\mathbf{k})\left[ \cos ^{2}(\theta _{\mathbf{k}%
})(\overline{c_{\mathbf{k}\uparrow }^{\dag }c_{\mathbf{k}\uparrow }}+%
\overline{c_{-\mathbf{k}\downarrow }^{\dag }c_{-\mathbf{k}\downarrow }}%
)+\sin ^{2}(\theta _{\mathbf{k}})(\overline{c_{\mathbf{k}\uparrow }c_{%
\mathbf{k}\uparrow }^{\dag }}+\overline{c_{-\mathbf{k}\downarrow }c_{-%
\mathbf{k}\downarrow }^{\dag }})\right]  \notag \\
& -\frac{1}{4}\beta g\sum_{\mathbf{k}^{\prime }}\sin (2\theta _{\mathbf{k}%
^{\prime }})e^{i\varphi _{\mathbf{k}^{\prime }}}(\overline{c_{\mathbf{k}%
^{\prime }\uparrow }^{\dag }c_{\mathbf{k}^{\prime }\uparrow }}-\overline{c_{-%
\mathbf{k}^{\prime }\downarrow }c_{-\mathbf{k}^{\prime }\downarrow }^{\dag }}%
)\times \sum_{\mathbf{k}}\sin (2\theta _{\mathbf{k}})e^{-i\varphi _{\mathbf{k%
}}}(\overline{c_{\mathbf{k}\uparrow }^{\dag }c_{\mathbf{k}\uparrow }}-%
\overline{c_{-\mathbf{k}\downarrow }c_{-\mathbf{k}\downarrow }^{\dag }})%
\text{{\Large ,}}  \label{SupS}
\end{align}%
where Eqs. (\ref{dOpr}), (\ref{Ctrans}), and (\ref{Zero}) have been used.

Substitution of Eq. (\ref{SupS}) into Eq. (\ref{Variation}) gives the
equations of order parameter,%
\begin{gather}
2\varepsilon (\mathbf{k})\sin (2\theta _{\mathbf{k}})+g\cos (2\theta _{%
\mathbf{k}})\sum_{\mathbf{k}^{\prime }}\cos (\varphi _{\mathbf{k}}-\varphi _{%
\mathbf{k}^{\prime }})\sin (2\theta _{\mathbf{k}^{\prime }})(\overline{c_{%
\mathbf{k}^{\prime }\uparrow }^{\dag }c_{\mathbf{k}^{\prime }\uparrow }}-%
\overline{c_{-\mathbf{k}^{\prime }\downarrow }c_{-\mathbf{k}^{\prime
}\downarrow }^{\dag }})=0, \\
\sin (2\theta _{\mathbf{k}})\sum_{\mathbf{k}^{\prime }}\sin (\varphi _{%
\mathbf{k}}-\varphi _{\mathbf{k}^{\prime }})\sin (2\theta _{\mathbf{k}%
^{\prime }})(\overline{c_{\mathbf{k}^{\prime }\uparrow }^{\dag }c_{\mathbf{k}%
^{\prime }\uparrow }}-\overline{c_{-\mathbf{k}^{\prime }\downarrow }c_{-%
\mathbf{k}^{\prime }\downarrow }^{\dag }})=0.
\end{gather}%
\end{widetext}
They have a trivial solution,%
\begin{equation}
\phi =0,  \label{STrv}
\end{equation}%
and a nontrivial solution,%
\begin{equation}
\left\{
\begin{array}{l}
\varphi _{\mathbf{k}}=\varphi _{0}=\mathrm{constant} \\
\varepsilon (\mathbf{k})\sin (2\theta _{\mathbf{k}})-\Lambda \cos (2\theta _{%
\mathbf{k}})=0,%
\end{array}%
\right.  \label{SNTrv}
\end{equation}%
where%
\begin{equation}
\Lambda =-ie^{i\varphi _{0}}g\sum_{\mathbf{k}}\left\langle c_{-\mathbf{k}%
\downarrow }c_{\mathbf{k}\uparrow }\right\rangle .
\end{equation}%
The trivial solution always remains zero, it does not change with
temperature. In contrast, the nontrivial one will depend on temperature, it
may be zero for some temperatures, and nonzero for the others. Without loss
of generality, we shall take $\varphi _{0}=\pi /2$, the nontrivial solution
is then simplified as%
\begin{gather}
\varepsilon (\mathbf{k})\sin (2\theta _{\mathbf{k}})-\Lambda \cos (2\theta _{%
\mathbf{k}})=0,  \label{SupTheta} \\
\Lambda =g\sum_{\mathbf{k}}\left\langle c_{-\mathbf{k}\downarrow }c_{\mathbf{%
k}\uparrow }\right\rangle ,  \label{SupLambda}
\end{gather}%
they are a pair of coupled equations to determine both $\theta _{\mathbf{k}}$
and $\Lambda $.

If $g=0$, Eq. (\ref{SupLambda}) shows that $\Lambda=0$, then Eq. (\ref%
{SupTheta}) comes back to the equation (\ref{IFGOrder}) for the ideal Fermi
gas. Physically, the BCS\ Hamiltonian of Eq. (\ref{BCS}) can describe only
the weak-coupling superconductivity. For the weak-coupling
superconductivity, the coupling strength $g$ is quite small, from Eqs. (\ref%
{SupTheta}) and (\ref{SupLambda}) it follows that $\Lambda$ and $\theta_{%
\mathbf{k}}$ are also small. This makes it easy for us to do further
approximations, as will be seen from below.

In order to solve Eqs. (\ref{SupTheta}) and (\ref{SupLambda}), one needs $%
\left\langle c_{-\mathbf{k}\downarrow }c_{\mathbf{k}\uparrow }\right\rangle $%
, it can be obtained by the corresponding retarded Green's function $\langle
\langle c_{\mathbf{k}\uparrow }|c_{-\mathbf{k}\downarrow }\rangle \rangle
_{\omega }$ defined with respect to $H^{\prime }(\phi ,c)$ \cite{Zubarev}.
As is well known, $\langle \langle c_{\mathbf{k}\uparrow }|c_{-\mathbf{k}%
\downarrow }\rangle \rangle _{\omega }$ satisfies the equation of motion,%
\begin{align}
\omega \langle \langle c_{\mathbf{k}\uparrow }|c_{-\mathbf{k}\downarrow
}\rangle \rangle _{\omega }& =\left\langle \left\{ c_{\mathbf{k}\uparrow
},c_{-\mathbf{k}\downarrow }\right\} \right\rangle  \notag \\
& -\langle \langle c_{\mathbf{k}\uparrow }|\left[ c_{-\mathbf{k}\downarrow
},H^{\prime }(\phi ,c)\right] \rangle \rangle _{\omega },  \label{GFMotion}
\end{align}%
where $\left\{ A,B\right\} $ denotes the anticommutator of $A$ and $B$.
Expanding $H^{\prime }(\phi ,c)$ into the power series of $\phi $,
\begin{widetext}
\begin{equation}
H^{\prime }(\phi ,c)=H(c)+i\left[ D(\phi ,c),H(c)\right] +\frac{i^{2}}{2!}%
\left[ D(\phi ,c),\left[ D(\phi ,c),H(c)\right] \right] +\cdots ,
\end{equation}%
and substituting it into Eq. (\ref{GFMotion}), we have%
\begin{align}
\omega \langle \langle c_{\mathbf{k}\uparrow }|c_{-\mathbf{k}\downarrow
}\rangle \rangle _{\omega }& =\left\langle \left\{ c_{\mathbf{k}\uparrow
},c_{-\mathbf{k}\downarrow }\right\} \right\rangle -\langle \langle c_{%
\mathbf{k}\uparrow }|\left[ c_{-\mathbf{k}\downarrow },H(c)\right] \rangle
\rangle _{\omega }-i\langle \langle c_{\mathbf{k}\uparrow }|\left[ c_{-%
\mathbf{k}\downarrow },\left[ D(\phi ,c),H(c)\right] \right] \rangle \rangle
_{\omega }  \notag \\
& -\frac{i^{2}}{2!}\langle \langle c_{\mathbf{k}\uparrow }|\left[ c_{-%
\mathbf{k}\downarrow },\left[ D(\phi ,c),[D(\phi ,c),H(c)]\right] \right]
\rangle \rangle _{\omega }+\cdots .
\end{align}%
As stated above, $\phi $ is small for the weak-coupling superconductivity,
it is thus rational to maintain only the zeroth-order term,%
\begin{equation}
\omega \langle \langle c_{\mathbf{k}\uparrow }|c_{-\mathbf{k}\downarrow
}\rangle \rangle _{\omega }=-\varepsilon (\mathbf{k})\langle \langle c_{%
\mathbf{k}\uparrow }|c_{-\mathbf{k}\downarrow }\rangle \rangle _{\omega
}+g\langle \langle c_{\mathbf{k}\uparrow }|c_{\mathbf{k}\uparrow }^{\dag
}\sum_{\mathbf{k}^{\prime }}c_{-\mathbf{k}^{\prime }\downarrow }c_{\mathbf{k}%
^{\prime }\uparrow }\rangle \rangle _{\omega }.
\end{equation}%
\end{widetext}
With regard to the second GF on the right-hand side, we do the factorization
as in Eq. (\ref{BCSFact}),
\begin{equation}
g\sum_{\mathbf{k}^{\prime }}\langle c_{-\mathbf{k}^{\prime }\downarrow }c_{%
\mathbf{k}^{\prime }\uparrow }\rangle =g\sum_{\mathbf{k}^{\prime }}\overline{%
d_{-\mathbf{k}^{\prime }\downarrow }d_{\mathbf{k}^{\prime }\uparrow }}%
=\Lambda ,
\end{equation}%
this leads to%
\begin{equation}
\left[ \omega +\varepsilon (\mathbf{k})\right] \langle \langle c_{\mathbf{k}%
\uparrow }|c_{-\mathbf{k}\downarrow }\rangle \rangle _{\omega }=\Lambda
\langle \langle c_{\mathbf{k}\uparrow }|c_{\mathbf{k}\uparrow }^{\dag
}\rangle \rangle _{\omega }.  \label{WGFMotion}
\end{equation}%
The GF on the right-hand side can be obtained by the same procedure as for $%
\langle \langle c_{\mathbf{k}\uparrow }|c_{-\mathbf{k}\downarrow }\rangle
\rangle _{\omega }$,%
\begin{equation}
\left[ \omega -\varepsilon (\mathbf{k})\right] \langle \langle c_{\mathbf{k}%
\uparrow }|c_{\mathbf{k}\uparrow }^{\dag }\rangle \rangle _{\omega
}=1+\Lambda ^{\dagger }\langle \langle c_{\mathbf{k}\uparrow }|c_{-\mathbf{k}%
\downarrow }\rangle \rangle _{\omega },
\end{equation}%
where
\begin{equation}
\Lambda ^{\dagger }=g\sum_{\mathbf{k}}\langle c_{\mathbf{k}\uparrow
}^{\dagger }c_{-\mathbf{k}\downarrow }^{\dagger }\rangle =\Lambda .
\label{DLambda}
\end{equation}%
From Eqs. (\ref{WGFMotion}--\ref{DLambda}), it follows that%
\begin{gather}
\langle \langle c_{\mathbf{k}\uparrow }|c_{\mathbf{k}\uparrow }^{\dag
}\rangle \rangle _{\omega }=\frac{\omega +\varepsilon (\mathbf{k})}{\omega
^{2}-\left[ \varepsilon ^{2}(\mathbf{k})+\Lambda ^{2}\right] },
\label{WGFMotion1} \\
\langle \langle c_{\mathbf{k}\uparrow }|c_{-\mathbf{k}\downarrow }\rangle
\rangle _{\omega }=\frac{\Lambda }{\omega ^{2}-\left[ \varepsilon ^{2}(%
\mathbf{k})+\Lambda ^{2}\right] },  \label{WGFMotion2} \\
\Lambda =-g\sum_{\mathbf{k}}\int_{-\infty }^{+\infty }\frac{\mathrm{d}\omega
}{\pi }f(\omega )\text{\textrm{Im}}\langle \langle c_{\mathbf{k}\uparrow
}|c_{-\mathbf{k}\downarrow }\rangle \rangle _{\omega }.  \label{DLambda1}
\end{gather}%
They form a self-consistent mean-field solution to the Green's functions
responsible for BCS superconductivity within the framework of the extended
ensemble theory. Comparing them with Eqs. (\ref{GMF1}--\ref{GMF3}), one sees
that the present solution is identical to the BCS solution except that the
statistical average is now defined with respect to $H^{\prime }(\phi ,c)$.
Thus far, the BCS mean-field results have been transplanted into the
extended ensemble theory, the paradox removed.

With the help of Eq. (\ref{WGFMotion2}), Eqs. (\ref{DLambda1}) and (\ref%
{SupTheta}) can be simplified as%
\begin{eqnarray}
\Lambda &=&\frac{1}{2}g\Lambda \sum_{\mathbf{k}}\frac{\tanh \left( \frac{1}{2%
}\beta \xi _{\mathbf{k}}\right) }{\xi _{\mathbf{k}}},  \label{GapEq} \\
\theta _{\mathbf{k}} &=&\left\{
\begin{array}{ll}
\frac{1}{2}\arcsin \left( \frac{\Lambda }{\xi _{\mathbf{k}}}\right) ,\text{ }
& \varepsilon (\mathbf{k})\geq 0 \\
-\frac{1}{2}\arcsin \left( \frac{\Lambda }{\xi _{\mathbf{k}}}\right) ,\text{
} & \varepsilon (\mathbf{k})<0,%
\end{array}%
\right.  \label{OrderEq}
\end{eqnarray}%
where%
\begin{equation}
\xi _{\mathbf{k}}=\sqrt{\varepsilon ^{2}(\mathbf{k})+\Lambda ^{2}}.
\end{equation}%
Eq. (\ref{OrderEq}) together with Eq. (\ref{GapEq}) gives the mean-field
solution to $\phi $ ($\phi _{\mathbf{k}}=i\theta _{\mathbf{k}}$), the order
parameter for BCS superconductivity.

Eq. (\ref{GapEq}) is familiar in the BCS mean-field theory \cite{BCS2,BCS3},
it has the nontrivial solution,
\begin{equation}
\Lambda =\left\{
\begin{array}{ll}
\mathrm{zero,}\text{ } & T\geq T_{c} \\
\text{\textrm{nonzero}}\mathrm{,}\text{ } & T<T_{c},%
\end{array}%
\right.  \label{PseudoOrder}
\end{equation}%
where $T_{c}$ is determined by the equation,
\begin{equation}
1=\frac{1}{2}g\sum_{\mathbf{k}}\frac{\tanh \left( \frac{1}{2}\beta
_{c}\varepsilon (\mathbf{k})\right) }{\varepsilon (\mathbf{k})},
\end{equation}%
with $\beta _{c}=1/k_{B}T_{c}$. From Eqs. (\ref{PseudoOrder}) and (\ref%
{OrderEq}), we find
\begin{equation}
\phi _{\mathbf{k}}=\left\{
\begin{array}{ll}
\mathrm{zero,}\text{ } & T\geq T_{c} \\
\text{\textrm{nonzero}}\mathrm{,}\text{ } & T<T_{c}.%
\end{array}%
\right.  \label{RealOrder}
\end{equation}%
Since $\phi $ is the order parameter of the system, Eq. (\ref{RealOrder})
indicates that $T_{c}$ is the transition temperature for BCS
superconductivity.

Compared with Eq. (\ref{RealOrder}), Eq. (\ref{PseudoOrder}) shows that the
energy gap $\Lambda $ behaviors with temperature just like an order
parameter. However, this behavior of the energy gap is not a property of
itself, but derives from the property of the order parameter, as can be seen
from the equation,%
\begin{equation}
\Lambda =g\sum_{\mathbf{k}}\mathrm{Tr}\big(c_{-\mathbf{k}\downarrow }c_{%
\mathbf{k}\uparrow }\rho \left( H^{\prime }(\phi ,c)\right) \big)=\left\{
\begin{array}{ll}
\mathrm{zero,}\text{ } & \phi =0 \\
\text{\textrm{nonzero}}\mathrm{,}\text{ } & \phi \neq 0,%
\end{array}%
\right.
\end{equation}%
which demonstrates that the forming of Cooper pairs is a direct consequence
of the gauge-symmetry breaking caused by the internal spontaneous field $%
\phi $. In the extended ensemble theory, the order parameter is a physical
quantity more fundamental than the energy gap.\ This picture is
significantly distinct from the BCS mean-field theory where the energy gap
is the most fundamental quantity that is responsible for the
superconductivity.

In sum, Eqs. (\ref{WGFMotion1}), (\ref{WGFMotion2}), (\ref{GapEq}), and (\ref%
{OrderEq}) constitute a self-consistent mean-field theory for BCS
superconductivity within the framework of the extended ensemble theory, its
nontrivial solution reproduces the BCS mean-field results on the
conventional low-$T_{c}$ superconductivity.

We are now confronted with the important task to prove the stability of this
nontrivial solution. By expanding the $S(\phi ,\beta )$ of Eq. (\ref{SupS})
into a Taylor or Volterra series in the neighborhood of the nontrivial
solution $\phi _{0}$, one finds
\begin{equation}
\Delta S=S(\phi ,\beta )-S(\phi _{0},\beta )=\delta ^{2}S+\delta
^{3}S+\delta ^{4}S+\cdots ,  \label{DeltaSBCS}
\end{equation}%
where $\delta ^{n}S$ represents the $n$th power term of the expansion of $%
S(\phi ,\beta )$, or the $n$th variation of $S$. Specifically, the second
variation $\delta ^{2}S$ has the form,
\begin{widetext}
\begin{align}
\delta ^{2}S& =\frac{1}{8}\beta g\Lambda ^{2}\sum_{\mathbf{k}}\sum_{\mathbf{k%
}^{\prime }}\frac{\tanh (\frac{1}{2}\beta \xi _{\mathbf{k}})}{\xi _{\mathbf{k%
}}}\frac{\tanh (\frac{1}{2}\beta \xi _{\mathbf{k}^{\prime }})}{\xi _{\mathbf{%
k}^{\prime }}}\left( \delta \varphi _{\mathbf{k}}-\delta \varphi _{\mathbf{k}%
^{\prime }}\right) ^{2}+2\beta \sum_{\mathbf{k}_{1}}\xi _{\mathbf{k}%
_{1}}\tanh (\frac{1}{2}\beta \xi _{\mathbf{k}_{1}})\left( \delta \theta _{%
\mathbf{k}_{1}}\right) ^{2}  \notag \\
& +\beta \left\{ \sum_{\mathbf{k}_{2}}2\xi _{\mathbf{k}_{2}}\tanh (\frac{1}{2%
}\beta \xi _{\mathbf{k}_{2}})\left( \delta \theta _{\mathbf{k}_{2}}\right)
^{2}-g\sum_{\mathbf{k}_{2}}\sum_{\mathbf{k}_{2}^{\prime }}\varepsilon (%
\mathbf{k}_{2})\frac{\tanh (\frac{1}{2}\beta \xi _{\mathbf{k}_{2}})}{\xi _{%
\mathbf{k}_{2}}}\varepsilon (\mathbf{k}_{2}^{\prime })\frac{\tanh (\frac{1}{2%
}\beta \xi _{\mathbf{k}_{2}^{\prime }})}{\xi _{\mathbf{k}_{2}^{\prime }}}%
\delta \theta _{\mathbf{k}_{2}}\delta \theta _{\mathbf{k}_{2}^{\prime
}}\right\} ,  \label{DeltaS2}
\end{align}%
\end{widetext}
where $\delta \varphi _{\mathbf{k}}$ and $\delta \theta _{\mathbf{k}}$
represent the variations of $\varphi _{\mathbf{k}}$ and $\theta _{\mathbf{k}%
} $ from the nontrivial solution $\phi _{0}$, respectively. Here, the set of
$\mathbf{k}$ has been separated into two subsets: the subset of $\mathbf{k}%
_{1}$ and the subset of $\mathbf{k}_{2}$, where $\mathbf{k}_{1}$ and $%
\mathbf{k}_{2}$ satisfy $\varepsilon (\mathbf{k}_{1})=0$ and $\varepsilon (%
\mathbf{k}_{2})\neq 0$, respectively. All the contributions from the subset
of $\mathbf{k}_{2}$ are included by the term within the curly brackets.

Let us prove that this term is definitely positive, it can be done by
proving that the eigenvalues of the matrix $M$,%
\begin{align}
M_{\mathbf{k}_{2}\mathbf{k}_{2}^{^{\prime }}}& =2\xi _{\mathbf{k}_{2}}\tanh (%
\frac{1}{2}\beta \xi _{\mathbf{k}_{2}})\delta _{\mathbf{k_{2}k}%
_{2}^{^{\prime }}}-g\varepsilon (\mathbf{k}_{2})\varepsilon (\mathbf{k}%
_{2}^{\prime })  \notag \\
& \times \frac{\tanh (\frac{1}{2}\beta \xi _{\mathbf{k}_{2}})}{\xi _{\mathbf{%
k}_{2}}}\frac{\tanh (\frac{1}{2}\beta \xi _{\mathbf{k}_{2}^{\prime }})}{\xi
_{\mathbf{k}_{2}^{\prime }}},  \label{Mkk}
\end{align}%
are all positive. The eigenvalue function for $M$ is,
\begin{equation}
|\Omega (\lambda )|=0,  \label{EigenValueEq}
\end{equation}%
where%
\begin{equation}
\Omega _{\mathbf{k}_{2}\mathbf{k}_{2}^{^{\prime }}}(\lambda )=\lambda \delta
_{\mathbf{k_{2}k}_{2}^{^{\prime }}}-M_{\mathbf{k}_{2}\mathbf{k}%
_{2}^{^{\prime }}}.
\end{equation}%
Observe that the second term on the right-hand side of Eq. (\ref{Mkk}) is a
product of the two factors corresponding to the row $\mathbf{k}_{2}$ and the
column $\mathbf{k}_{2}^{^{\prime }}$ respectively, we can reduce Eq. (\ref%
{EigenValueEq}) into
\begin{equation}
|\widetilde{\Omega }(\lambda )|=0,  \label{EigenValueEq1}
\end{equation}%
where
\begin{equation}
\widetilde{\Omega }_{\mathbf{k}_{2}\mathbf{k}_{2}^{^{\prime }}}(\lambda )=%
\frac{\lambda -2\xi _{\mathbf{k}_{2}}\tanh (\frac{1}{2}\beta \xi _{\mathbf{k}%
_{2}})}{g\left[ \varepsilon (\mathbf{k}_{2})\tanh (\frac{1}{2}\beta \xi _{%
\mathbf{k}_{2}})/\xi _{\mathbf{k}_{2}}\right] ^{2}}\delta _{\mathbf{k_{2}k}%
_{2}^{^{\prime }}}+1.
\end{equation}%
Obviously,%
\begin{equation}
\lambda =2\xi _{\mathbf{k}_{2}}\tanh (\frac{1}{2}\beta \xi _{\mathbf{k}%
_{2}})>0  \label{LambdaRoot}
\end{equation}%
is an eigenvalue of Eq. (\ref{EigenValueEq1}) because at least two rows ($%
\pm \mathbf{k}_{2}$) of the determinant $|\widetilde{\Omega }(\lambda )|$
are identical. Apart from those eigenvalues, there are possibly other ones,
they satisfy the equation,
\begin{equation}
1=g\sum_{\mathbf{k}_{2}}\frac{\left[ \varepsilon (\mathbf{k}_{2})\tanh (%
\frac{1}{2}\beta \xi _{\mathbf{k}_{2}})/\xi _{\mathbf{k}_{2}}\right] ^{2}}{%
2\xi _{\mathbf{k}_{2}}\tanh (\frac{1}{2}\beta \xi _{\mathbf{k}_{2}})-\lambda
}.  \label{EigenValueEq2}
\end{equation}%
To derive this equation from Eq. (\ref{EigenValueEq1}), it is sufficient to
heed that every element of $\widetilde{\Omega }(\lambda )$ contains the
number $1$. If $\lambda \leq 0$, one has%
\begin{equation}
g\sum_{\mathbf{k}_{2}}\frac{\left[ \varepsilon (\mathbf{k}_{2})\tanh (\frac{1%
}{2}\beta \xi _{\mathbf{k}_{2}})/\xi _{\mathbf{k}_{2}}\right] ^{2}}{2\xi _{%
\mathbf{k}_{2}}\tanh (\frac{1}{2}\beta \xi _{\mathbf{k}_{2}})-\lambda }<%
\frac{1}{2}g\sum\limits_{\mathbf{k}}\frac{\tanh (\frac{1}{2}\beta \xi _{%
\mathbf{k}})}{\xi _{\mathbf{k}}}.  \label{Inequ}
\end{equation}%
Eqs. (\ref{EigenValueEq2}) and (\ref{Inequ}) lead us to%
\begin{equation}
\frac{1}{2}g\sum\limits_{\mathbf{k}}\frac{\tanh \left( \frac{1}{2}\beta \xi
_{\mathbf{k}}\right) }{\xi _{\mathbf{k}}}>1.
\end{equation}%
This inequality contradicts the result of Eq. (\ref{GapEq}),%
\begin{equation}
\frac{1}{2}g\sum\limits_{\mathbf{k}}\frac{\tanh (\frac{1}{2}\beta \xi _{%
\mathbf{k}})}{\xi _{\mathbf{k}}}\leq 1.
\end{equation}%
That is to say, the root of Eq. (\ref{EigenValueEq2}) must be greater than
zero if it has any. Combination of this result with Eq. (\ref{LambdaRoot})
demonstrates that all the eigenvalues of the matrix $M$ are definitely
positive, which ends our proof for the positivity of the term within the
curly brackets of Eq. (\ref{DeltaS2}).

This positivity implies that%
\begin{equation}
\delta ^{2}S\geq 0.
\end{equation}%
When $\delta ^{2}S>0$, we know from Eq. (\ref{DeltaSBCS}) that $\Delta S>0$,
it meets the requirement of Eq. (\ref{Increment}). However, if $\delta
^{2}S=0$, we must generally examine $\delta ^{3}S$, $\delta ^{4}S$, or even
more higher variations of $S$ to check whether $\Delta S\geq 0$, which is
obviously complicated to handle. Fortunately, we need not do that in the
case considered now. When $\delta ^{2}S=0$, Eq. (\ref{DeltaS2}) shows that%
\begin{equation}
\delta \theta _{\mathbf{k}_{2}}=0,\text{ }\mathrm{if}\text{ }T\geq T_{c};%
\text{ }\mathrm{and}\text{ }\left\{
\begin{array}{l}
\delta \varphi _{\mathbf{k}}=\delta \varphi _{\mathbf{k}^{\prime }} \\
\delta \theta _{\mathbf{k}_{1}}=0 \\
\delta \theta _{\mathbf{k}_{2}}=0,%
\end{array}%
\right. \text{ }\mathrm{if}\text{ }T<T_{c}.
\end{equation}%
With them, one can easily verify that $S(\phi ,\beta )=S(\phi _{0},\beta )$.
Namely, $\Delta S=0$ if $\delta ^{2}S=0$.

To sum up, $\Delta S>0$ if $\delta ^{2}S>0$, and $\Delta S=0$ if $\delta
^{2}S=0$. Since $\delta ^{2}S\geq 0$, we conclude that the requirement $%
\Delta S\geq 0$ is satisfied, the nontrivial solution is stable.%
\begin{widetext}%

As regards the trivial solution, $\phi =0$ ($\Lambda =0$), it is identical
to the nontrivial one when $T\geq T_{c}$, and thus is stable at $T\geq T_{c}$%
. If $T<T_{c}$, we find
\begin{equation}
\delta ^{2}S=\beta \left\{ \sum_{\mathbf{k}_{2}}2\varepsilon (\mathbf{k}%
_{2})\tanh \left( \frac{1}{2}\beta \varepsilon (\mathbf{k}_{2})\right)
\left( \delta \theta _{\mathbf{k}_{2}}\right) ^{2}-g\sum_{\mathbf{k}%
_{2}}\sum_{\mathbf{k}_{2}^{\prime }}\tanh \left( \frac{1}{2}\beta
\varepsilon (\mathbf{k}_{2})\right) \tanh \left( \frac{1}{2}\beta
\varepsilon (\mathbf{k}_{2}^{\prime })\right) \delta \theta _{\mathbf{k}%
_{2}}\delta \theta _{\mathbf{k}_{2}^{\prime }}\right\} .  \label{NDS2}
\end{equation}%
\end{widetext}
Accordingly, it has positive eigenvalues,%
\begin{equation}
\lambda =2\varepsilon (\mathbf{k}_{2})\tanh \left( \frac{1}{2}\beta
\varepsilon (\mathbf{k}_{2})\right) >0,
\end{equation}%
In addition, we also have
\begin{equation}
1=g\sum_{\mathbf{k}_{2}}\frac{\left[ \tanh \left( \frac{1}{2}\beta
\varepsilon (\mathbf{k}_{2})\right) \right] ^{2}}{2\varepsilon (\mathbf{k}%
_{2})\tanh \left( \frac{1}{2}\beta \varepsilon (\mathbf{k}_{2})\right)
-\lambda }.
\end{equation}%
In contrast to Eq. (\ref{EigenValueEq2}), this equation has a negative root
of $\lambda $ when $T<T_{c}$. To see it clearly, let us consider, for
instance, the zero temperature case,%
\begin{equation}
1=2g\sum_{\varepsilon (\mathbf{k})>0}\frac{1}{2\varepsilon (\mathbf{k}%
)-\lambda }=2g\mathcal{N}(0)\int_{0}^{\hbar \omega _{D}}\mathrm{d}%
\varepsilon \frac{1}{2\varepsilon -\lambda },
\end{equation}%
where, as usual, $\mathcal{N}(0)$ denotes the density of states at the Fermi
level, and $\omega _{D}$ the Debye frequency. Apparently, it has the
solution,%
\begin{equation}
\lambda =-\frac{2\hbar \omega _{D}}{\exp (\frac{1}{g\mathcal{N}(0)})-1}<0.
\end{equation}%
The existence of both the positive and negative eigenvalues implies that the
trivial solution is a saddle point at $T<T_{c}$. Therefore, the term within
the curly brackets of Eq. (\ref{NDS2}) is indefinite, that is, it can change
in sign. As a consequence, the trivial solution, or rather the normal phase,
is unstable at $T<T_{c}$. That is just the Cooper instability \cite%
{BCS1,BCS2,BCS3}.

Of the two solutions, the nontrivial is the only one that is stable within
the whole temperature range.

With the nontrivial solution obtained, other physical quantities can be
easily expressed and calculated in terms of the two Green's functions $%
\langle\langle c_{\mathbf{k}\uparrow}|c_{\mathbf{k}\uparrow}^{\dag}\rangle%
\rangle_{\omega}$ and $\langle\langle c_{\mathbf{k}\uparrow}|c_{-\mathbf{k}%
\downarrow}\rangle\rangle_{\omega}$, as was done in Refs. \cite{BCS3}, \cite%
{Rickayzen} and \cite{Ambegaokar}, with the results unchanged and the same
as the BCS theory.

So far, BCS superconductivity has been interpreted within the extended
ensemble theory. In this interpretation, the order parameter for BCS
superconductivity evolves with temperature according to the principle of
least entropy. At high temperatures ($T\geq T_{c}$), it is zero and stable,
the system Hamiltonian realizes the representation with perfect gauge
symmetry, the resulting phase is normal and disordered. As temperature
decreases and goes below the critical temperature ($T<T_{c}$), the zero
solution becomes unstable, instead, there arises a new stable solution of
the order parameter, it is nonzero. At the same time, the system Hamiltonian
transforms into a new representation with gauge symmetry broken, the
electrons are hence formed into Cooper pairs, and the resulting phase gets
superconducting. Thus and so, the extended ensemble theory answers why, when
and how the electrons can coagulate to form Cooper pairs and
superconductivity, the problem posed by Born and Fucks \cite{Born}.

Lastly, it should be noted that the BCS and Landau theories are unified into
a single formalism within the framework of the extended ensemble theory.

\section{Bose-Einstein Condensation \label{BECIdeal}}

After the discussion of BCS superconductivity, we proceed now to study
Bose-Einstein condensation (BEC). We shall concentrate on three systems: the
ideal Bose gas, the photon gas in a black body, and the ideal phonon gas in
a solid body. They are the most important cases of Bose systems: the first
is a system with the conservation of particles, but the other two are not;
the second belongs to gauge fields whereas the third does not. In addition,
we shall also discuss the quantization of Dirac field from a standpoint of
the extended ensemble theory.

\subsection{The Ideal Bose Gas}

Let us begin with the ideal Bose gas. Its Hamiltonian reads as follows,
\begin{equation}
H(b)=\sum_{\mathbf{k}}(\epsilon _{\mathbf{k}}-\mu )b_{\mathbf{k}}^{\dag }b_{%
\mathbf{k}},  \label{Hibg}
\end{equation}%
where $b_{\mathbf{k}}$ ($b_{\mathbf{k}}^{\dag }$) is the annihilation
(creation) operator for the bosons with a momentum $\hbar \mathbf{k}$, and $%
\epsilon _{\mathbf{k}}=\hbar ^{2}\mathbf{k}^{2}/\left( 2m\right) $ the
single-particle energy, and $\mu $ the chemical potential. This Hamiltonian
is gauge invariant,%
\begin{equation}
G(\vartheta ,b)H(b)G^{\dag }(\vartheta ,b)=H(b),
\end{equation}%
where
\begin{equation}
G(\vartheta ,b)=e^{-i\vartheta \sum_{\mathbf{k}}b_{\mathbf{k}}^{\dag }b_{%
\mathbf{k}}},\text{ }\vartheta \in \lbrack 0,2\pi ).
\end{equation}%
The invariance has an important consequence,
\begin{equation}
\left\langle b_{\mathbf{k}}\right\rangle =\mathrm{Tr}\big(b_{\mathbf{k}}\rho
(H(b))\big)=0.  \label{BECGauge}
\end{equation}%
It signifies that the condensation amplitude will be zero forever if the
gauge symmetry does not break down.

In order to examine whether the gauge symmetry can break down spontaneously
or not, one needs, following the hypotheses in Sec. \ref{EET}, to consider
the phase-transition operator,
\begin{equation}
D(\eta,b)=\sum_{\mathbf{k}}(\eta_{\mathbf{k}}^{\dag}b_{\mathbf{k}}+\eta_{%
\mathbf{k}}b_{\mathbf{k}}^{\dag}),  \label{BECPTO}
\end{equation}
where the internal field $\eta$ is the order parameter for BEC. The reason
for introducing Eq. (\ref{BECPTO}) can be seen from comparing Eqs. (\ref%
{BECGauge}) and (\ref{BECPTO}) with Eqs. (\ref{Zero}) and (\ref{Dfactor})
respectively.

We shall consider first the finite volume case where $V$ and $N$ are both
finite, and then the thermodynamic limit case where $V\rightarrow+\infty$,
and $N\rightarrow+\infty$, but $N/V$ is finite. As will be seen, the two
cases are significantly different.

\subsubsection{The Finite Volume Case}

In this case, the wave vector $\mathbf{k}$ is discrete. The entropy of the
system can be written as
\begin{equation}
S(\eta,\beta)=S(0,\beta)+\beta\sum_{\mathbf{k}}(\epsilon_{\mathbf{k}}-\mu
)\eta_{\mathbf{k}}^{\dag}\eta_{\mathbf{k}},  \label{BECS}
\end{equation}
where
\begin{equation}
S(0,\beta)=-\mathrm{Tr}\big(\ln\left( \rho(H(b))\right) \rho(H(b))\big)\!
\end{equation}
is independent of the order parameter $\eta$.

According to Eqs. (\ref{Variation}) and (\ref{Increment}), we have
\begin{gather}
\frac{\partial S}{\partial \eta _{\mathbf{k}}}=0,  \label{OrderParam0} \\
\Delta S=\beta \sum_{\mathbf{k}}(\epsilon _{\mathbf{k}}-\mu )\delta \eta _{%
\mathbf{k}}^{\dag }\delta \eta _{\mathbf{k}}\geq 0,  \label{BECDelta}
\end{gather}%
where $\delta \eta _{\mathbf{k}}$ represents the variation of $\eta _{%
\mathbf{k}}$ from the solution given by Eq. (\ref{OrderParam0}). Eq. (\ref%
{BECDelta}) shows that the chemical potential $\mu $ must satisfy the
following condition,%
\begin{equation}
\mu \leq \epsilon _{\mathbf{k}}.
\end{equation}%
As $\epsilon _{\mathbf{k}}\geq 0$, this condition is equivalent to
\begin{equation}
\mu \leq 0.  \label{StableCond}
\end{equation}%
This inequality is the physical condition for the stability of the system.

In combination with Eq. (\ref{BECS}), Eq. (\ref{OrderParam0}) gives us the
equation of order parameter,
\begin{equation}
(\epsilon _{\mathbf{k}}-\mu )\eta _{\mathbf{k}}=0.
\end{equation}%
It has two solutions: the trivial one,%
\begin{equation}
\eta _{\mathbf{k}}=0,
\end{equation}%
and the nontrivial one,%
\begin{equation}
\eta _{\mathbf{k}}=\eta _{0}\delta _{\mathbf{k,}0},\text{ }\mathrm{only}%
\text{ }\mathrm{if}\text{ }\mu =0.
\end{equation}

For the nontrivial solution, $\mu =0$, it causes a contradiction,%
\begin{eqnarray}
\eta _{0}^{+}\eta _{0} &=&N-\sum_{\mathbf{k}}\mathrm{Tr}\big(b_{\mathbf{k}%
}^{\dag }b_{\mathbf{k}}\rho (H(b))\big)  \notag \\
&=&N-\sum_{\mathbf{k}}\frac{1}{e^{\beta \epsilon _{\mathbf{k}}}-1},
\label{Lesb0}
\end{eqnarray}%
where the left-hand side can not be negative, but the right-hand side is
negative because the term of $\mathbf{k}=0$ is infinite. Therefore, the
nontrivial solution must be discarded.

With regard to the trivial solution, it corresponds to the normal phase of
the system. It does not cause any contradiction,%
\begin{equation}
N=\sum_{\mathbf{k}}\mathrm{Tr}\big(b_{\mathbf{k}}^{\dag }b_{\mathbf{k}}\rho
(H(b))\big)=\sum_{\mathbf{k}}\frac{1}{e^{\beta \left( \epsilon _{\mathbf{k}%
}-\mu \right) }-1},
\end{equation}%
this equation can be satisfied at any temperature. In fact, it is just the
equation for the chemical potential $\mu $, its solution will give $\mu =\mu
(T)$. Because $N$ is finite, $\mu $ will be less than zero irrespective of
temperature, i.e.,%
\begin{equation}
\mu (T)<0,\text{ for }T>0.
\end{equation}%
It implies that the normal phase of the system is always stable,%
\begin{equation}
\Delta S>0.
\end{equation}%
This proves that there is no phase transition in the finite volume case, the
system will stay in its normal phase forever.

\subsubsection{The Thermodynamic Limit Case \label{TLC}}

In this case, $V\rightarrow +\infty $, we should use the density of entropy
instead of the entropy itself,
\begin{eqnarray}
\hspace{-0.3in}s(\eta ,\beta ) &=&\lim_{V\rightarrow +\infty }\frac{S(\eta
,\beta )}{V}  \notag \\
&=&s(0,\beta )+\beta \lim_{V\rightarrow +\infty }\frac{1}{V}\sum_{\mathbf{k}%
}(\epsilon _{\mathbf{k}}-\mu )\eta _{\mathbf{k}}^{\dagger }\eta _{\mathbf{k}%
},  \label{sdensity}
\end{eqnarray}%
where
\begin{equation}
s(0,\beta )=-\lim_{V\rightarrow +\infty }\frac{1}{V}\mathrm{Tr}\big(\ln
\left( \rho (H(b))\right) \rho (H(b))\big)
\end{equation}%
is a term irrelevant to $\eta $. The limit $V\rightarrow +\infty $ can be
achieved as follows.

As usual, suppose that the system is box normalized, i.e.,
\begin{equation}
k_{x(y,z)}=n_{x(y,z)}\frac{2\pi }{l_{x(y,z)}},\text{\ }n_{x(y,z)}\in
\mathbb{Z}
,  \label{Quantum}
\end{equation}%
where $%
\mathbb{Z}
$ denotes the set of integers, and $l_{x}$, $l_{y}$, and $l_{z}$ the
dimensions of the box along the $x$, $y$ and $z$ directions respectively ($%
V=l_{x}l_{y}l_{z}$). As is well known, each $\mathbf{k}$ corresponds to an
eigenfunction, all the eigenfunctions constitute a Hilbert space. Here, the
Hilbert space is a complete space of integrable functions, with the inner
product defined as a Lebesgue integral. It should be pointed out that the
inner product must be defined as a Lebesgue rather than Riemann integral
because the former can ensure the completeness of an integrable function
space whereas the latter can not \cite{Hewitt}. This implies that all the
problems relevant to a Hilbert space of integrable functions must be treated
with Lebesgue theory. On the other hand, Lebesgue measure and integration is
itself the mathematical foundation of probability theory \cite%
{Kolmogorov,Feller} and statistical physics \cite{Georgii}. We shall
therefore perform the thermodynamic limit according to the Lebesgue theory
of integration.

Formally, Lebesgue integration includes both the sum over discrete numbers
and the integral on a continuous region. When $V$ is finite, a sum over $%
\mathbf{k}$ is a Lebesgue integral of discrete form. It will transform into
a Lebesgue integral of continuous form as $V\rightarrow +\infty $. To show
the transformation, let us consider, e.g., the sum over $\mathbf{k}$ in Eq. (%
\ref{sdensity}). In the first place, we write it into an explicit form of
Lebesgue integral,
\begin{widetext}
\begin{equation}
\frac{1}{V}\sum_{\mathbf{k}}(\epsilon _{\mathbf{k}}-\mu )\eta _{\mathbf{k}%
}^{\dagger }\eta _{\mathbf{k}}=\frac{1}{\left( 2\pi \right) ^{3}}\mathbf{%
\sum_{\mathbf{k}}}(\epsilon _{\mathbf{k}}-\mu )\eta _{\mathbf{k}}^{\dag
}\eta _{\mathbf{k}}m(A_{\mathbf{k}}\cap
\mathbb{R}
^{3})\mathbf{=}\frac{1}{\left( 2\pi \right) ^{3}}\int\limits_{E}h(\mathbf{p})%
\mathrm{d}\mathbf{p},  \label{LI}
\end{equation}%
\end{widetext}
where%
\begin{gather}
\mathbb{R}
^{3}=\left( -\infty ,+\infty \right) \times \left( -\infty ,+\infty \right)
\times \left( -\infty ,+\infty \right) , \\
m(A_{\mathbf{k}}\cap
\mathbb{R}
^{3})=\frac{\left( 2\pi \right) ^{3}}{V}, \\
A_{\mathbf{k}}=[k_{x}\mathbf{,}k_{x}\mathbf{+}\frac{2\pi }{l_{x}})\times
\lbrack k_{y}\mathbf{,}k_{y}\mathbf{+}\frac{2\pi }{l_{y}})\times \lbrack
k_{z}\mathbf{,}k_{z}\mathbf{+}\frac{2\pi }{l_{z}}),  \label{ASET} \\
h(\mathbf{p})=\mathbf{\sum_{\mathbf{k}}}(\epsilon _{\mathbf{k}}-\mu )\eta _{%
\mathbf{k}}^{\dag }\eta _{\mathbf{k}}\chi _{A_{\mathbf{k}}}(\mathbf{p}),
\label{Stepf} \\
\chi _{A_{\mathbf{k}}}(\mathbf{p})=\left\{
\begin{array}{ll}
1, & \mathbf{p}\in A_{\mathbf{k}} \\
0, & \mathbf{p}\notin A_{\mathbf{k}}.%
\end{array}%
\right.  \label{Characterf}
\end{gather}%
As usual, $\chi _{A_{\mathbf{k}}}(\mathbf{p})$ represents the characteristic
function of the measurable set $A_{\mathbf{k}}$, and $m(A_{\mathbf{k}}\cap
\mathbb{R}
^{3})$ the Lebesgue measure of the set $A_{\mathbf{k}}$. From Eqs. (\ref%
{ASET}--\ref{Characterf}), it follows that
\begin{eqnarray}
\lim_{V\rightarrow +\infty }h(\mathbf{p}) &=&(\epsilon _{\mathbf{p}}-\mu
)\eta _{\mathbf{p}}^{\dag }\eta _{\mathbf{p}}  \notag \\
&\equiv &[\epsilon (\mathbf{p})-\mu ]\eta ^{\dag }(\mathbf{p})\eta (\mathbf{p%
}).  \label{VLimit}
\end{eqnarray}%
Now, in the second place, let us take the thermodynamic limit on both the
sides of Eq. (\ref{LI}),
\begin{widetext}%
\begin{equation}
\lim_{V\rightarrow +\infty }\frac{1}{V}\sum_{\mathbf{k}}(\epsilon _{\mathbf{k%
}}-\mu )\eta _{\mathbf{k}}^{\dagger }\eta _{\mathbf{k}}=\frac{1}{\left( 2\pi
\right) ^{3}}\lim_{V\rightarrow +\infty }\int\limits_{%
\mathbb{R}
^{3}}h(\mathbf{p})\mathrm{d}\mathbf{p}=\frac{1}{\left( 2\pi \right) ^{3}}%
\int\limits_{%
\mathbb{R}
^{3}}\lim_{V\rightarrow +\infty }h(\mathbf{p})\mathrm{d}\mathbf{p}=\frac{1}{%
\left( 2\pi \right) ^{3}}\int\limits_{%
\mathbb{R}
^{3}}[\epsilon (\mathbf{k})-\mu ]\eta ^{\dag }(\mathbf{k})\eta (\mathbf{k})%
\mathrm{d}\mathbf{k}.
\end{equation}%
\end{widetext}
In such a natural way, a discrete sum over $\mathbf{k}$ will transform into
a continuous Lebesgue integral in the thermodynamic limit.

With the help of the above equation, one arrives at%
\begin{gather}
\left[ \epsilon (\mathbf{k})-\mu \right] \eta (\mathbf{k})=0,
\label{BECCOrder} \\
\Delta s=\beta \frac{1}{\left( 2\pi \right) ^{3}}\int\limits_{%
\mathbb{R}
^{3}}[\epsilon (\mathbf{k})-\mu ]\delta \eta ^{\dag }(\mathbf{k})\delta \eta
(\mathbf{k})\mathrm{d}\mathbf{k}\geq 0.  \label{BECCDelta}
\end{gather}%
The second equation means that the chemical potential must be less than or
equal to zero,%
\begin{equation}
\mu \leq 0,  \label{BECMu}
\end{equation}%
it is the condition for the stability of the system. The first equation is
equivalent to
\begin{equation}
\lbrack \epsilon (\mathbf{k})-\mu ]\eta ^{\dag }(\mathbf{k})\eta (\mathbf{k}%
)=0,  \label{Eta}
\end{equation}%
which can be easily verified with respect to the two cases: $\eta ^{\dag }(%
\mathbf{k})=0$ and $\eta ^{\dag }(\mathbf{k})\neq 0$. Eq. (\ref{Eta}) has
two solutions: the trivial one,%
\begin{equation}
\eta (\mathbf{k})=0,
\end{equation}%
and the nontrivial one,%
\begin{equation}
\eta (\mathbf{k})=\xi \sqrt{\delta (\mathbf{k})},\text{ only if }\mu =0,
\end{equation}%
where $\xi $ is a complex which does not depend on $\mathbf{k}$, and $\delta
(\mathbf{k})$ the Dirac $\delta $ function.

For the trivial solution, the system is in its normal phase, this phase can
exist only at high temperatures, which can be easily deduced from the
equation of chemical potential,%
\begin{eqnarray}
n &\equiv &\lim_{V\rightarrow +\infty }\frac{N}{V}  \notag \\
&=&\lim_{V\rightarrow +\infty }\frac{1}{V}\sum_{\mathbf{k}}\mathrm{Tr}\big(%
b_{\mathbf{k}}^{\dag }b_{\mathbf{k}}\rho (H(b))\big)  \notag \\
&=&\frac{1}{\left( 2\pi \right) ^{3}}\int\limits_{%
\mathbb{R}
^{3}}\frac{1}{e^{\beta (\epsilon \left( \mathbf{k}\right) -\mu )}-1}\mathrm{d%
}\mathbf{k}  \notag \\
&=&\frac{(2m)^{3/2}}{4\pi ^{2}\hbar ^{3}}\int_{0^{+}}^{+\infty }\mathrm{d}%
\epsilon \frac{\epsilon ^{1/2}}{e^{\beta (\epsilon -\mu )}-1},  \label{Lesb1}
\end{eqnarray}%
where we have reduced the Lebesgue integral into an improper Riemann
integral in the end. Because, as shown by Eq. (\ref{BECMu}), $\mu $ must be
less than or equal to zero, the above equation can not hold if $T<T_{c}$
where $T_{c}$ is determined by
\begin{equation}
n=\frac{(2m)^{3/2}}{4\pi ^{2}\hbar ^{3}}\int_{0^{+}}^{+\infty }\mathrm{d}%
\epsilon \frac{\epsilon ^{1/2}}{e^{\beta _{c}\epsilon }-1}.  \label{Lesb2}
\end{equation}%
As is well known, this gives%
\begin{equation}
T_{c}=\frac{2\pi \hbar ^{2}}{mk_{B}}\left( \frac{n}{\zeta (3/2}\right)
^{2/3},
\end{equation}%
where $\zeta (x)$ denotes the Riemann zeta function.

For the nontrivial solution, $\mu $ must be zero, the order parameter is
determined by%
\begin{eqnarray}
n &=&\frac{1}{\left( 2\pi \right) ^{3}}\int\limits_{%
\mathbb{R}
^{3}}\eta ^{\dagger }(\mathbf{k})\eta (\mathbf{k})\mathrm{d}\mathbf{k+}\frac{%
1}{\left( 2\pi \right) ^{3}}\int\limits_{%
\mathbb{R}
^{3}}\frac{1}{e^{\beta \epsilon \left( \mathbf{k}\right) }-1}\mathrm{d}%
\mathbf{k}  \notag \\
&=&\frac{1}{\left( 2\pi \right) ^{3}}\xi ^{\dagger }\xi \mathbf{+}\frac{%
(2m)^{3/2}}{4\pi ^{2}\hbar ^{3}}\int_{0^{+}}^{+\infty }\mathrm{d}\epsilon
\frac{\epsilon ^{1/2}}{e^{\beta \epsilon }-1},  \label{Lesb3}
\end{eqnarray}%
where we have used the fact that the contribution from the state of $\mathbf{%
k}=0$ is zero when reducing the Lebesgue integral into the improper Riemann
integral in the end. Because $\xi ^{\dagger }\xi \geq 0$, the above equation
can not hold if $T>T_{c}$.

From those discussions, it follows that%
\begin{equation}
\eta =\left\{
\begin{array}{ll}
\mathrm{zero,}\text{ } & T\geq T_{c} \\
\text{\textrm{nonzero}}\mathrm{,}\text{ } & T<T_{c}.%
\end{array}%
\right.
\end{equation}%
That is to say, a phase transition will happen at $T=T_{c}$. Because $\Delta
s\geq 0$ when $\mu \leq 0$, the normal phase ($\eta =0$) exists and is
stable at $T\geq T_{c}$, the ordered phase ($\eta \neq 0$) exists and is
stable at $T<T_{c}$. Here, we note that it is a natural consequence of the
extended ensemble theory that the chemical potential of the ideal Bose gas
is fixed at zero ($\mu =0$) in the ordered phase ($T<T_{c}$), in contrast to
the Einstein's treatment where it is fixed by physical insight \cite%
{Einstein1,Einstein2}.

Obviously, the above results are the same as the BEC given by Einstein \cite%
{Einstein1,Einstein2} if $\xi ^{\dagger }\xi /(2\pi )^{3}$ were regarded as
the density of the bosons condensed onto the state of $\mathbf{k}=0$.
Therefore, the transition happening at $T=T_{c}$ must be identified
physically as the Bose-Einstein condensation. It should be pointed out that,
generally, the quantity $\xi ^{\dagger }\xi /(2\pi )^{3}$ can not be
interpreted as the density of condensed bosons, as will be clarified in Sec. %
\ref{WIBG}.

So far, we have presented a new description to the Bose-Einstein
condensation of the ideal Bose gas. In comparison with the Einstein's
description \cite{Einstein1,Einstein2}, we now describe the BEC with two new
conceptions: spontaneous symmetry breaking and Lebesgue integration. Within
this new description, the BEC happens simultaneously with the spontaneous
breaking of the gauge symmetry; the integration is defined and performed in
Lebesgue way. That is fundamentally different from Einstein's description
\cite{Einstein1,Einstein2} where the BEC happens with no symmetry breaking,
and the integration is done in Riemann way.

Here, it is worth emphasizing the important role played by Lebesgue
integration in the study of the BEC of the ideal Bose gas. As has been seen,
there arises the Bose distribution function,
\begin{equation}
f(\epsilon \left( \mathbf{k}\right) )=\frac{1}{e^{\beta \epsilon \left(
\mathbf{k}\right) }-1},  \label{FSingular}
\end{equation}%
which is unbounded above on the set $E$. Because the integrand of a Riemann
integral must be bounded everywhere, such a function is not Riemann
integrable and can not be handled with Riemann integration. Fortunately,
that makes no trouble. As pointed above, it is, in fact, unnecessary for us
to handle the Bose distribution function with Riemann integration; what
really confronts us in statistical physics is Lebesgue other than Riemann
integration. For Lebesgue integration, it permits more general functions as
integrands, and treats bounded and unbounded functions on an equal footing.
In the sense of Lebesgue integration, the Bose distribution function is
integrable, which can be shown as follows,
\begin{widetext}%
\begin{eqnarray}
\lim_{V\rightarrow +\infty }\frac{1}{V}\sum_{\mathbf{k}}\frac{1}{e^{\beta
\epsilon _{\mathbf{k}}}-1} &=&\frac{1}{\left( 2\pi \right) ^{3}}\int\limits_{%
\mathbb{R}
^{3}}\frac{1}{e^{\beta \epsilon \left( \mathbf{k}\right) }-1}\mathrm{d}%
\mathbf{k=}\frac{1}{\left( 2\pi \right) ^{3}}\int\limits_{E_{0}}\frac{1}{%
e^{\beta \epsilon \left( \mathbf{k}\right) }-1}\mathrm{d}\mathbf{k+}\frac{1}{%
\left( 2\pi \right) ^{3}}\int\limits_{%
\mathbb{R}
^{3}\backslash E_{0}}\frac{1}{e^{\beta \epsilon \left( \mathbf{k}\right) }-1}%
\mathrm{d}\mathbf{k}  \notag \\
&=&\frac{1}{\left( 2\pi \right) ^{3}}\int\limits_{%
\mathbb{R}
^{3}\backslash E_{0}}\frac{1}{e^{\beta \epsilon \left( \mathbf{k}\right) }-1}%
\mathrm{d}\mathbf{k}=\frac{(2m)^{3/2}}{4\pi ^{2}\hbar ^{3}}%
\int_{0^{+}}^{+\infty }\mathrm{d}\epsilon \frac{\epsilon ^{1/2}}{e^{\beta
\epsilon }-1},  \label{BOSE1}
\end{eqnarray}%
\end{widetext}
where the set $E_{0}\equiv \{\mathbf{k}=0\}$ includes one point only. $E_{0}$
is a null set, i.e., its measure is zero, thus, the integral on it vanishes.
The rest part, i.e., the integral on $%
\mathbb{R}
^{3}\backslash E_{0}$, reduces finally into an improper Riemann integral. As
is well known, the integral on a null set always vanishes irrespective of
the integrand values on this set, even if they are infinite. Thereby, the
integral on any single state equals zero, the ground state ($\mathbf{k}=0$)
is not more advantageous than the other states ($\mathbf{k}\neq 0$). In
other words, the singularity of $f(\epsilon \left( \mathbf{k}\right) )$ on
the point $\mathbf{k}=0$ has no special meaning, both in mathematics and in
physics.

This aspect makes the present treatment quite distinct from that of
Einstein, where the integration is handled in Riemann way, with a special
treatment given to the ground state of $\mathbf{k}=0$. Intuitively, Einstein
thought that the bosons would gather in the ground state if $T<T_{c}$, and
thus the contribution from this state is nonzero \cite{Einstein1,Einstein2},
that is,
\begin{eqnarray}
n &=&\lim_{V\rightarrow +\infty }\frac{1}{V}\sum_{\mathbf{k}}\frac{1}{%
e^{\beta \epsilon _{\mathbf{k}}}-1}  \notag \\
&=&n_{0}+\frac{(2m)^{3/2}}{4\pi ^{2}\hbar ^{3}}\int_{0^{+}}^{+\infty }%
\mathrm{d}\epsilon \frac{\epsilon ^{1/2}}{e^{\beta \epsilon }-1},
\label{BOSE2}
\end{eqnarray}%
where $n_{0}>0$ is the particle density gathering in the ground state.
Comparing Eqs. (\ref{Lesb3}), (\ref{BOSE1}) and (\ref{BOSE2}) shows more
clearly the difference between the present treatment and Einstein's. From
this difference it follows that the quantity $\xi ^{\dagger }\xi /(2\pi
)^{3} $ needs not to be interpreted as the density of the bosons condensed
onto the state of $\mathbf{k}=0$. In fact, Eq. (\ref{Lesb3}) is physically
just an equation of order parameter. Here, it should be noted that the
Bogoliubov theory of weakly imperfect Bose gas \cite{Bogoliubov,Zagrebnov}
is based on Eq. (\ref{BOSE2}) rather than Eq. (\ref{BOSE1}). At last, why
did Einstein not use Lebesgue integration? there is a historical reason:
Until 1925 when Einstein \cite{Einstein1,Einstein2} studied the BEC,
Lebesgue integration had not yet been introduced to probability theory by
Kolmogorov, who did it in 1933 \cite{Kolmogorov}.

Also, it is significant to compare the two Lebesgue integrals represented in
Eq. (\ref{Lesb0}) and Eq. (\ref{Lesb3}): Both integrands are unbounded on $%
\mathbb{R}
^{3}$; however, the former is infinite on a set with a nonzero measure ($%
m(A_{\mathbf{k=0}}\cap
\mathbb{R}
^{3})=\left( 2\pi \right) ^{3}/V$), the latter is finite almost everywhere,
or infinite just on a null set ($m\left( E_{0}\cap
\mathbb{R}
^{3}\right) =0$). This difference leads to the result that the nontrivial
solution must be discarded in the finite volume case whereas it has physical
sense in the thermodynamic limit. In a word, the ideal Bose gas can produce
BEC in the thermodynamic limit, but it can not in the finite volume case.
That is a rigorous conclusion deduced from the extended ensemble theory.

As another instance, the ideal Bose gas shows unambiguously that Gibbs
ensemble theory needs some kind of extension. Otherwise, the conservation of
particles will be broken when $T<T_{c}$, as can be seen by comparing Eqs. (%
\ref{BOSE1}) and (\ref{BOSE2}).

As the only one of exactly solvable quantum phase transition, the BEC in the
ideal Bose gas demonstrates that the extended ensemble theory postulated in
Sec. \ref{EET} is feasible in physics for us to describe phase transitions.

Lastly, we would like to stress that the condition of Eq. (\ref{BECMu}),
i.e., $\mu \leq 0$, is not a mathematical limit but just a physical
requirement. Like the case of $\mu \leq 0$, the extended ensemble theory is
also well defined mathematically for $\mu >0$, as is shown in Appendix \ref%
{FSum}. Removing the limitation on the chemical potential is physically
reasonable, it enables the conservation of particles to be handled freely,
and makes it easy to describe the BEC in the interacting Bose gas, as can be
seen in Sec. \ref{WIBG}.

\textbf{Remark 1}: The ideal Bose gas is a touchstone for the theory of
phase transitions because it is the only one of the many-body systems that
can show a quantum phase transition and can be solved rigorously. Any theory
of phase transitions must be examined by the ideal Bose gas. It can not be
accepted in physics if it can not give a rigorous solution to the ideal Bose
gas; and it can be further applied to the interacting Bose gas only after it
has given a rigorous solution to the ideal Bose gas.

\textbf{Remark 2}: In 1926, Uhlenbeck \cite{Uhlenbeck} raised an objection
to Einstein's viewpoint on the condensation, he argued that the ideal Bose
gas can not produce BEC if the volume of the system remains finite. This
objection caused an exciting debate during the van der Waals Century
Conference in November 1937 \cite{Pais}. Evidently, the present theory has
given the exact answer to the debate: Uhlenbeck and Einstein are both right,
which is also in accordance with the agreement reached finally in the
conference.

\textbf{Remark 3}: In the momentum space used here, the condensed phase is
shown to be homogeneous. In Sec. \ref{WIBG}, using the real space, we shall
prove further that the condensed phase can only be homogeneous. There can
not exist any supercurrent or quantized vortex in the ideal Bose gas.

\textbf{Remark 4}: The rigorous solution for the ideal Bose gas shows that
the extended ensemble theory holds in the critical region.

\subsection{The Photon Gas in a Black Body}

For the photon gas in a black body, its Hamiltonian reads,
\begin{equation}
H(a)=\sum_{\mathbf{k}\nu }\omega _{\mathbf{k}}\left( a_{\mathbf{k}\nu
}^{\dag }a_{\mathbf{k}\nu }+\frac{1}{2}\right) ,  \label{Hphoton}
\end{equation}%
where Coulomb gauge is used, $a_{\mathbf{k}\nu }$ ($a_{\mathbf{k}\nu }^{\dag
}$) is the destruction (construction) operator for the photons with momentum
$\mathbf{k}$ and polarization $\nu $, and $\omega _{\mathbf{k}}=\hbar c%
\mathbf{k}$ represents the photon energy with $c$ the speed of light.

Analogous to the ideal Bose gas, this system has the phase-transition
operator,
\begin{equation}
D(\xi,a)=\sum_{\mathbf{k}\nu}(\xi_{\mathbf{k}\nu}^{\dag}a_{\mathbf{k}%
\nu}+\xi_{\mathbf{k}\nu}a_{\mathbf{k}\nu}^{\dag}),
\end{equation}
where $\xi$ is the order parameter for BEC.

From $H(a)$\ and $D(\xi ,a)$, it follows that%
\begin{gather}
\omega _{\mathbf{k}}\xi _{\mathbf{k}\nu }=0, \\
\Delta S=\beta \sum_{\mathbf{k,}\sigma }\omega _{\mathbf{k}}\delta \xi _{%
\mathbf{k},\sigma }^{\dagger }\delta \xi _{\mathbf{k},\sigma }.
\end{gather}%
The first is the equation of motion of order parameter, it has the solution,%
\begin{equation}
\xi _{\mathbf{k}\nu }=\lambda _{\nu }\delta _{\mathbf{k,}0},
\end{equation}%
where $\lambda _{\nu }$ is a complex constant which does not depends on $T$.
Owing to the fact: $\omega _{\mathbf{k}}\geq 0$, this solution is stable,%
\begin{equation}
\Delta S\geq 0.  \label{EDS}
\end{equation}%
To determine $\lambda _{\nu }$, it is sufficient to consider the initial
state at $\beta =0$ ($T\rightarrow +\infty $). Physically, any system must
stay in its disordered phase at $\beta =0$. Therefore, $\lambda _{\nu }$
must be zero at $\beta =0$. Since $\lambda _{\nu }$ does not depends on $%
\beta $, it is always zero. That is to say,
\begin{equation}
\xi \left( \beta \right) =0,\text{ for }\beta \geq 0.
\end{equation}%
This equation shows that the photon gas in a black body can not produce BEC,
and must always stay in its normal phase. Eq. (\ref{EDS}) demonstrates that
the normal phase is stable. This result is also valid for the thermodynamic
limit case.

In short, there can not arise BEC in the photon gas in a black body, the
system stays in its normal phase forever, Planck's radiation law holds at
any temperature. Obviously, this conclusion agrees completely with the
experiment.

\subsection{The Ideal Phonon Gas in a Solid Body \label{IPGSB}}

As the third system, let us consider the ideal phonon gas in a solid body,
whose Hamiltonian reads,
\begin{equation}
H(d)=\sum_{\mathbf{k}\sigma }\omega _{\mathbf{k\sigma }}\left( d_{\mathbf{k}%
\sigma }^{\dag }d_{\mathbf{k}\sigma }+\frac{1}{2}\right) ,  \label{Hphonon}
\end{equation}%
where $d_{\mathbf{k}\sigma }$ ($d_{\mathbf{k}\sigma }^{\dag }$) is the
destruction (construction) operator for the phonons with momentum $\mathbf{k}
$ and polarization $\sigma $, $\omega _{\mathbf{k\sigma }}=\hbar c_{\sigma }%
\mathbf{k}$ represents the phonon energy with $c_{\sigma }$ the velocity of
sound.

By comparing Eq. (\ref{Hphonon}) with Eq. (\ref{Hphoton}), one concludes at
once that the ideal phonon gas in a solid body can not produce BEC either,
and must also stay in its normal phase forever.

\subsection{Quantization of Dirac Field}

At last, let us consider the quantization of Dirac field from the standpoint
of the extended ensemble theory hypothesized in Sec. \ref{EET}. If one
quantizes Dirac field using Bose-Einstein statistics, he has the following
Hamiltonian \cite{Greiner},%
\begin{equation}
H(b;d)=\sum_{\mathbf{p,}\sigma}\omega_{\mathbf{p}}\left( b_{\mathbf{p}%
,\sigma}^{\dagger}b_{\mathbf{p},\sigma}-d_{\mathbf{p},\sigma}^{\dagger }d_{%
\mathbf{p},\sigma}\right) ,
\end{equation}
where $\mathbf{p}$ and $\sigma$ denote momentum and spin, respectively; $b_{%
\mathbf{p},\sigma}$ ($b_{\mathbf{p},\sigma}^{\dagger}$) and $d_{\mathbf{p}%
\sigma}$ ($d_{\mathbf{p}\sigma}^{\dag}$) are bosonic destruction
(construction) operators; and
\begin{equation}
\omega(\mathbf{p})=\sqrt{\left( c\mathbf{p}\right) ^{2}+\left( mc^{2}\right)
^{2}}
\end{equation}
with $m$ and $c$ being the electron mass and the speed of light,
respectively.

Evidently, this system has the phase-transition operator,%
\begin{align}
D(\eta ,b;\xi ,d)& =\sum_{\mathbf{p},\sigma }(\eta _{\mathbf{p}\sigma
}^{\dag }b_{\mathbf{p}\sigma }+\xi _{\mathbf{p}\sigma }^{\dag }d_{\mathbf{p}%
\sigma }  \notag \\
& +\eta _{\mathbf{p}\sigma }b_{\mathbf{p}\sigma }^{\dag }+\xi _{\mathbf{p}%
\sigma }d_{\mathbf{p}\sigma }^{\dag }),
\end{align}%
where $\eta $ and $\xi $ are the order parameters for BEC.

Eqs. (\ref{Variation}) and (\ref{Increment}) imply that%
\begin{gather}
\eta _{\mathbf{p}\sigma }=0,  \label{Dirac1} \\
\xi _{\mathbf{p}\sigma }=0,  \label{Dirac2} \\
\Delta S=\beta \sum_{\mathbf{p,}\sigma }\omega _{\mathbf{p}}(\delta \eta _{%
\mathbf{p},\sigma }^{\dagger }\delta \eta _{\mathbf{p},\sigma }-\delta \xi _{%
\mathbf{p},\sigma }^{\dagger }\delta \xi _{\mathbf{p},\sigma }).
\label{Dirac3}
\end{gather}%
Eqs. (\ref{Dirac1}) and (\ref{Dirac2}) show that the system should always
stay in its normal phase, but Eq. (\ref{Dirac3}) demonstrates that this
normal phase corresponds to a saddle point, and is thus instable. Therefore,
if Dirac field were quantized using Bose-Einstein statistics, the system
would be instable and collapse. In other words, it is not admissible to
quantize Dirac field using Bose-Einstein statistics. This conclusion is in
accordance with the spin-statistics theorem in quantum field theory \cite%
{Greiner}.

On the other hand, if Dirac field is quantized using Fermi-Dirac statistics,
its Hamiltonian will read as follows \cite{Greiner},%
\begin{equation}
H(b;d)=\sum_{\mathbf{p,}\sigma }\omega _{\mathbf{p}}(b_{\mathbf{p},\sigma
}^{\dagger }b_{\mathbf{p},\sigma }+d_{\mathbf{p},\sigma }^{\dagger }d_{%
\mathbf{p},\sigma }),  \label{EPRP}
\end{equation}%
now $b_{\mathbf{p},\sigma }$ ($b_{\mathbf{p},\sigma }^{\dagger }$) and $d_{%
\mathbf{p}\sigma }$ ($d_{\mathbf{p}\sigma }^{\dag }$) are fermionic
destruction (construction) operators for electrons and positrons. According
to Sec. \ref{EET}, its phase-transition operator is%
\begin{align}
D(\phi ,b;\zeta ,d)& =\sum_{\mathbf{k}}[(\phi _{\mathbf{k}}b_{-\mathbf{k}%
\downarrow }b_{\mathbf{k}\uparrow }+\phi _{\mathbf{k}}^{\dag }b_{\mathbf{k}%
\uparrow }^{\dag }b_{-\mathbf{k}\downarrow }^{\dag })  \notag \\
& +(\zeta _{\mathbf{k}}d_{-\mathbf{k}\downarrow }d_{\mathbf{k}\uparrow
}+\zeta _{\mathbf{k}}^{\dag }d_{\mathbf{k}\uparrow }^{\dag }d_{-\mathbf{k}%
\downarrow }^{\dag })],
\end{align}%
where $\phi $ and $\zeta $ are order parameters. Following the arguments of
Sec. \ref{IFG}, one recognizes immediately that the system is always stable
in its normal phase. This holds even in the original representation of the
Hamiltonian \cite{Greiner},
\begin{equation}
H(c)=\sum_{\mathbf{p}}\omega _{\mathbf{p}}\left( \sum_{s=1}^{2}c_{\mathbf{p}%
,s}^{\dagger }c_{\mathbf{p},s}-\sum_{s=3}^{4}c_{\mathbf{p},s}^{\dagger }c_{%
\mathbf{p},s}\right) ,  \label{EHPR}
\end{equation}%
where
\begin{equation}
c_{\mathbf{p},s}=\left\{
\begin{array}{ll}
b_{\mathbf{p},s}, & s=1,2 \\
d_{\mathbf{p},s-2}^{\dagger },\text{ } & s=3,4,%
\end{array}%
\right.
\end{equation}%
\vspace*{-0.3cm}%
\begin{align}
D(\phi ,\zeta ,c)& =\sum_{\mathbf{k}}[(\phi _{\mathbf{k}}c_{-\mathbf{k,}2}c_{%
\mathbf{k,}1}+\phi _{\mathbf{k}}^{\dag }c_{\mathbf{k,}1}^{\dag }c_{-\mathbf{k%
},2}^{\dag })  \notag \\
& +(\zeta _{\mathbf{k}}c_{-\mathbf{k},4}c_{\mathbf{k},3}+\zeta _{\mathbf{k}%
}^{\dag }c_{\mathbf{k},3}^{\dag }c_{-\mathbf{k},4}^{\dag })].
\end{align}%
In contrast to the electron-positron representation of Eq. (\ref{EPRP}), a
minus sign appears in the original representation of Eq. (\ref{EHPR}), but
it can not influence the stability of the system. Anyway, there is no
problem when Dirac field is quantized using Fermi-Dirac statistics.

In this section, the systems concerned are free boson fields, they can all
be solved rigorously. We shall proceed to the interacting systems in Sec. %
\ref{SPT}. Generally, they can not be solved rigorously, approximations are
needed.\vspace*{0.2in}

\section{Physical Origination for Phase Transitions \label{POPT}}

Before proceeding forward to the interacting Bose systems, it is worth
making a research into the physical origination for phase transitions.

Again, let us take BCS superconductivity as an instance. According to Eq. (%
\ref{Average}), the statistical average of a one-body operator can be
reformulated as
\begin{eqnarray}
\int d\mathbf{r}\langle \psi ^{\dagger }(\mathbf{r})f(\mathbf{r})\psi (%
\mathbf{r})\rangle &=&\int \mathrm{d}\mathbf{r}\mathrm{Tr}\big(\widetilde{%
\psi }^{\dagger }(\mathbf{r})f(\mathbf{r})\widetilde{\psi }(\mathbf{r})\rho
(H(c))\big)  \notag \\
&\equiv &\int \mathrm{d}\mathbf{r}\overline{\widetilde{\psi }^{\dagger }(%
\mathbf{r})f(\mathbf{r})\widetilde{\psi }(\mathbf{r})},  \label{Single}
\end{eqnarray}%
where $\psi (\mathbf{r})$ is the electron field, $f(\mathbf{r})$ stands for
the one-body operator in Schr\"{o}dinger picture, and
\begin{equation}
\widetilde{\psi }(\mathbf{r})=e^{-iD(\phi ,c)}\psi (\mathbf{r})e^{iD(\phi
,c)}.
\end{equation}%
After the phase transition ($\phi \neq 0$), the electron field $\widetilde{%
\psi }(\mathbf{r})$ is separated into two fields,
\begin{equation}
\widetilde{\psi }(\mathbf{r})=\psi (\mathbf{r})+\varphi (\mathbf{r}),
\label{PsiPrime}
\end{equation}%
where
\begin{equation}
\varphi (\mathbf{r})=e^{-iD(\phi ,c)}\psi (\mathbf{r})e^{iD(\phi ,c)}-\psi (%
\mathbf{r}).
\end{equation}%
Substituting Eq. (\ref{PsiPrime}) into Eq. (\ref{Single}), we obtain
\begin{widetext}
\begin{equation}
\int \mathrm{d}\mathbf{r}\langle \psi ^{\dagger }(\mathbf{r})f(\mathbf{r}%
)\psi (\mathbf{r})\rangle =\int \mathrm{d}\mathbf{r}\overline{[\psi
^{\dagger }(\mathbf{r})+\varphi ^{\dagger }(\mathbf{r})]f(\mathbf{r})[\psi (%
\mathbf{r})+\varphi (\mathbf{r})]}\mathbf{.}  \label{Pattern1}
\end{equation}%
The right-hand side shows that there exists interference, as usual, it is
described by the cross terms,
\begin{equation}
\overline{\psi ^{\dagger }(\mathbf{r})f(\mathbf{r})\varphi (\mathbf{r})}+%
\overline{\varphi ^{\dagger }(\mathbf{r})f(\mathbf{r})\psi (\mathbf{r})}.
\label{CrossTerms}
\end{equation}%
\end{widetext}
In terminology of quantum optics, it represents the single-particle
interference. This single-particle interference will vanish if the system
goes into the normal phase where $\varphi (\mathbf{r})=0$. A special case is
the interference appearing in the particle density where $f(\mathbf{r})=1$,
\begin{equation}
\overline{\psi ^{\dagger }(\mathbf{r})\varphi (\mathbf{r})}+\overline{%
\varphi ^{\dagger }(\mathbf{r})\psi (\mathbf{r})},  \label{CrossTerms1}
\end{equation}%
this form of interference is very familiar in optics, Eq. (\ref{CrossTerms})
being the general case.

In particular, an abnormal form of single-particle interference can appear
in the electron-pair amplitude,
\begin{equation}
\langle \psi _{\downarrow }(\mathbf{r})\psi _{\uparrow }(\mathbf{r})\rangle =%
\overline{[\psi _{\downarrow }(\mathbf{r})+\varphi _{\downarrow }(\mathbf{r}%
)][\psi _{\uparrow }(\mathbf{r})+\varphi _{\uparrow }(\mathbf{r})]},
\end{equation}%
it is described by the abnormal cross terms%
\begin{equation}
\overline{\psi _{\downarrow }(\mathbf{r})\varphi _{\uparrow }(\mathbf{r})}+%
\overline{\varphi _{\downarrow }(\mathbf{r})\psi _{\uparrow }(\mathbf{r})}+%
\overline{\varphi _{\downarrow }(\mathbf{r})\varphi _{\uparrow }(\mathbf{r})}%
,
\end{equation}%
which have no counterparts in optics, in contrast to Eq. (\ref{CrossTerms1}%
). Observe Eq. (\ref{Ctrans}), one recognizes that the field $\varphi (%
\mathbf{r})$ contains two parts,%
\begin{equation}
\varphi (\mathbf{r})=\varphi ^{\left( +\right) }(\mathbf{r})+\varphi
^{\left( -\right) }(\mathbf{r}),
\end{equation}%
where $\varphi ^{\left( +\right) }(\mathbf{r})$ represents the creation
part,
\begin{equation}
\varphi ^{\left( +\right) }(\mathbf{r})=\left[
\begin{array}{l}
\phi _{\downarrow }^{\dagger }(\mathbf{r}) \\
\phi _{\uparrow }^{\dagger }(\mathbf{r})%
\end{array}%
\right] ,
\end{equation}%
with $\phi _{\uparrow }^{\dagger }(\mathbf{r})$ and $\phi _{\downarrow
}^{\dagger }(\mathbf{r})$ being connected with the operators $c_{\mathbf{k}%
\uparrow }^{\dagger }$ and $c_{-\mathbf{k}\downarrow }^{\dagger }$,
respectively; and $\varphi ^{\left( -\right) }(\mathbf{r})$ the annihilation
part, which is connected with the operators $c_{\mathbf{k}\uparrow }$ and $%
c_{-\mathbf{k}\downarrow }$. This separation of two parts leads us to
\begin{widetext}
\begin{equation}
\overline{\psi _{\downarrow }(\mathbf{r})\varphi _{\uparrow }(\mathbf{r})}+%
\overline{\varphi _{\downarrow }(\mathbf{r})\psi _{\uparrow }(\mathbf{r})}+%
\overline{\varphi _{\downarrow }(\mathbf{r})\varphi _{\uparrow }(\mathbf{r})}%
=\overline{\psi _{\downarrow }(\mathbf{r})\phi _{\downarrow }^{\dagger }(%
\mathbf{r})}+\overline{\phi _{\uparrow }^{\dagger }(\mathbf{r})\psi
_{\uparrow }(\mathbf{r})}+\overline{\varphi _{\downarrow }^{\left( -\right)
}(\mathbf{r})\phi _{\downarrow }^{\dagger }(\mathbf{r})}+\overline{\phi
_{\uparrow }^{\dagger }(\mathbf{r})\varphi _{\uparrow }^{\left( -\right) }(%
\mathbf{r})}.
\end{equation}%
Clearly, there exists interference on the right-hand side, this interference
induce the abnormal interference on the left-hand side. The abnormal
interference has an important consequence, that is, the electrons are formed
into Cooper pairs in the superconducting phase. Evidently, the abnormal
interference vanishes in the normal phase.

Also, there exists the two-particle interference in the superconducting
phase, which arises from the statistical average of a two-body operator,%
\begin{equation}
\frac{1}{2}\iint \mathrm{d}\mathbf{r}\mathrm{d}\mathbf{r}^{\prime }\langle
\psi ^{\dagger }(\mathbf{r})\psi ^{\dagger }(\mathbf{r}^{\prime })v(\mathbf{r%
}-\mathbf{r}^{\prime })\psi (\mathbf{r}^{\prime })\psi (\mathbf{r})\rangle =%
\frac{1}{2}\iint \mathrm{d}\mathbf{r\mathrm{d}\mathbf{r}^{\prime }}v(\mathbf{%
r}-\mathbf{r}^{\prime })\overline{\widetilde{\psi }^{\dagger }(\mathbf{r})%
\widetilde{\psi }^{\dagger }(\mathbf{r}^{\prime })\widetilde{\psi }(\mathbf{r%
}^{\prime })\widetilde{\psi }(\mathbf{r})}\mathbf{,}  \label{Pattern2}
\end{equation}%
\end{widetext}
where $v(\mathbf{r}-\mathbf{r}^{\prime })$ stands for the two-body operator
in Schr\"{o}dinger picture.

In the sense of Eqs. (\ref{Pattern1}) and (\ref{Pattern2}), the
superconducting phase can be identified as a form of interference pattern.
Physically, the onset of an interference pattern will make a system more
structured, and thus increases its degree of order and decreases its
entropy. That is why the superconducting phase has a higher degree of order
and a less amount of entropy than the normal phase.

From the above discussions, it follows that there will appear interference
in the ordered phase whereas there will not in the normal (disordered)
phase. Obviously, the same conclusion holds for other phase transitions.
Therefore, a phase transition amounts to the onset of a form of
interference. As well-known, interference is a characteristic property of
wave, we thus conclude that phase transitions originate physically from the
wave nature of matter.

Physically, any system will go in its disordered phase at sufficiently high
temperatures, or as $T\rightarrow+\infty$. Because order parameter equals
zero in disordered phase, the extended ensemble theory should ensure that
there exists a zero solution of order parameter, at least at high
temperatures. So it does, indeed.

Upon a transformation, Eq. (\ref{Entropy}) can be expressed as
\begin{widetext}
\begin{equation}
S(\phi ,\beta )=-\mathrm{Tr}\Big(\!\ln \!\!\big(\rho (e^{-iD(\phi
,c)}H\left( c\right) e^{iD(\phi ,c)})\big)\rho (H(c))\Big).
\end{equation}%
Considering the expansion,%
\begin{equation}
e^{-iD(\phi ,c)}H\left( c\right) e^{iD(\phi ,c)}=H\left( c\right) -i[D(\phi
,c),H\left( c\right) ]+\frac{\left( -i\right) ^{2}}{2}[D(\phi ,c),[D(\phi
,c),H\left( c\right) ]]+\cdots ,
\end{equation}%
we obtain%
\begin{equation}
S(\phi ,\beta )=-\mathrm{Tr}\Big(\ln \!\big(\rho \left( H\left( c\right)
\right) \big)\rho (H(c))\Big)+\beta \frac{\left( -i\right) ^{2}}{2}\mathrm{Tr%
}\Big([D(\phi ,c),[D(\phi ,c),H\left( c\right) ]]\rho (H(c))\Big)+\cdots ,
\end{equation}%
the linear term of $\phi $ vanishes.

Apart from an irrelevant term, this equation shows that the powers of $\phi $
are equal to or higher than the $2$nd in the expansion of $S(\phi ,\beta )$.
As a result of $\delta S=0$, the powers of $\phi $ are equal to or higher
than the $1$st in the equation of order parameter. It implies that there
always exists a zero solution of $\phi $ at any temperature, i.e., the
trivial solution. In fact, this solution is a stable solution as $%
T\rightarrow +\infty $,%
\begin{equation}
\Delta S=S(\delta \phi ,\beta )-S(0,\beta )=-\mathrm{Tr}\Big(\ln \!\big(\rho
(H(c))\big)\left[ \rho (H^{\prime }(\delta \phi ,c))-\rho (H(c))\right] \Big)%
=0,\text{ }T\rightarrow +\infty .
\end{equation}%
\end{widetext}
which satisfies the criterion of Eq. (\ref{Increment}). Of course, this
conclusion is also valid for other phase transitions.

To summarize, there exists a zero solution of order parameter at any
temperature in the extended ensemble theory, this solution becomes stable as
$T\rightarrow +\infty $. Therefore, a system always stays in its normal
phase at sufficiently high temperatures, this high-temperature phase is
disordered and structureless. As temperature decreases, there can arise
spontaneously the quantum interference of matter waves, which makes the
system structured and transform into its ordered phase. That is the physical
picture for phase transitions within the framework of the extended ensemble
theory.

\textbf{Remark}: In Ref. \cite{Yang5}, Yang suggested an important concept,
i.e., off-diagonal long-range order (ODLRO), to describe quantum phase
transitions. Physically, ODLRO is a generalization of the concept of Copper
pairing. With this generalized concept, he developed many beautiful
theorems, which are characteristics of quantum phases. Regrettably, in that
paper, there is no discussion of the properties of the Hamiltonian that is
needed to ensure the existence of ODLRO at low temperatures. Evidently,
there exists no ODLRO if a system Hamiltonian is in its symmetric
representation. The requirement can be satisfied if the system Hamiltonian
takes its asymmetric, or broken-symmetry, representation. Therefore, those
theorems in Ref. \cite{Yang5} still hold in the extended ensemble theory. As
a matter of fact, the ODLRO's of the reduced density matrices $\rho _{1}$
and $\rho _{2}$ \cite{Yang5}, e.g., $\mathrm{Tr}(c_{i\sigma }\rho (H^{\prime
}(\phi ,c))c_{j\sigma }^{\dagger })$ and $\mathrm{Tr}(c_{k\sigma }c_{l\sigma
}\rho (H^{\prime }(\phi ,c))c_{j\sigma }^{\dagger }c_{i\sigma }^{\dagger })$%
, result physically from the single-particle and two-particle interferences,
respectively. That is to say, off-diagonal long-range order is a physical
manifestation of the quantum interference of matter waves.

\section{\textbf{Interacting Bose Systems} \label{SPT}}

In this section, we focus our attention on the interacting Bose systems. We
shall first study the weakly interacting Bose gas, and then the double-well
potential systems where structural phase transitions, Goldstone bosons, and
Higgs mechanism are concerned.\vspace*{0.2in}

\subsection{Weakly Interacting Bose Gas \label{WIBG}}

As usual, the Hamiltonian for an interacting Bose gas reads,
\begin{widetext}
\begin{equation}
H[\psi ]=\int \mathrm{d}\mathbf{r}\left[ \frac{\hbar ^{2}}{2m}\nabla \psi
^{\dagger }(\mathbf{r})\cdot \nabla \psi (\mathbf{r})-\mu \psi ^{\dagger }(%
\mathbf{r})\psi (\mathbf{r})\right] +\frac{1}{2}\iint \mathrm{d}\mathbf{r}%
\mathrm{d}\mathbf{r}^{\prime }\psi ^{\dagger }(\mathbf{r})\psi ^{\dagger }(%
\mathbf{r}^{\prime })v(\mathbf{r}-\mathbf{r}^{\prime })\psi (\mathbf{r}%
^{\prime })\psi (\mathbf{r}),
\end{equation}%
where $\psi (\mathbf{r})$ represents the boson-field operator, and $v(%
\mathbf{r}-\mathbf{r}^{\prime })$ the interaction. Expressed in terms of $%
\psi (\mathbf{r})$ and $\psi ^{\dagger }(\mathbf{r})$, the phase-transition
operator for BEC becomes,
\begin{equation}
D\left[ \eta ,\psi \right] =\int \mathrm{d}\mathbf{r}\left[ \eta ^{\dagger }(%
\mathbf{r})\psi (\mathbf{r})+\eta (\mathbf{r})\psi ^{\dagger }(\mathbf{r})%
\right] .
\end{equation}

With $H[\psi ]$ and $D\left[ \eta ,\psi \right] $, the entropy of the system
can be obtained,
\begin{align}
S\left[ \eta ,\beta \right] & =-\mathrm{Tr}\big(\ln \left( \rho (H\left[
\psi \right] )\right) \!\rho (H\left[ \psi \right] )\big)  \notag \\
& +\beta \bigg\{\int \mathrm{d}\mathbf{r}\left[ \frac{\hbar ^{2}}{2m}\nabla
\eta ^{\dagger }(\mathbf{r})\cdot \nabla \eta (\mathbf{r})-\mu \eta
^{\dagger }(\mathbf{r})\eta (\mathbf{r})\right] +\iint \mathrm{d}\mathbf{r}%
\mathrm{d}\mathbf{r}^{\prime }\eta ^{\dagger }(\mathbf{r})\eta (\mathbf{r})v(%
\mathbf{r}-\mathbf{r}^{\prime })\overline{\psi ^{\dagger }(\mathbf{r}%
^{\prime })\psi (\mathbf{r}^{\prime })}  \notag \\
& +\iint \mathrm{d}\mathbf{r}\mathrm{d}\mathbf{r}^{\prime }\eta ^{\dagger }(%
\mathbf{r})\eta (\mathbf{r}^{\prime })v(\mathbf{r}-\mathbf{r}^{\prime })%
\overline{\psi ^{\dagger }(\mathbf{r}^{\prime })\psi (\mathbf{r})}+\frac{1}{2%
}\iint \mathrm{d}\mathbf{r}\mathrm{d}\mathbf{r}^{\prime }\eta ^{\dagger }(%
\mathbf{r})\eta (\mathbf{r})v(\mathbf{r}-\mathbf{r}^{\prime })\eta ^{\dagger
}(\mathbf{r}^{\prime })\eta (\mathbf{r}^{\prime })\bigg\},  \label{IBECS}
\end{align}%
where $\overline{F}$ denotes the statistical average of the operator $F$
with respect to $H[\psi ]$. Its variation yields the equation of order
parameter,%
\begin{equation}
\left( -\frac{\hbar ^{2}}{2m}\nabla ^{2}-\mu \right) \eta (\mathbf{r})+\int d%
\mathbf{r}^{\prime }\eta (\mathbf{r})v(\mathbf{r}-\mathbf{r}^{\prime })%
\overline{\psi ^{\dagger }(\mathbf{r}^{\prime })\psi (\mathbf{r}^{\prime })}%
+\int \mathrm{d}\mathbf{r}^{\prime }\eta (\mathbf{r}^{\prime })v(\mathbf{r}-%
\mathbf{r}^{\prime })\overline{\psi ^{\dagger }(\mathbf{r}^{\prime })\psi (%
\mathbf{r})}+\int \mathrm{d}\mathbf{r}^{\prime }\eta (\mathbf{r})v(\mathbf{r}%
-\mathbf{r}^{\prime })\eta ^{\dagger }(\mathbf{r}^{\prime })\eta (\mathbf{r}%
^{\prime })=0.  \label{IOP}
\end{equation}%
That is a generalized Ginzburg-Landau or Gross-Pitaevskii equation. In the
simple case where $v(\mathbf{r}-\mathbf{r}^{\prime })=g\delta (\mathbf{r}-%
\mathbf{r}^{\prime })$, it reduces to%
\begin{equation}
-\frac{\hbar ^{2}}{2m}\nabla ^{2}\eta (\mathbf{r})+\left[ 2g\overline{\psi
^{\dagger }(\mathbf{r})\psi (\mathbf{r})}-\mu \right] \eta (\mathbf{r}%
)+g\left\vert \eta (\mathbf{r})\right\vert ^{2}\eta (\mathbf{r})=0,
\end{equation}%
\end{widetext}
which is the standard Ginzburg-Landau \cite{GL} or Gross-Pitaevskii \cite%
{GP1,GP2} equation. In addition to Eq. (\ref{IOP}), one also needs%
\begin{equation}
\int \mathrm{d}\mathbf{r}\eta ^{\dagger }(\mathbf{r})\eta (\mathbf{r})+\int
\mathrm{d}\mathbf{r}\overline{\psi ^{\dagger }(\mathbf{r})\psi (\mathbf{r})}%
=N,  \label{ICP}
\end{equation}%
which is the equation of chemical potential. Eq. (\ref{IOP}) and Eq. (\ref%
{ICP}) constitute the two basic equations for an interacting Bose system.

For simplicity, let us first consider the homogeneous solution,%
\begin{equation}
\eta (\mathbf{r})=\xi ,  \label{HomoKsi}
\end{equation}%
where $\xi $ is independent of $\mathbf{r}$. Accordingly, Eqs. (\ref{IBECS})
and (\ref{IOP}) get simplified as%
\begin{align}
S\left[ \eta ,\beta \right] & =-\mathrm{Tr}\big(\ln \left( \rho (H\left[
\psi \right] )\right) \!\rho (H\left[ \psi \right] )\big)  \notag \\
& +\beta \int \mathrm{d}\mathbf{r}[-\widetilde{\mu }\xi ^{\dagger }\xi +%
\frac{1}{2}v(0)(\xi ^{\dagger }\xi )^{2}],  \label{IBECS1}
\end{align}%
\vspace*{-0.3in}%
\begin{equation}
-\widetilde{\mu }\xi +v(0)\left\vert \xi \right\vert ^{2}\xi =0,
\label{IOP1}
\end{equation}%
where%
\begin{align}
\widetilde{\mu }& =\mu -\int \frac{\mathrm{d}\mathbf{k}}{\left( 2\pi \right)
^{3}}\left[ v(0)+v(-\mathbf{k})\right]  \notag \\
& \times \left( -\frac{1}{\beta \hbar }\right) \sum_{n}e^{i\omega _{n}\eta }%
\mathcal{G}(\mathbf{k},i\omega _{n}),  \label{RMuW} \\
v(\mathbf{k})& =\int \mathrm{d}\mathbf{r}v(\mathbf{r})e^{-i\mathbf{k}\cdot
\mathbf{r}},
\end{align}%
$\mathcal{G}(\mathbf{k},i\omega _{n})$ being the single-particle Green's
function defined with respect to $H[\psi ]$.

Eq. (\ref{IBECS1}) has the typical form of landau free energy \cite{Landau}.
From Eqs. (\ref{IBECS1}), (\ref{IOP1}) and (\ref{Increment}), it follows
that the interaction between two particles must be repulsive for an
interacting Bose system to be stable, i.e., $v(\mathbf{r}-\mathbf{r}^{\prime
})>0$. Otherwise, $v(0)<0$, the system will collapse.

Suppose that the repulsive interaction is weak, we can then treat it as a
perturbation. As a technique of perturbation, we shall employ GF, and adopt
the self-consistent Hartree-Fock approximation as in Ref. \cite{Fetter},\
\begin{align}
\hspace{-1cm}\mathcal{G}(\mathbf{k},i\omega_{n}) & =\frac{1}{i\omega
_{n}-\hbar^{-1}\left[ \varepsilon\left( \mathbf{k}\right) -\mu-\hbar
\Sigma^{\star}\left( \mathbf{k}\right) \right] },  \label{IBECHF1} \\
\hspace{-1cm}\hbar\Sigma^{\star}\left( \mathbf{k}\right) & =\int \frac{%
\mathrm{d}\mathbf{k}^{\prime}}{\left( 2\pi\right) ^{3}}\left[ v(0)+v(\mathbf{%
k}-\mathbf{k}^{\prime})\right]  \notag \\
& \times\left( -\frac{1}{\beta\hbar}\right) \sum_{n}e^{i\omega_{n}\eta }%
\mathcal{G}(\mathbf{k}^{\prime},i\omega_{n}),  \label{IBECHF2}
\end{align}
where $\hbar\Sigma^{\star}\left( \mathbf{k}\right) $ represents the proper
self-energy.\ By use of Eq. (\ref{RMuW}), $\mathcal{G}(\mathbf{k},i\omega
_{n})$ can be expressed as%
\begin{equation}
\mathcal{G}(\mathbf{k},i\omega_{n})=\frac{1}{i\omega_{n}-\hbar^{-1}\left[
\widetilde{\varepsilon}\left( \mathbf{k}\right) -\widetilde{\mu}\right] },
\label{IBECGF}
\end{equation}
where%
\begin{align}
\widetilde{\varepsilon}\left( \mathbf{k}\right) & =\varepsilon\left( \mathbf{%
k}\right) +\int\frac{\mathrm{d}\mathbf{k}^{\prime}}{\left( 2\pi\right) ^{3}}%
\left[ v(\mathbf{k}-\mathbf{k}^{\prime})-v(-\mathbf{k}^{\prime})\right]
\notag \\
& \times\left( -\frac{1}{\beta\hbar}\right) \sum_{n}e^{i\omega_{n}\eta }%
\mathcal{G}(\mathbf{k}^{\prime},i\omega_{n}).  \label{REnergy}
\end{align}
Eq. (\ref{IBECGF}) shows that $\widetilde{\varepsilon}\left( \mathbf{k}%
\right) $ and $\widetilde{\mu}$ are the renormalized energy and chemical
potential, respectively.

With the help of Eqs. (\ref{HomoKsi}) and (\ref{IBECGF}), Eq. (\ref{ICP})
can be reduced as%
\begin{equation}
\xi^{\dagger}\xi+\int\frac{\mathrm{d}\mathbf{k}}{\left( 2\pi\right) ^{3}}%
\left( -\frac{1}{\beta\hbar}\right) \sum_{n}e^{i\omega_{n}\eta}\mathcal{G}(%
\mathbf{k},i\omega_{n})=n,  \label{ICP1}
\end{equation}
where $n=N/V$ is the particle density of the system.

When $\widetilde{\mu}<0$, Eq. (\ref{IOP1}) has the only solution $\xi=0$. As
shown by Eq. (\ref{IBECS1}), it is a stable solution. In this case, the
system is in its normal phase. The renormalized chemical potential $%
\widetilde{\mu}$ is given by Eq. (\ref{ICP1}),
\begin{equation}
\mathcal{P}\int_{-\infty}^{+\infty}\mathrm{d}\omega\mathcal{N}(\omega ,%
\widetilde{\mu})\frac{1}{e^{\beta\omega}-1}=n,  \label{CPGTc}
\end{equation}
where $\mathcal{P}$ denotes the principal value introduced in Appendix \ref%
{FSum}, and $\mathcal{N}(\omega,\widetilde{\mu})$ the density of states,
\begin{equation}
\mathcal{N}(\omega,\widetilde{\mu})=-\int\frac{\mathrm{d}\mathbf{k}}{\left(
2\pi\right) ^{3}}\mathrm{Im}\mathcal{G}(\mathbf{k},\omega+i0^{+}).
\label{IBECDOS}
\end{equation}
If $\widetilde{\mu}>0$, Eqs. (\ref{IBECS1}) and (\ref{IOP1}) show that the
system will transform into a condensation phase where
\begin{equation}
\left\vert \xi\right\vert ^{2}=\frac{\widetilde{\mu}}{v(0)}.
\end{equation}
Substitution of it into Eq. (\ref{ICP1}) yields
\begin{equation}
\frac{\widetilde{\mu}}{v(0)}+\mathcal{P}\int_{-\infty}^{+\infty}\mathrm{d}%
\omega\mathcal{N}(\omega,\widetilde{\mu})\frac{1}{e^{\beta\omega}-1}=n,
\label{CPLTc}
\end{equation}
which gives the renormalized chemical potential $\widetilde{\mu}$ at $%
T<T_{c} $, where $T_{c}$ is the critical temperature for BEC,%
\begin{equation}
\mathcal{P}\int_{-\infty}^{+\infty}\mathrm{d}\omega\mathcal{N}(\omega ,0)%
\frac{1}{e^{\beta_{c}\omega}-1}=n,  \label{ITc}
\end{equation}
that is to say, $\widetilde{\mu}(T_{c})=0$. In sum, Eqs. (\ref{CPGTc}) and (%
\ref{CPLTc}) yield together the renormalized chemical potential of the
system at $T>T_{c}$ and $T<T_{c}$, respectively.

In order to deduce $\widetilde{\mu }$ and $\xi $ in more detail, let us
suppose further that $v(\mathbf{r}-\mathbf{r}^{\prime })$ is of short range.
Following Ref. \cite{Fetter}, we can approximate $v(\mathbf{k})$ as%
\begin{equation}
v(\mathbf{k})=v(0)\left[ 1-\frac{1}{6}(ka)^{2}\right] ,  \label{IBECSR}
\end{equation}%
where%
\begin{equation}
a^{2}=\frac{\int d\mathbf{r\,r}^{2}v(\mathbf{r})}{\int d\mathbf{r\,}v(%
\mathbf{r})}.
\end{equation}

Inserting Eq. (\ref{IBECSR}) into Eq. (\ref{REnergy}) and Eq. (\ref{RMuW})
leads us to%
\begin{align}
\widetilde{\varepsilon }\left( \mathbf{k}\right) & =\frac{\hbar ^{2}\mathbf{k%
}^{2}}{2m^{\star }}, \\
\widetilde{\mu }& =\mu -\left[ 2v(0)-\frac{v(0)nma^{2}}{3\hbar ^{2}}\frac{%
m^{\star }}{m}\right]  \notag \\
& \times \mathcal{P}\int_{-\infty }^{+\infty }\mathrm{d}\omega \mathcal{N}%
(\omega ,\widetilde{\mu })\frac{\left( \omega +\widetilde{\mu }\right) }{%
e^{\beta \omega }-1},  \label{IApprox2}
\end{align}%
where $m^{\star }$ is the renormalized mass,%
\begin{equation}
\frac{1}{m^{\star }}=\frac{1}{m}\left[ 1-\frac{v(0)nma^{2}}{3\hbar ^{2}}%
\left( 1-\frac{\left\vert \xi \right\vert ^{2}}{n}\right) \right] ,
\label{IApprox3}
\end{equation}%
and $\mathcal{N}(\omega ,\widetilde{\mu })$ the density of states defined by
Eq. (\ref{IBECDOS}),
\begin{equation}
\mathcal{N}(\omega ,\widetilde{\mu })=\left\{
\begin{array}{ll}
\frac{1}{4\pi ^{2}\hbar ^{3}}(2m^{\star })^{3/2}\left( \omega +\widetilde{%
\mu }\right) ^{1/2},\text{ } & \omega \geq -\widetilde{\mu } \\
0,\text{ } & \omega <-\widetilde{\mu }.%
\end{array}%
\right.  \label{IApprox4}
\end{equation}

From Eqs. (\ref{ITc}), (\ref{IApprox3}) and (\ref{IApprox4}), it follows that%
\begin{equation}
\frac{T_{c}}{T_{0}}=1-\frac{v(0)nma^{2}}{3\hbar^{2}},  \label{LowTc}
\end{equation}
where $T_{0}$ stands for the critical temperature of the BEC happening in
the corresponding ideal Bose gas,
\begin{equation}
T_{0}=\frac{2\pi\hbar^{2}}{mk_{B}}\left( \frac{n}{\zeta\left( 3/2\right) }%
\right) ^{2/3}.
\end{equation}
Eq. (\ref{LowTc}) manifests that a repulsive interaction will lower the
transition temperature of BEC,%
\begin{equation}
\gamma\equiv\frac{\Delta T}{T_{0}}=\frac{T_{c}-T_{0}}{T_{0}}=-\frac {%
v(0)nma^{2}}{3\hbar^{2}}<0.
\end{equation}

Also, we have by Eq. (\ref{CPGTc}),%
\begin{equation}
\frac{2}{\sqrt{\pi }\zeta \left( 3/2\right) }t^{3/2}\int_{0}^{+\infty }%
\mathrm{d}x\frac{x^{1/2}}{e^{x-\overline{\mu }/t}-1}=1,  \label{MuGTc}
\end{equation}%
where $t=T/T_{c}$ and $\overline{\mu }=\widetilde{\mu }/\left(
k_{B}T_{c}\right) $. This equation gives the $\widetilde{\mu }$ at $T>T_{c}$%
. If $T<T_{c}$, Eq. (\ref{CPLTc}) must be used in place of Eq. (\ref{CPGTc}),%
\begin{align}
1& =y+\frac{2}{\sqrt{\pi }\zeta \left( 3/2\right) }\left( \frac{1+\gamma }{%
1+\gamma -\gamma y}\right) ^{3/2}  \notag \\
& \times \mathcal{P}\int_{0}^{+\infty }\mathrm{d}x\frac{x^{1/2}}{e^{\left(
x-\nu y\right) /t}-1},  \label{MuLTc}
\end{align}%
where $y=\widetilde{\mu }/\left( v(0)n\right) $, and $\nu =v(0)n/\left(
k_{B}T_{c}\right) $. This equation will yield both the $\widetilde{\mu }$
and $\left\vert \xi \right\vert ^{2}=\widetilde{\mu }/v(0)$ at $T<T_{c}$.

\begin{figure*}[hbtp]
\includegraphics[scale=0.6,angle=-90]{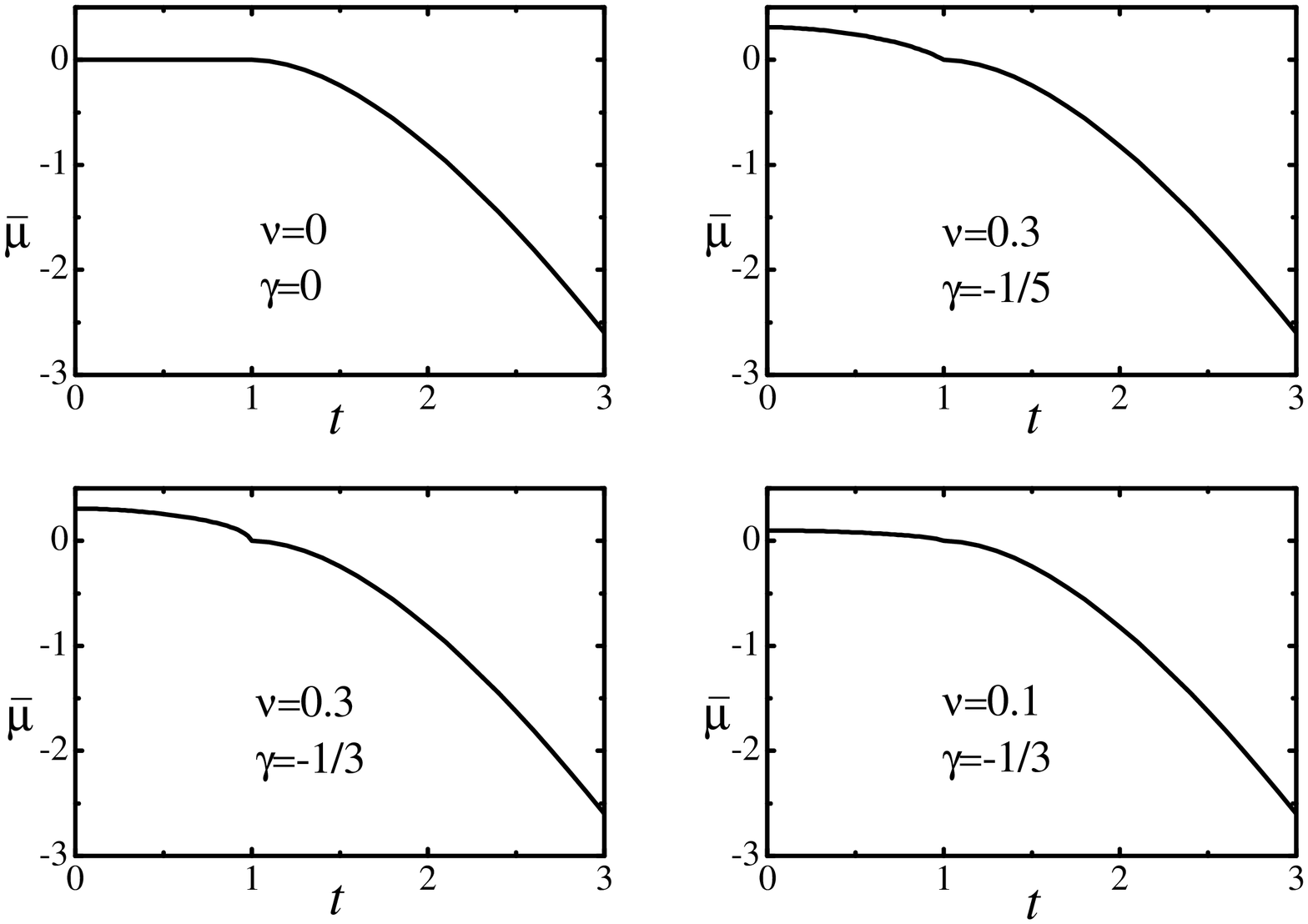}
\caption{
The chemical potential of the system versus
temperature, where $\overline{\mu}=\widetilde{\mu}/\left(
k_{B}T_{c}\right) $, $t=T/T_{c}$, $\nu=v(0)n/\left( k_{B}T_{c}\right) $,
and $\gamma=\Delta T/T_{0}=-v(0)nma^{2}/\left( 3\hbar^{2}\right) $.
For the sake of contrast, the chemical potential of the ideal Bose gas
($\nu=0$, $\gamma=0$) is also plotted.
}
\end{figure*}%

The numerical results of $\widetilde{\mu }$ and $\left\vert \xi \right\vert
^{2}$ are shown in Figs. 4 and 5, respectively. We observe that the
renormalized chemical potential $\widetilde{\mu }$ is a monotonically
decreasing function of temperature, at $T=T_{c}$ itself continuous but its
derivative not. This agrees with the $T\geq T_{c}$ result given by Ref. \cite%
{Fetter} and the references therein. When $T<T_{c}$, the chemical potential $%
\widetilde{\mu }$ is set zero in Ref. \cite{Fetter}, as Einstein \cite%
{Einstein1,Einstein2} did in the ideal Bose gas. In the extended ensemble
theory, there is no mathematical limit on $\widetilde{\mu }$, it will change
with temperature even if $T<T_{c}$ so as to ensure the conservation of
particles, as pointed out in Appendix \ref{FSum}. The behavior of $%
\left\vert \xi \right\vert ^{2}$ shown in Fig. 5 is just expected for an
order parameter.

\begin{figure*}[htbp]
\includegraphics[scale=0.6,angle=-90]{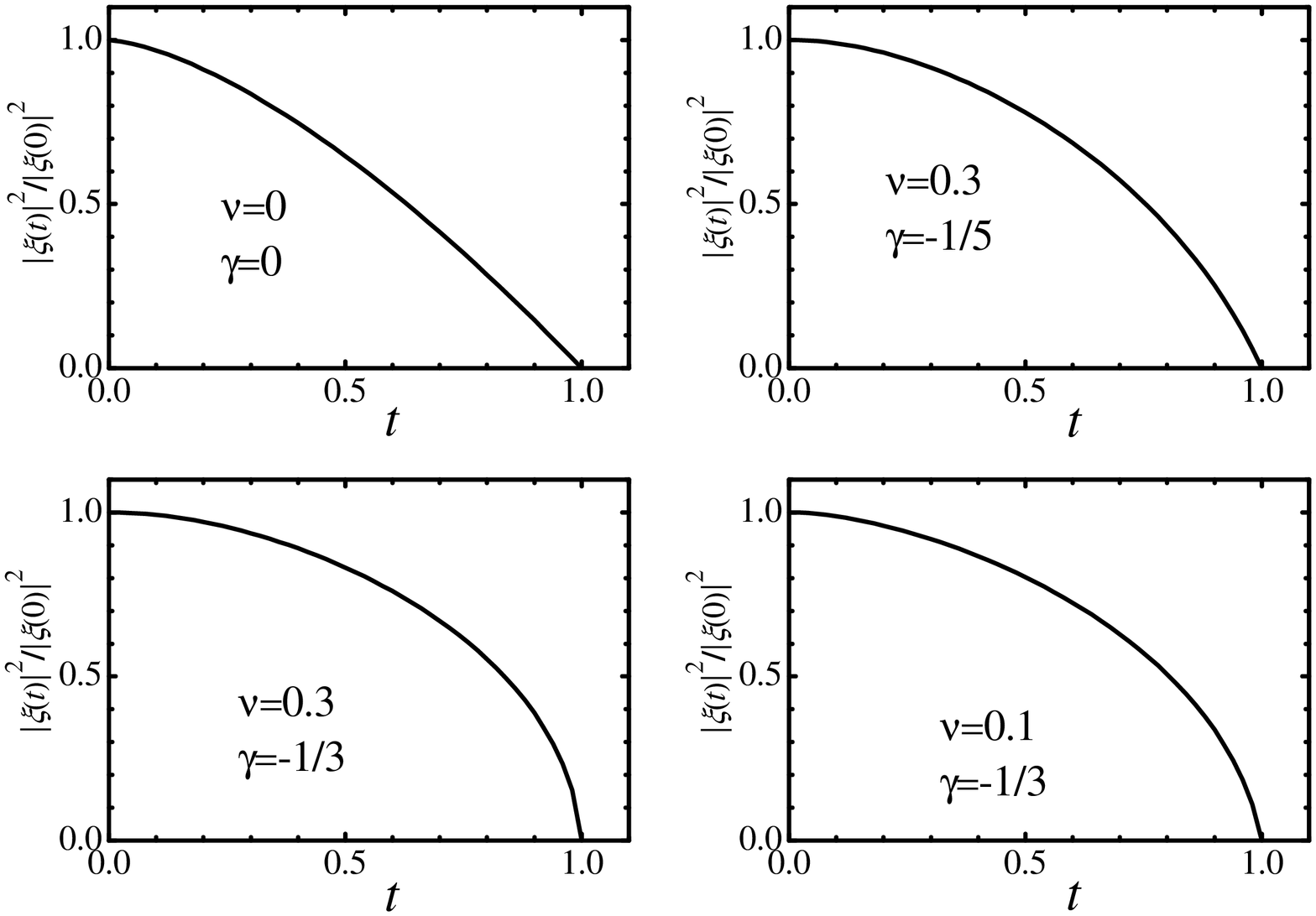}
\caption{
The order parameter for BEC versus temperature,
where $t=T/T_{c}$, $\nu=v(0)n/\left( k_{B}T_{c}\right) $, and $\gamma=\Delta
T/T_{0}=-v(0)nma^{2}/\left( 3\hbar^{2}\right) $.
For the sake of contrast, the order parameter for the ideal Bose gas
($\nu=0$, $\gamma=0$) is also plotted.
}
\end{figure*}%

The internal energy of the system, $E$, can be derived conveniently from $%
E=\langle H\rangle +\mu N$,
\begin{align}
\varepsilon & =\nu +\frac{2}{\sqrt{\pi }\zeta \left( 3/2\right) }  \notag \\
& \times \int_{0}^{+\infty }\mathrm{d}x\frac{x^{3/2}}{e^{\left( x-\overline{%
\mu }\right) /t}-1},\text{ }T\geq T_{c},  \label{EGTc} \\
\varepsilon & =\nu -\frac{\overline{\mu }^{2}}{2\nu }+\frac{2}{\sqrt{\pi }%
\zeta \left( 3/2\right) }\left( \frac{1+\gamma }{1+\gamma -\gamma y}\right)
^{5/2}  \notag \\
& \times \mathcal{P}\int_{0}^{+\infty }\mathrm{d}x\frac{x^{3/2}}{e^{\left( x-%
\overline{\mu }\right) /t}-1},\text{ }T\leq T_{c},  \label{ELTc}
\end{align}%
where%
\begin{equation}
\varepsilon =\frac{1}{k_{B}T_{c}}\frac{E}{N}.
\end{equation}%
Combination of Eqs. (\ref{MuGTc}) and (\ref{EGTc}) gives $C_{V}$, the
specific heat at constant volume,
\begin{equation}
\frac{C_{V}}{Nk_{B}}=\frac{15}{4}\frac{g_{5/2}(z)}{g_{3/2}(z)}-\frac{9}{4}%
\frac{g_{3/2}(z)}{g_{1/2}(z)},\text{ }T\geq T_{c},
\end{equation}%
where%
\begin{gather}
z=e^{\overline{\mu }/t}, \\
g_{n}(z)=\frac{1}{\Gamma (n)}\int_{0}^{+\infty }\mathrm{d}x\frac{x^{n-1}}{%
z^{-1}e^{x}-1}.
\end{gather}%
This result is the same in form as that for the ideal Bose gas, which is
familiar in standard books on quantum statistical mechanics. If $T\leq T_{c}$%
, $C_{V}$ must be calculated from Eqs. (\ref{MuLTc}) and (\ref{ELTc}).

\begin{figure*}[htbp]
\hspace*{-0.6cm}\includegraphics[scale=0.6,angle=-90]{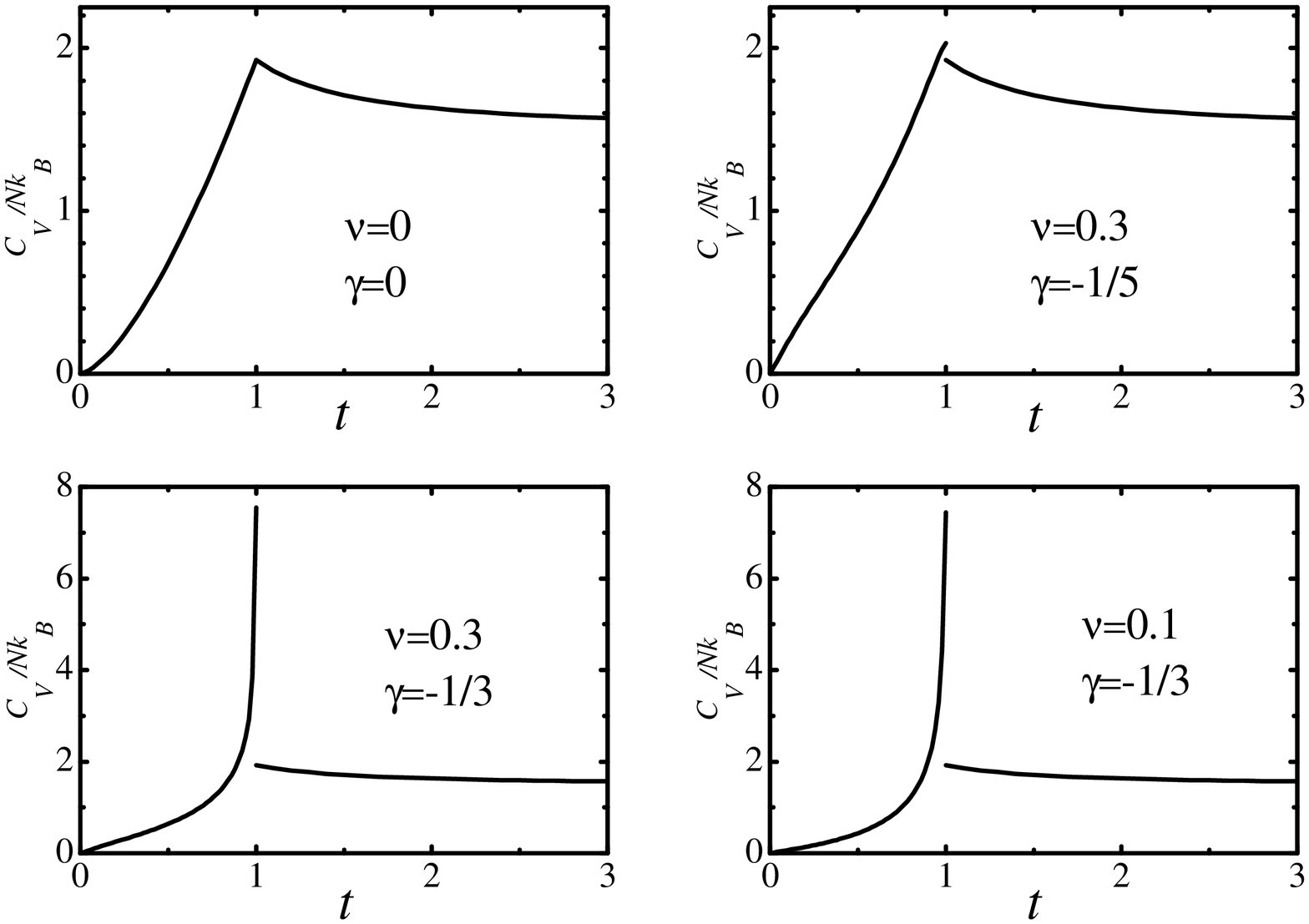}
\caption{
The specific heat of a weakly interacting Bose gas
versus temperature, where $t=T/T_{c}$, $\nu=v(0)n/\left( k_{B}T_{c}\right) $,
and $\gamma=\Delta T/T_{0}=-v(0)nma^{2}/\left( 3\hbar^{2}\right) $.
For the sake of contrast, the specific heat of the ideal Bose gas
($\nu=0$, $\gamma=0$) is also plotted.
}
\end{figure*}%

The numerical results of $C_{V}$ are shown in Fig. 6. The shape of $C_{V}$
shows that the BEC occurring in the interacting Bose gas is a
\textquotedblleft $\lambda $\textquotedblright -transition, as is naturally
expected. In contrast to the ideal Bose gas, the weakly interacting Bose gas
manifests itself with two new features: First, $C_{V}$ is discontinuous at $%
T=T_{c}$; and second, $C_{V}$ is linear in temperature when $T\ll T_{c}$.
Evidently, this theoretical result for $C_{V}$ meets the requirement of the
third law of thermodynamics: $C_{V}\rightarrow 0$ as $T\rightarrow 0$. The
discontinuity at $T=T_{c}$ is obviously due to the discontinuity of the
derivatives of $\widetilde{\mu }$ and $\left\vert \xi \right\vert ^{2}$ with
respect to $T$. As to the linearity at low temperatures, it can be explained
as follows.

Observing that
\begin{widetext}%
\begin{equation}
\mathcal{P}\int_{0}^{+\infty }\mathrm{d}x\frac{x^{1/2}}{e^{\left( x-\nu
y\right) /t}-1}=-\int_{0}^{\nu y}x^{1/2}\mathrm{d}x+t\int_{0}^{+\infty }%
\mathrm{d}x\frac{\left( 2\nu y+tx\right) ^{1/2}}{e^{x+\nu y/t}-1}%
+t\int_{0}^{\nu y/t}\mathrm{d}x\frac{\left( \nu y+tx\right) ^{1/2}-\left(
\nu y-tx\right) ^{1/2}}{e^{x}-1},
\end{equation}%
we obtain%
\begin{equation}
\mathcal{P}\int_{0}^{+\infty }\mathrm{d}x\frac{x^{1/2}}{e^{\left( x-\nu
y\right) /t}-1}=-\frac{2}{3}\left( \nu y\right) ^{3/2}+\zeta (2)\left( \nu
y\right) ^{-1/2}t^{2},  \label{X12LTc}
\end{equation}%
\end{widetext}
to the second power of $t$ as $t\ll 1$. Inserting it into Eq. (\ref{MuLTc}),
we have
\begin{align}
y& =1-\frac{2}{\sqrt{\pi }\zeta \left( 3/2\right) }\left( \frac{1+\gamma }{%
1+\gamma -\gamma y}\right) ^{3/2}  \notag \\
& \times \left[ -\frac{2}{3}\left( \nu y\right) ^{3/2}+\zeta (2)\left( \nu
y\right) ^{-1/2}t^{2}\right] ,  \label{YLTc}
\end{align}%
it is an iterative equation of $y$. At absolute zero ($t=0$), it reduces to%
\begin{equation}
y_{0}=1+\frac{4}{3\sqrt{\pi }\zeta \left( 3/2\right) }\left( \frac{1+\gamma
}{1+\gamma -\gamma y_{0}}\right) ^{3/2}\left( \nu y_{0}\right) ^{3/2}.
\end{equation}%
This gives the zeroth order solution of $y$, which we denoted by $y_{0}$.
Iterating of Eq. (\ref{YLTc}) to the second power of $t$ leads to%
\begin{equation}
y=y_{0}-\frac{2\zeta (2)}{\sqrt{\pi }\zeta \left( 3/2\right) }\left( \frac{%
1+\gamma }{1+\gamma -\gamma y_{0}}\right) ^{3/2}\left( \nu y_{0}\right)
^{-1/2}t^{2},  \label{YLTc1}
\end{equation}%
which shows that $y$, i.e., $\widetilde{\mu }$ and $\left\vert \xi
\right\vert ^{2}$, decreases as $T^{2}$ in the low temperature region ($T\ll
T_{c}$), as can also be seen directly from Figs. 4 and 5. With the same
procedure, one finds%
\begin{equation}
\mathcal{P}\int_{0}^{+\infty }\mathrm{d}x\frac{x^{3/2}}{e^{\left( x-\nu
y\right) /t}-1}=-\frac{2}{5}\left( \nu y\right) ^{5/2}+3\zeta (2)\left( \nu
y\right) ^{1/2}t^{2}.  \label{X32LTc}
\end{equation}%
From Eqs. (\ref{YLTc1}), (\ref{X32LTc}), and (\ref{ELTc}), it follows that $%
\varepsilon $ increases as $T^{2}$. Therefore, the specific heat $C_{V}$
will be linear in temperature $T$ when $T\ll T_{c}$.

If one expands Eqs. (\ref{X12LTc}) and (\ref{X32LTc}) further to the fourth
power of $t$, he arrives at%
\begin{equation}
C_{V}=a(V)T+b(V)T^{3},  \label{CvTV}
\end{equation}
where the expansion coefficients $a(V)$ and $b(V)$ are functions of $V$.
Appendix \ref{TAWIBSLT} shows that this expansion does not depend on the
approximations used here, it is valid provided that the interaction is weak.
Therefore, it is a fundamental property of the weakly interacting Bose gas
that the specific heat $C_{V}$ vanishes linearly as $T\rightarrow0$.

In connection with the $\lambda$-transition occurring in liquid $^{4}\mathrm{%
He}$, obviously, the shape of $C_{V}$ agrees well with that of the specific
heat observed experimentally \cite{Keesom}. However, the latter is $C_{P}$,
the specific heat at constant pressure rather than at constant volume. As is
well known from thermodynamics, they differ from each other by a
temperature-dependent term,%
\begin{equation}
C_{P}=C_{V}+T\left( \frac{\partial p}{\partial T}\right) _{V}\left( \frac{%
\partial V}{\partial T}\right) _{P}.  \label{CpCvDiffer}
\end{equation}
For solids at low temperatures,%
\begin{equation}
C_{P}\approx C_{V},
\end{equation}
because
\begin{equation}
\left( \frac{\partial V}{\partial T}\right) _{P}\approx0.
\end{equation}
That is why the Debye $T^{3}$ law for the phonons, the $T$ law for the
electron gases in normal metals, and the exponential law for BCS
superconductors are observed experimentally in $C_{P}$ albeit they are the
laws of $C_{V}$ at low temperatures. In other words, the theoretical result
of $C_{V}$ for a solid can be verified straightforwardly by the experiments.
However, for a liquid, the situation is different, both the volume and
pressure are much more sensitive to temperature. Therefore, the second term
on the right-hand side of Eq. (\ref{CpCvDiffer}) can not be neglected any
longer, the difference between $C_{P}$ and $C_{V}$ becomes important and
should be taken into account. As a result, to compare the theory with the
experiment for a liquid, the experimental data for $C_{V}$ or the
theoretical result for $C_{P}$ are needed. Regrettably, $C_{V}$ is difficult
to measure experimentally, and $C_{P}$ is difficult to calculate
theoretically. It is thus hard to compare the theoretical result with the
experimental data. Fortunately, $C_{V}$ tends to zero linearly as $%
T\rightarrow0$ for a weakly interacting Bose gas, we can compare the
theoretical result with the experimental data in the limit $T\rightarrow0$.
That can be seen clearly from the following analyses.

According to the third law of thermodynamics, we have%
\begin{equation}
S_{th}=S_{th}(T,V)=\int_{0}^{T}C_{V}(T^{\prime},V)\frac{\mathrm{d}T^{\prime}%
}{T^{\prime}},
\end{equation}
where, as pointed in Sec. \ref{EET}, $S_{th}$ represents the thermodynamical
entropy. Substituting Eq. (\ref{CvTV}) into this equation, we obtain%
\begin{equation}
S_{th}=a(V)T+\frac{1}{3}b(V)T^{3},
\end{equation}
at low temperatures. Making use of the Maxwell relations,
\begin{gather}
\left( \frac{\partial P}{\partial T}\right) _{V}=\left( \frac{\partial S_{th}%
}{\partial V}\right) _{T}, \\
\left( \frac{\partial V}{\partial T}\right) _{P}=-\left( \frac{\partial
S_{th}}{\partial P}\right) _{T},
\end{gather}
one has%
\begin{align}
\hspace{-1cm}\left( \frac{\partial P}{\partial T}\right) _{V} & =a^{\prime
}(V)T+\frac{1}{3}b^{\prime}(V)T^{3}, \\
\hspace{-1cm}\left( \frac{\partial V}{\partial T}\right) _{P} & =-\left(
\frac{\partial V}{\partial P}\right) _{T}\left[ a^{\prime}(V)T+\frac{1}{3}%
b^{\prime}(V)T^{3}\right] .
\end{align}
Substituting them into Eq. (\ref{CpCvDiffer}) leads us to%
\begin{equation}
C_{P}=a(V)T+\widetilde{b}(V)T^{3},  \label{CpLLTc}
\end{equation}
where the terms with power higher than cube are omitted, and
\begin{equation}
\widetilde{b}(V)=b(V)-\left[ a^{\prime}(V)\right] ^{2}\left( \frac{\partial V%
}{\partial P}\right) _{T}.  \label{SoundContr}
\end{equation}
Eq. (\ref{CpLLTc}) presents the form of $C_{P}$ at low temperatures within
the present theory. It contains a linear and a cubic term of $T$, the former
includes only the contribution from $C_{V}$, and the latter includes both
the contributions from $C_{V}$ and the second term on the right-hand side of
Eq. (\ref{CpCvDiffer}). Heeding that $S_{th}=0$ at $T=0$, we have \cite%
{Lifshitz}%
\begin{equation}
u^{2}=-\frac{V^{2}}{mN}\left( \frac{\partial P}{\partial V}\right) _{T=0},
\end{equation}
where $u$ denotes the velocity of sound. This implies that at low
temperatures the velocity of sound is approximate to%
\begin{equation}
u^{2}\approx-\frac{V^{2}}{mN}\left( \frac{\partial P}{\partial V}\right)
_{T}.
\end{equation}
That is to say, the factor $(\partial V/\partial P)_{T}$ on the right-hand
side of Eq. (\ref{SoundContr}) represents physically the effect of sound.
Therefore, the cubic term of $C_{P}$ includes the contribution from the
sound, which is in accordance with the viewpoint of Landau on quantum Bose
liquid $\mathrm{He}$ II \cite{Landau1,Landau2}. This $T^{3}$ law has already
been observed experimentally in liquid $^{4}\mathrm{He}$ at low temperatures
\cite{Keesom}, but it is not an intrinsic property of the BEC because sound
exists in all fluids. The intrinsic property of the BEC is the linear term
of $C_{P}$, it comes purely from $C_{V}$, and is characteristic of a weakly
interacting Bose gas at $T\ll T_{c}$, as mentioned above. There is no such
linear term in the Landau theory of quantum Bose liquid \cite%
{Landau1,Landau2}. To our knowledge, this linear behavior of $C_{P}$ has not
yet been reported experimentally for liquid $^{4}\mathrm{He}$, it is thus a
prediction of the present theory. Eq. (\ref{CpLLTc}) indicates that the
temperature $T$ has to go much lower for $C_{P}$ to show the linear behavior
than to show the $T^{3}$ law. We believe that it would be observed if the
measuring temperature is lowered enough, and if liquid $^{4}\mathrm{He}$
could be regarded as a weakly interacting Bose gas and the $\lambda$%
-transition were a Bose-Einstein condensation.

\textbf{Remark}: In general, $\left\vert \xi \right\vert ^{2}$ can not be
interpreted as the density of condensed particles. This can be seen from Eq.
(\ref{MuLTc}), it shows that $y>1$ at $T=0\mathrm{K}$, that is,
\begin{equation}
\left\vert \xi \right\vert ^{2}>n
\end{equation}%
at absolute zero. Only in the case of the ideal Bose gas, $\left\vert \xi
\right\vert ^{2}$ can not be greater than the density of particles, i.e.,
\begin{equation}
\left\vert \xi \right\vert ^{2}\leq n
\end{equation}%
at any temperature. Physically, $\xi $ is the order parameter and internal
spontaneous field of the system, there is no \textit{a priori} reason why it
must be interpreted using the the particle density. In fact, any statistical
average of observable, including the particle density, will depend
explicitly on $\xi $ when $T<T_{c}$.

Now, let us discuss the correlation of the BEC to the superfluidity. As
usual, we set%
\begin{equation}
\eta (\mathbf{r})=\sqrt{\rho (\mathbf{r})}e^{i\varphi (\mathbf{r})},
\end{equation}%
where $\rho (\mathbf{r})=\left\vert \eta (\mathbf{r})\right\vert ^{2}$ and $%
\varphi (\mathbf{r})=\arg (\eta (\mathbf{r}))$. With this, we have%
\begin{eqnarray}
\mathbf{j}(\mathbf{r}) &=&-\frac{i\hbar }{2m}\langle (\nabla -\nabla
^{\prime })\psi ^{\dagger }(\mathbf{r})\psi (\mathbf{r}^{\prime })\rangle |_{%
\mathbf{r}^{\prime }=\mathbf{r}}  \notag \\
&=&\frac{\hbar }{m}\rho (\mathbf{r})\nabla \varphi (\mathbf{r}),
\end{eqnarray}%
obviously, $\mathbf{j}(\mathbf{r})$ is the supercurrent density. If $\varphi
(\mathbf{r})$, the phase of the BEC order parameter, is independent of $%
\mathbf{r}$, there is no supercurrent, $\mathbf{j}(\mathbf{r})=0$;
otherwise, $\mathbf{j}(\mathbf{r})\neq 0$. The supercurrent velocity $%
\mathbf{v}$ is given by
\begin{equation}
\mathbf{v}(\mathbf{r})=\frac{\hbar }{m}\nabla \varphi (\mathbf{r}).
\end{equation}%
Since the order parameter $\eta (\mathbf{r})$ is a single-valued function of
the position $\mathbf{r}$, we have%
\begin{equation}
\oint \mathbf{v}(\mathbf{r})\cdot \mathrm{d}\mathbf{r}=l\frac{h}{m},\text{ }%
l\in
\mathbb{Z}
.
\end{equation}%
That is to say, there can exist quantized vortices in the condensed phase if
$\varphi (\mathbf{r})$ depends on $\mathbf{r}$. We remark that there exists
a critical velocity $v_{c}$, the velocity of the supercurrent can not be
greater than it, i.e., $v\leq v_{c}$. That is because the increase of the
supercurrent velocity is unfavorable to the decrease of the entropy of the
system; the larger the $v$ is, the greater the entropy will be, which can be
easily seen from the first integral over $\mathbf{r}$ on the right-hand side
of Eq. (\ref{IBECS}) with, e.g., $\eta (\mathbf{r})=\sqrt{\rho }e^{i\mathbf{k%
}\cdot \mathbf{r}}$ where $\rho =\mathrm{constant}$.

Finally, we shall prove that there can not exist any supercurrent or
quantized vortex in the ideal Bose gas.

For the ideal Bose gas, $v(\mathbf{r}-\mathbf{r}^{\prime })=0$, Eq. (\ref%
{IOP}) reduces to%
\begin{equation}
-\frac{\hbar ^{2}}{2m}\nabla ^{2}\eta (\mathbf{r})=\mu \eta (\mathbf{r}),
\label{Keig}
\end{equation}%
obviously, it is the eigenvalue equation of the kinetic energy operator.
This equation can be rewritten as%
\begin{equation}
\frac{1}{2m}\int \mathrm{d}\mathbf{r\,}[-i\hbar \nabla \eta (\mathbf{r}%
)]^{\dagger }\cdot \lbrack -i\hbar \nabla \eta (\mathbf{r})]=\mu \int
\mathrm{d}\mathbf{r\,}\eta ^{\dagger }(\mathbf{r})\eta (\mathbf{r}).
\end{equation}%
It indicates that, for any nontrivial eigenfunction, the chemical potential $%
\mu $ can not be negative, i.e., $\mu \geq 0$. Therefore, there can only
exist the trivial solution, $\eta (\mathbf{r})=0$, when $\mu <0$. The
nontrivial solution, $\eta (\mathbf{r})\neq 0$, can exist only when $\mu
\geq 0$ for the ideal Bose gas.

To discuss the stabilities of those solutions of Eq. (\ref{Keig}), one needs
to investigate whether $\Delta S\geq 0$. From Eq. (\ref{IBECS}), we find
\begin{equation}
\Delta S=\beta \int \mathrm{d}\mathbf{r}\left[ \frac{\hbar ^{2}}{2m}\nabla
\delta \eta ^{\dagger }(\mathbf{r})\cdot \nabla \delta \eta (\mathbf{r})-\mu
\delta \eta ^{\dagger }(\mathbf{r})\delta \eta (\mathbf{r})\right] ,
\end{equation}%
where $\delta \eta (\mathbf{r})$ represents the variation of the order
parameter $\eta (\mathbf{r})$ from any one of the solutions of Eq. (\ref%
{Keig}). Evidently, $\Delta S\geq 0$ when $\mu \leq 0$, this means that the
system is stable when $\mu \leq 0$. Since $\eta (\mathbf{r})=0$ when $\mu <0$
and $\eta (\mathbf{r})=\mathrm{constant}$ when $\mu =0$, the normal phase
and the homogeneous condensed phase are both stable, which is in accordance
with the results of Sec. \ref{TLC}. As to the case of $\mu >0$, the solution
of Eq. (\ref{Keig}) is inhomogeneous, which is just the case we are now
interested in because an inhomogeneous solution can produce supercurrent and
quantized vortices. To determine whether there can exist any supercurrent
and quantized vortex in the condensed phase, we first reformulate the above
equation as follows,%
\begin{equation}
\Delta S=\beta \int \mathrm{d}\mathbf{r}\left[ \delta \eta ^{\dagger }(%
\mathbf{r})\left( -\frac{\hbar ^{2}}{2m}\nabla ^{2}\right) \delta \eta (%
\mathbf{r})-\mu \delta \eta ^{\dagger }(\mathbf{r})\delta \eta (\mathbf{r})%
\right] ,
\end{equation}%
and then examine the eigenvalue problem,
\begin{eqnarray}
-\frac{\hbar ^{2}}{2m}\nabla ^{2}\delta \eta (\mathbf{r}) &=&E\delta \eta (%
\mathbf{r}),  \label{Bessel} \\
\delta \eta (\mathbf{r})|_{r\rightarrow +\infty } &=&0.  \label{boundary}
\end{eqnarray}%
Eq. (\ref{boundary}) is the boundary condition obeyed by the variation $%
\delta \eta (\mathbf{r})$. Obviously, this eigenvalue problem always has
solutions when $E>0$, e.g.,%
\begin{equation}
\delta \eta (\mathbf{r})=j_{l}(kr),\text{ }k=\sqrt{\frac{2m}{\hbar ^{2}}E},
\end{equation}%
where $j_{l}(z)$ is the spherical Bessel function. Accompanying those
solutions, $\Delta S$ can be reexpressed as%
\begin{equation}
\Delta S=\beta \int \mathrm{d}\mathbf{r}(E-\mu )\delta \eta ^{\dagger }(%
\mathbf{r})\delta \eta (\mathbf{r}),
\end{equation}%
clearly,%
\begin{equation}
\left\{
\begin{array}{ll}
\Delta S>0,\text{ } & E>\mu \\
\Delta S=0,\text{ } & E=\mu \\
\Delta S<0,\text{ } & E<\mu .%
\end{array}%
\right.
\end{equation}%
It shows that any solution of Eq. (\ref{Keig}) for $\mu >0$ belongs to the
saddle points of the system entropy. In other words, any inhomogeneous
condensed phase is physically instable, needless to say, the supercurrent
and quantized vortex. This proves that there can not exist any supercurrent
or quantized vortex in the ideal Bose gas.

By the way, the above analyses demonstrate that the condensed phase of the
ideal Bose gas must be homogeneous. There is no inhomogeneous condensation
in the ideal Bose gas. This result is very natural, it confirms the
Einstein's deep physical insight on BEC, again.

\subsection{Double-well Potential and the BEC in Configuration Space}

From Sec. \ref{IPGSB}, it is learned that an ideal phonon gas cannot produce
BEC. However, the ideal phonon gas is simply a harmonic approximation to an
actual solid. Generally, the interatomic interaction is anharmonic. It thus
raises the question as to whether an anharmonic system can produce BEC. We
intend to discuss the question in this subsection.

To that end, it is helpful to make a representation transformation to the
formulation used in Sec. \ref{IPGSB}. There, what we used is the Fock space
where all the observables are expressed in terms of creation and
annihilation operators. In the following, we prefer to use the phase space
where all the observables are expressed in terms of generalized coordinates
and momenta. As is well known, the two spaces are equivalent and can be
transformed into each other according to such a rule \cite{Berezin},
\begin{equation}
\left\{
\begin{array}{l}
q_{i}=\sqrt{\frac{\hbar }{2m_{i}\omega _{i}}}(a_{i}+a_{i}^{\dagger }) \\
p_{i}=-i\sqrt{\frac{m_{i}\hbar \omega _{i}}{2}}(a_{i}-a_{i}^{\dagger }),%
\end{array}%
\right.  \label{Transform1}
\end{equation}%
and
\begin{equation}
\left\{
\begin{array}{l}
a_{i}=\sqrt{\frac{m_{i}\omega _{i}}{2\hbar }}q_{i}+i\frac{1}{\sqrt{%
2m_{i}\hbar \omega _{i}}}p_{i} \\
a_{i}^{\dagger }=\sqrt{\frac{m_{i}\omega _{i}}{2\hbar }}q_{i}-i\frac{1}{%
\sqrt{2m_{i}\hbar \omega _{i}}}p_{i},%
\end{array}%
\right.  \label{Transform2}
\end{equation}%
where $q_{i}$ and $p_{i}$ are the $i$th pair of coordinate and momentum; $%
a_{i}$ and $a_{i}^{\dagger }$ the corresponding annihilation and creation
operators; $m_{i}$ the effective mass; and $\omega _{i}>0$ the $i$th
parameter, which can generally take any positive value, and particularly the
natural frequency if $q_{i}$ and $p_{i}$ constitute a harmonic oscillator.
Along with Eqs. (\ref{Transform1}) and (\ref{Transform2}), the main
quantities concerned are transformed as follows,%
\begin{align}
F(a)& \Longleftrightarrow F(q,p), \\
H(a)& \Longleftrightarrow H(q,p), \\
D(\xi ,a)& \Longleftrightarrow D(\eta ,\zeta ,q,p),  \label{PTD} \\
S(\xi ,\beta )& \Longleftrightarrow S(\eta ,\zeta ,\beta ),
\label{PSEntropy}
\end{align}%
where%
\begin{gather}
D(\xi ,a)=\sum\limits_{i}(\xi _{i}^{\dag }a_{i}+\xi _{i}a_{i}^{\dag }), \\
D(\eta ,\zeta ,q,p)=\frac{1}{\hbar }\sum\limits_{i}(\eta _{i}q_{i}+\zeta
_{i}p_{i}), \\
\left\{
\begin{array}{l}
\eta _{i}=\sqrt{\frac{m_{i}\hbar \omega _{i}}{2}}(\xi _{i}^{\dag }+\xi _{i})
\\
\zeta _{i}=i\sqrt{\frac{\hbar }{2m_{i}\omega _{i}}}(\xi _{i}^{\dag }-\xi
_{i}).%
\end{array}%
\right.
\end{gather}%
Here, $\xi $ represents the order parameter for BEC in Fock space, $\eta
_{i} $ and $\zeta _{i}$ the corresponding order parameters in phase space.

Under the action of $D(\eta,\zeta,q,p)$, the statistical averages of $q_{i} $
and $p_{i}$ will change with temperature as
\begin{equation}
\left\{
\begin{array}{l}
\left\langle q_{i}\right\rangle =-\zeta_{i}+\mathrm{Tr}\big(q_{i}\rho\left(
H(q,p)\right) \big) \\
\left\langle p_{i}\right\rangle =\eta_{i}+\mathrm{Tr}\big(p_{i}\rho\left(
H(q,p)\right) \big).%
\end{array}
\right.  \label{Distortions}
\end{equation}
If the system produces a BEC, i.e., $\eta_{i}$ and/or $\zeta_{i}$ change
from zero to nonzero with the decreasing of temperature, the average
positions of the particles will redistribute in phase space. In this sense,
the BEC is said to be the phase-space BEC, it changes the distribution of
the system in phase space. If, however, the BEC only causes $\eta_{i}$ (or $%
\zeta_{i}$) to change from zero to nonzero, the particles will redistribute
merely in momentum space (or configuration space), the subspace of phase
space. In such a case, the BEC is said to be the momentum-space BEC (or the
configuration-space BEC), it changes only the distribution of the system in
momentum space (or configuration space). Obviously, once a BEC changes the
distribution of a system in configuration space ($\zeta_{i}\neq0$, $%
\eta_{i}=0$ or not), it induces a structural phase transition (SPT) \cite%
{Blinc}. In this connection, it can be said that a SPT is just an instance
of BEC.

Commonly, a system Hamiltonian has the form,
\begin{equation}
H(q,p)=\sum_{i}\frac{p_{i}^{2}}{2m_{i}}+\sum_{i}u\left( q_{i}\right) +\frac{1%
}{2}\sum_{i,j}v(q_{i}-q_{j}),  \label{SPHamiltonian}
\end{equation}%
where $u\left( q_{i}\right) $ and $v(q_{i}-q_{j})$ stand for the single- and
two-particle potentials, respectively. One can easily verify, using Eqs. (%
\ref{Variation}), (\ref{Increment}), (\ref{PTD}) and (\ref{PSEntropy}), that
$\eta _{i}$ must be zero at any temperature, i.e., there cannot arise the
momentum-space BEC. In consequence, the system with such a form of
Hamiltonian can, at most, produce a configuration-space BEC. Therefore, a
SPT is commonly a configuration-space BEC. In the following, we shall
confine our attention within this kind of SPT. The phase-transition operator
$D(\eta ,\zeta ,q,p)$ can now be simplified as
\begin{equation}
D(\zeta ,p)=\frac{1}{\hbar }\sum\limits_{i}\zeta _{i}p_{i},  \label{PSD}
\end{equation}%
for $\eta _{i}$ vanishes forever.

In the field of SPT, the Hamiltonian of Eq. (\ref{SPHamiltonian}) is made
concrete as \cite{Blinc},
\begin{align}
H(q,p)& =\sum_{i}\frac{p_{i}^{2}}{2m}+\sum_{i}\left( \frac{1}{2}\mu
^{2}q_{i}^{2}+\frac{1}{4}gq_{i}^{4}\right)  \notag \\
& +\sum_{\langle ij\rangle }\frac{1}{2}\lambda (q_{i}-q_{j})^{2},
\label{PSHamiltonian}
\end{align}%
where $\mu ^{2}$, $g$, and $\lambda $ are constants, and $\langle ij\rangle $
denotes the nearest neighbor sites. The system gets anharmonic if $\mu
^{2}<0 $ or $g\neq 0$. The terms included by the second sum represent the
single-particle potential that arises from an underlying sublattice of
atoms, which do not participate actively in the phase transition, and the
one included by the third sum represents the elastic coupling between
nearest neighbors, which is described by a harmonic potential with the
coupling constant $\lambda \geq 0$. Obviously, $H(q,p)$ has the discrete
symmetry,
\begin{equation}
q_{i}\Longrightarrow q_{i}^{\prime }=e^{i\frac{\pi }{\hbar }%
\sum\limits_{i}\left( \frac{1}{2}p_{i}^{2}+\frac{1}{2}q_{i}^{2}\right)
}q_{i}e^{-i\frac{\pi }{\hbar }\sum\limits_{i}\left( \frac{1}{2}p_{i}^{2}+%
\frac{1}{2}q_{i}^{2}\right) }=-q_{i},
\end{equation}%
it leads us to
\begin{equation}
\overline{q_{i}}=0,  \label{QBar}
\end{equation}%
where
\begin{equation}
\overline{F(q,p)}=\mathrm{Tr}\big(F(q,p)\rho \left( H(q,p)\right) \big).
\end{equation}%
This discrete symmetry will break down if a SPT takes place.

Let us begin with the simple case of $\lambda=0$. Making use of Eq. (\ref%
{PSD}), we have
\begin{widetext}%
\begin{equation}
S(\zeta,\beta)=-\overline{\ln\left( \rho\left( H(q,p)\right) \right) }%
+\beta\sum_{i}\left\{ \frac{1}{2}\left[ \mu^{2}+3g\left( \overline {q_{i}^{2}%
}-\overline{q_{i}}^{2}\right) \right] \zeta_{i}^{2}+\frac{1}{4}%
g\zeta_{i}^{4}\right\} .  \label{PSS}
\end{equation}%
\end{widetext}
It shows that the entropy $S(\zeta,\beta)$ has the typical form of Landau
free energy \cite{Landau}. From Eqs. (\ref{PSS}), (\ref{Variation}), and (%
\ref{Increment}), one can draw three conclusions.

\begin{enumerate}
\item The case of $g<0$. The system is instable and will collapse because
the entropy is unbounded from below with respect to the variation of $%
\zeta_{i}$.

\item The case of $g=0$. (a) If $\mu ^{2}>0$, the trivial solution $\zeta
_{i}=0$ is the only minimizer of $S(\zeta ,\beta )$. That is to say, a
system consisting of harmonic oscillators, or rather an ideal phonon gas,
can not produce SPT, it will stay in the normal phase forever, which is in
accordance with the result of Sec. \ref{IPGSB}. (b) If $\mu ^{2}=0$, $\zeta
_{i}$ can take a temperature-independent value, it represents, as indicated
by Eqs. (\ref{Distortions}) and (\ref{QBar}), the equilibrium position of $%
q_{i}$, i.e., $\left\langle q_{i}\right\rangle =-\zeta _{i}$. As the initial
equilibrium position of $q_{i}$ at $\beta =0$, $-\zeta _{i}$ can be chosen
as zero. Therefore, a system composed of free distinguishable particles can
not produce SPT, as is just expected. (c) If $\mu ^{2}<0$, $\zeta _{i}=0$ is
the only solution, but it is a maximizer of $S(\zeta ,\beta )$, the system
is instable and will collapse. This result is quite natural since all
particles go in a reversed harmonic potential.

\item The case of $g>0$. (a) If $\mu^{2}\geq0$, $\zeta_{i}=0$ is the only
solution, and it is a minimizer of $S(\zeta,\beta)$. There is no phase
transition in the system, the potential is a single well. (b) If $\mu^{2}<0$%
, the potential gets double welled, and it must be determined by the
position fluctuation $\overline{q_{i}^{2}}-\overline{q_{i}}^{2}$ whether the
system can produce a phase transition. Evidently, a transition can occur if
the fluctuation is such a decreasing function of temperature that $\mu
^{2}+3g(\overline{q_{i}^{2}}-\overline{q_{i}}^{2})$ can change in sign, from
positive to negative. If the fluctuation is so strong that $\mu^{2}+3g[%
\overline{(q_{i}^{2})}-\left( \overline{q_{i}}\right) ^{2}]>0$ up to zero
temperature, the system has to stay in the normal phase forever.
\end{enumerate}

From the above discussions, it follows that a double-well potential where $%
\mu^{2}<0$ and $g>0$ is the only possible case for the system to produce a
structural phase transition under the condition of $\lambda=0$. One can
easily verify that this conclusion is still valid for $\lambda>0$.

We are now confronted with performing the evaluation of the position
fluctuation $\overline{q_{i}^{2}}-\overline{q_{i}}^{2}$ so as to ascertain
whether a phase transition can occur or not if $\mu ^{2}<0$ and $g>0$. To
this end, we need to perform a representation transformation to the
Hamiltonian of Eq. (\ref{PSHamiltonian}) because, as well-known, this form
of Hamiltonian has no good basic part and is thus unfavorable for doing
approximations. Complying with quantum field theory, the transformation can
be achieved as follows,
\begin{equation}
\left\{
\begin{array}{l}
q_{i}\implies \phi _{i}=e^{\frac{i}{\hbar }\sum\limits_{j}\nu p_{j}}q_{i}e^{-%
\frac{i}{\hbar }\sum\limits_{j}\nu p_{j}}=q_{i}+\nu \\
p_{i}\implies \pi _{i}=e^{\frac{i}{\hbar }\sum\limits_{j}\nu p_{j}}p_{i}e^{-%
\frac{i}{\hbar }\sum\limits_{j}\nu p_{j}}=p_{i},%
\end{array}%
\right.  \label{qpt}
\end{equation}%
where
\begin{equation}
\nu =\sqrt{-\mu ^{2}/g}.
\end{equation}%
In the sense of classical mechanics, the above transformation amounts to
performing a translation so that the new coordinate $\phi _{i}$ is
referenced from $q_{i}=-\nu $, one of the two minimizers of the double-well
potential. Along with Eq. (\ref{qpt}), the new representation $H(\phi ,\pi )$
of the Hamiltonian becomes%
\begin{align}
H(\phi ,\pi )& =e^{\frac{i}{\hbar }\sum\limits_{j}\nu p_{j}}H(q,p)e^{-\frac{i%
}{\hbar }\sum\limits_{j}\nu p_{j}}  \notag \\
& =\sum_{i}\left( \frac{p_{i}^{2}}{2m}-\mu ^{2}q_{i}^{2}\right)
+\sum_{i}(g\nu q_{i}^{3}+\frac{1}{4}gq_{i}^{4})  \notag \\
& -\sum_{i}\frac{1}{4}g\nu ^{4}.
\end{align}%
The first sum on the right-hand side represents harmonic oscillators with
natural frequencies $\omega _{i}=\sqrt{-2\mu ^{2}/m}$, it constitutes a
basic Hamiltonian which is favorable for approximation handling, the second
the interaction, and the third an unimportant constant.

With $H(\phi,\pi)$, one finds
\begin{equation}
\overline{q_{i}^{2}}-\overline{q_{i}}^{2}=\widetilde{\phi_{i}^{2}}-%
\widetilde{\phi_{i}}^{2}=\widetilde{q_{i}^{2}}-\widetilde{q_{i}}^{2},
\label{qfluctuation}
\end{equation}
where $\widetilde{F}$ stands for the statistical average with respect to $%
H(\phi,\pi)$,%
\begin{equation}
\widetilde{F}=\mathrm{Tr}\big(F(q,p)\rho(\widetilde{H}(q,p))\big),
\end{equation}
where $\widetilde{H}(q,p)\equiv H(\phi,\pi)$. According to Eq. (\ref%
{qfluctuation}), we now just need to evaluate the fluctuation $\widetilde{%
q_{i}^{2}}-\widetilde{q_{i}}^{2}$, it can be worked out by the retarded
Green's function $\langle\langle q_{i}|q_{i}\rangle\rangle_{\omega}$ which
is defined with respect to $\widetilde{H}(q,p)$ \cite{Zubarev}. The Green's
function satisfies the following equation of motion,
\begin{align}
\hbar\omega\langle\langle q_{i}|q_{i}\rangle\rangle_{\omega} & =\langle
\lbrack q_{i},q_{i}]\rangle+\langle\langle\lbrack q_{i},\widetilde {H}%
(q,p)]|q_{i}\rangle\rangle_{\omega}  \notag \\
& =\frac{i\hbar}{m}\langle\langle p_{i}|q_{i}\rangle\rangle_{\omega}.
\end{align}
Also,%
\begin{align}
\hbar\omega\langle\langle p_{i}|q_{i}\rangle\rangle_{\omega} & =\langle
\lbrack p_{i},q_{i}]\rangle+\langle\langle\lbrack p_{i},\widetilde {H}%
(q,p)]|q_{i}\rangle\rangle_{\omega}  \notag \\
& =-i\hbar\big[[1-2\mu^{2}\langle\langle q_{i}|q_{i}\rangle\rangle_{\omega
}+3g\nu\langle\langle q_{i}^{2}|q_{i}\rangle\rangle_{\omega}  \notag \\
& +g\langle\langle q_{i}^{3}|q_{i}\rangle\rangle_{\omega}\big].
\end{align}
We would like to truncate the GF chain with the following decoupling
approximation,%
\begin{gather}
\langle\langle q_{i}^{2}|q_{i}\rangle\rangle_{\omega}=\widetilde{q_{i}}%
\langle\langle q_{i}|q_{i}\rangle\rangle_{\omega}=-\nu\langle\langle
q_{i}|q_{i}\rangle\rangle_{\omega}, \\
\langle\langle q_{i}^{3}|q_{i}\rangle\rangle_{\omega}=\widetilde{q_{i}^{2}}%
\langle\langle q_{i}|q_{i}\rangle\rangle_{\omega}=[(\overline{q_{i}^{2}}-%
\overline{q_{i}}^{2})+\nu^{2}]\langle\langle q_{i}|q_{i}\rangle
\rangle_{\omega},
\end{gather}
where we have used the result $\widetilde{\phi_{i}}=\widetilde{q_{i}}+\nu=0 $%
, which follows from Eq. (\ref{QBar}).\ This approximation leads us to%
\begin{equation}
\langle\langle q_{i}|q_{i}\rangle\rangle_{\omega}=\frac{1}{m}\frac{1}{%
\omega^{2}-\omega_{s}^{2}},
\end{equation}
where
\begin{equation}
\omega_{s}=\left[ \frac{g}{m}(\overline{q_{i}^{2}}-\overline{q_{i}}^{2})%
\right] ^{1/2}  \label{Omegas}
\end{equation}
represents the renormalized frequency. According to the
fluctuation-dissipation theorem \cite{Zubarev}, the fluctuation $\overline {%
q_{i}^{2}}-\overline{q_{i}}^{2}$ can be obtained by%
\begin{equation}
\overline{q_{i}^{2}}-\overline{q_{i}}^{2}=\mathcal{P}\int_{-\infty}^{+\infty
}d\omega\frac{1}{e^{\beta\hbar\omega}-1}\left[ -\frac{\hbar}{\pi}\mathrm{Im}%
\langle\langle q_{i}|q_{i}\rangle\rangle_{\omega+i0}\right] ,
\end{equation}
it gives rise to%
\begin{equation}
\overline{q_{i}^{2}}-\overline{q_{i}}^{2}=\frac{\hbar}{m\omega_{s}}\left(
\frac{1}{2}+\frac{1}{e^{\beta\hbar\omega_{s}}-1}\right) .
\label{Fluctuation}
\end{equation}
Eq. (\ref{Fluctuation}) and Eq. (\ref{Omegas}) constitute a self-consistent
approximation to the fluctuation $\overline{q_{i}^{2}}-\overline{q_{i}}^{2}$%
. Evidently, the first term on the right-hand side of Eq. (\ref{Fluctuation}%
) represents the zero-point fluctuation. Eq. (\ref{Fluctuation}) shows that
the fluctuation gets weaker and weaker when temperature goes lower and
lower. At zero temperature, one has%
\begin{equation}
\overline{q_{i}^{2}}-\overline{q_{i}}^{2}=\frac{1}{\sqrt[3]{4mg}}\hbar
^{2/3}\sim O(\hbar^{2/3}),
\end{equation}
and at high temperatures,
\begin{equation}
\overline{q_{i}^{2}}-\overline{q_{i}}^{2}=\frac{1}{\sqrt{g\beta}}\sim
O(T^{1/2}).
\end{equation}
They imply that there exists a finite temperature $T_{c}$ such that%
\begin{equation}
\left\{
\begin{array}{ll}
\mu^{2}+3g(\overline{q_{i}^{2}}-\overline{q_{i}}^{2})>0,\text{ } & T>T_{c}
\\
\mu^{2}+3g(\overline{q_{i}^{2}}-\overline{q_{i}}^{2})=0,\text{ } & T=T_{c}
\\
\mu^{2}+3g(\overline{q_{i}^{2}}-\overline{q_{i}}^{2})<0,\text{ } & T<T_{c}.%
\end{array}
\right.  \label{SignCh}
\end{equation}
From Eq. (\ref{SignCh}) and Eq. (\ref{PSS}), it follows that a SPT will
occur at $T_{c}$,
\begin{equation}
\zeta_{i}=\left\{
\begin{array}{ll}
0,\text{ } & T\geq T_{c} \\
\pm\left[ -\frac{\mu^{2}}{g}-\frac{3\hbar}{m\omega_{s}}\left( \frac{1}{2}+%
\frac{1}{e^{\beta\hbar\omega_{s}}-1}\right) \right] ^{1/2},\text{ } &
T<T_{c}.%
\end{array}
\right.  \label{Zetai}
\end{equation}
Specifically, $\zeta_{i}=\pm\sqrt{-\mu^{2}/g}$ at zero temperature if the
zero-point fluctuation is neglected, i.e., the order parameter will equal
one of the two minimizers of the double-well potential at zero temperature
when the zero-point fluctuation is left out of consideration.

Combining Eqs. (\ref{Zetai}), (\ref{Distortions}), and (\ref{QBar}), we have
\begin{equation}
\left\langle q_{i}\right\rangle =\left\{
\begin{array}{ll}
0,\text{ } & T\geq T_{c} \\
\mp \left[ -\frac{\mu ^{2}}{g}-\frac{3\hbar }{m\omega _{s}}\left( \frac{1}{2}%
+\frac{1}{e^{\beta \hbar \omega _{s}}-1}\right) \right] ^{1/2},\text{ } &
T<T_{c}.%
\end{array}%
\right.
\end{equation}%
It shows that for each site the distortion $\left\langle q_{i}\right\rangle $
can take independently either of the two values at $T<T_{c}$, there is no
correlation between the distortions of different sites, therefore, the new
phase below $T_{c}$ is a structural glass. Physically, that is because we
have discarded the coupling between nearest neighbors ($\lambda =0$). If it
is taken into account ($\lambda >0$), one can easily verify that the new
phase below the transition temperature will be ferrodistortive:
\begin{equation}
\left\langle q_{i}\right\rangle =\left\langle q_{j}\right\rangle ,
\end{equation}%
where $i$ and $j$ denote any two different sites. This result is in
agreement with the classical mean-field theory \cite{Blinc}.

In conclusion, we find that the double-well potential system described by
the Hamiltonian of Eq. (\ref{PSHamiltonian}) can produce a structural phase
transition. Mechanically, that is because the position fluctuations of
particles will decrease with temperature. In the high-temperature regime,
the fluctuations are relatively strong, the system can only stay in its
normal phase. The fluctuations will get weaker and weaker when temperature
goes lower and lower, and finally the system undergoes a transition at a
finite temperature and then turns into a new ordered phase in the
low-temperature regime. We learn from here that the position fluctuations of
particles play an important role in structural phase transitions. Also, we
see that configuration space is very convenient for describing SPT, in
contrast to the Fock space.

\subsection{Goldstone Bosons, Ginzburg-Landau Equations, and Higgs Mechanism}

Obviously, the preceding theory for the discrete case can be extended
straightforwardly to the continuum case. Let us consider the so-called $O(N)$%
-symmetric vector model \cite{Justin,Tsvelik},
\begin{align}
H\left[ \psi ,\pi \right] & =\int \mathrm{d}\mathbf{x}\text{ }\Big[\frac{1}{2%
}\pi ^{\dagger }\pi +\frac{1}{2}\nabla \psi ^{\dagger }\cdot \nabla \psi
\notag \\
& +\frac{m^{2}}{2}\psi ^{\dagger }\psi +\frac{g}{4}(\psi ^{\dagger }\psi
)^{2}\Big],
\end{align}%
where $m^{2}<0$ and $g>0$ are the two constant parameters for a double-well
potential, $\psi (\mathbf{x})$ is usually called Higgs field, it represents
a real-valued vector field with $N$ components, and $\pi (\mathbf{x})$ the
corresponding momentum,
\begin{equation}
\lbrack \psi _{i}(\mathbf{x}),\pi _{j}(\mathbf{x}^{\prime })]=i\delta
_{ij}\delta (\mathbf{x}-\mathbf{x}^{\prime }).
\end{equation}%
As is well known, this double-well Hamiltonian is the typical model for SSB
in quantum field theory \cite{Justin,Tsvelik,Muller}. The arguments given in
the preceding subsection are still valid. They reveal that the physical
reason why the $O(N)$ symmetry will break down is that the fluctuation of
the field $\psi (\mathbf{x})$ decreases monotonically with temperature, and
that this SSB belongs essentially to the BEC happening in configuration
space.

In the condensation phase, the system Hamiltonian takes the form,
\begin{equation}
H^{\prime }\left[ \varphi ,\psi ,\pi \right] =e^{iD\left[ \varphi ,\pi %
\right] }H\left[ \psi ,\pi \right] e^{-iD\left[ \varphi ,\pi \right] },
\label{HPhiPsiPI}
\end{equation}%
where $\varphi (\mathbf{x})$ is the order parameter for Higgs field, and $D%
\left[ \varphi ,\pi \right] $ the phase-transition operator,
\begin{widetext}
\begin{equation}
D\left[ \varphi ,\pi \right] =\int \mathrm{d}\mathbf{x}\text{ }\varphi
^{\dagger }(\mathbf{x})\pi (\mathbf{x}).
\end{equation}%
In more detail, Eq. (\ref{HPhiPsiPI}) can be written as%
\begin{equation}
H^{\prime }\left[ \varphi ,\psi ,\pi \right] =\int \mathrm{d}\mathbf{x}\text{
}\left\{ \frac{1}{2}\pi ^{\dagger }\pi +\frac{1}{2}\nabla (\psi ^{\dagger
}+\varphi ^{\dagger })\cdot \nabla \left( \psi +\varphi \right) +\frac{m^{2}%
}{2}(\psi ^{\dagger }+\varphi ^{\dagger })\left( \psi +\varphi \right) +%
\frac{g}{4}[(\psi ^{\dagger }+\varphi ^{\dagger })\left( \psi +\varphi
\right) ]^{2}\right\} .  \label{HPhiPsiPi1}
\end{equation}%
\end{widetext}%

At absolute zero, if one neglects the zero-point fluctuation of the field $%
\psi $, i.e., $\overline{\psi ^{\dagger }\psi }\approx 0$, then he has the
stable solution,
\begin{equation}
\varphi ^{\dagger }\varphi =\nu ^{2},
\end{equation}%
where%
\begin{equation}
\nu =\sqrt{-m^{2}/g}
\end{equation}%
is the minimizer of the double-well potential. Without loss of generality,
we shall take
\begin{equation}
\varphi ^{\dagger }=\left[ \nu ,0,...,0\right] .
\end{equation}%
Substituting it into Eq. (\ref{HPhiPsiPi1}) leads us to
\begin{widetext}%
\begin{equation}
H^{\prime }\left[ \varphi ,\psi ,\pi \right] =\int \mathrm{d}\mathbf{x}\text{
}\left[ \frac{1}{2}\pi ^{\dagger }\pi +\frac{1}{2}\nabla \psi ^{\dagger
}\cdot \nabla \psi -m^{2}\psi _{1}^{2}+g\nu \psi _{1}\psi ^{\dagger }\psi +%
\frac{g}{4}(\psi ^{\dagger }\psi )^{2}-\frac{g\nu ^{4}}{4}\right] .
\end{equation}%
\end{widetext}
That is the zero-temperature representation of the system Hamiltonian, it
shows that the first component field $\psi _{1}(\mathbf{x})$ gets a mass of $%
-2m^{2}$, the others are massless. As well-known, those massless particles
are the so-called Goldstone bosons \cite{Goldstone}. From here, one can see
how Goldstone bosons are produced in a natural way by spontaneous symmetry
breaking within the framework of the extended ensemble theory. Of course, he
should also recognize that the zero-point fluctuation can impose a little
influence on the Goldstone bosons.

At last, it should be emphasized that, in the $O(N)$-symmetric vector model,
Goldstone bosons originate physically from the Bose-Einstein condensation of
Higgs field.

If a gauge field is coupled with the Higgs field $\psi(\mathbf{x})$, what
will happen to the gauge field? can it produce BEC as the Higgs field $\psi(%
\mathbf{x})$? To clarify this problem, let us consider, for simplicity, the
coupling of a complex-valued (or two-component) Higgs field $\psi (\mathbf{x}%
)$ and an electromagnetic field $\mathbf{A}(\mathbf{x})$,
\begin{widetext}%
\begin{equation}
H\left[ \psi,\pi;\mathbf{A,E}\right] =\int\mathrm{d}\mathbf{x}\text{ }\left[
\pi^{\dagger}\pi+\left( \nabla+ie\mathbf{A}\right) \psi^{\dagger
}\cdot\left( \nabla-ie\mathbf{A}\right) \psi+m^{2}\psi^{\dagger}\psi
+g(\psi^{\dagger}\psi)^{2}+\frac{1}{2}\mathbf{E}^{2}+\frac{1}{2}\left(
\nabla\times\mathbf{A}\right) ^{2}\right] ,
\end{equation}%
\end{widetext}
where $e$ is the charge of the Higgs field $\psi(\mathbf{x})$, and $\mathbf{E%
}(\mathbf{x})$ the electric field strength which plays the role of the
canonical momentum corresponding to the canonical coordinate $\mathbf{A}(%
\mathbf{x})$,%
\begin{equation}
\lbrack\mathrm{A}_{i}(\mathbf{x}),-\mathrm{E}_{j}(\mathbf{x}^{\prime
})]=i\delta_{ij}\delta(\mathbf{x}-\mathbf{x}^{\prime}).
\end{equation}
Here, the temporal gauge has been used to perform a canonical quantization
to the electromagnetic field, an Abelian gauge field. The reason for
choosing this gauge is that it can be easily applied to quantize non-Abelian
gauge fields, and transformed into other gauges through the path-integral
formalism introduced by Faddeev and Popov \cite{Faddeev}.

For this coupled system, BEC can be explored through the phase-transition
operator,
\begin{equation}
D\left[ \varphi,\pi\mathbf{;\Lambda},\mathbf{E}\right] =\int\mathrm{d}%
\mathbf{x}\text{ }\left( \varphi\pi+\varphi^{\dagger}\pi^{\dagger }-\mathbf{%
\Lambda}\cdot\mathbf{E}\right) ,  \label{DPiLambda}
\end{equation}
where $\varphi(\mathbf{x})$ and $\mathbf{\Lambda}(\mathbf{x})$ are the order
parameters for the Higgs field $\psi(\mathbf{x})$ and gauge field $\mathbf{A}%
(\mathbf{x})$, respectively. With $D\left[ \varphi,\pi \mathbf{;\Lambda},%
\mathbf{E}\right] $, one can obtain the entropy of the system,
\begin{widetext}%
\begin{align}
S\left[ \varphi,\mathbf{\Lambda,}\beta\right] & =-\mathrm{Tr}\big(\ln (\rho(H%
\left[ \psi,\pi;\mathbf{A,E}\right] ))\rho(H\left[ \psi ,\pi;\mathbf{A,E}%
\right] )\big)+\int\mathrm{d}\mathbf{x}\Big[\left( \nabla+ie\mathbf{\Lambda}%
\right) \varphi^{\dagger}\cdot\left( \nabla -ie\mathbf{\Lambda}\right)
\varphi  \notag \\
& +(m^{2}+e^{2}\overline{\mathbf{A}\cdot\mathbf{A}}+4g\overline{%
\psi^{\dagger }\psi})\varphi^{\dagger}\varphi+g\left(
\varphi^{\dagger}\varphi\right) ^{2}+e^{2}\overline{\psi^{\dagger}\psi}%
\mathbf{\Lambda}^{2}+\frac{1}{2}\left( \nabla\times\mathbf{\Lambda}\right)
^{2}\Big],  \label{CoupledPhiLambda}
\end{align}
which shows that the two order parameters $\varphi$ and $\mathbf{\Lambda}$
are coupled together. Eq. (\ref{CoupledPhiLambda}) reminds us of the Landau
free energy for superconductivity \cite{GL,Lifshitz}, indeed, they are both
similar in form.

Variation of $S$ with respect to $\varphi$ and $\mathbf{\Lambda}$ yields%
\begin{gather}
-\left( \nabla-ie\mathbf{\Lambda}\right) ^{2}\varphi+(m^{2}+e^{2}\overline{%
\mathbf{A}\cdot\mathbf{A}}+4g\overline{\psi^{\dagger}\psi}%
)\varphi+2g\varphi^{\dagger}\varphi\varphi=0,  \label{GL1} \\
\nabla\times\nabla\times\mathbf{\Lambda}=-ie\left(
\varphi^{\dagger}\nabla\varphi-\varphi\nabla\varphi^{\dagger}\right)
-2e^{2}(\varphi^{\dagger }\varphi+\overline{\psi^{\dagger}\psi})\mathbf{%
\Lambda.}  \label{GL2}
\end{gather}%
\end{widetext}
This form of two coupled nonlinear equations is known as Ginzburg-Landau
equations \cite{GL,Lifshitz}. Observing that%
\begin{equation}
\langle\mathbf{B}\rangle=-\nabla\times\mathbf{\Lambda,}
\end{equation}
where $\mathbf{B}$ denotes the magnetic field strength, one can specify%
\begin{equation}
\nabla\cdot\mathbf{\Lambda}=0,  \label{CoulombLambda}
\end{equation}
which indicates that there are only two independent order parameters for the
electromagnetic field. Eq. (\ref{CoulombLambda}) can be regarded as the
Coulomb gauge to the vector order parameter $\mathbf{\Lambda}$.

From Eqs. (\ref{CoupledPhiLambda}--\ref{CoulombLambda}), one can easily
verify that there is no BEC for the electromagnetic field $\mathbf{A}$ if it
is not coupled with the Higgs field $\psi $, or it is coupled with $\psi $
but the latter has not condensed yet ($\varphi =0$). This also confirms the
conclusion of Sec. \ref{IPGSB}: an ideal photon gas, or a free
electromagnetic filed, can not produce BEC. Once the Higgs field $\psi $
gets condensed ($\varphi \neq 0$), the electromagnetic field $\mathbf{A}$
must condense down ($\mathbf{\Lambda }\neq 0$) simultaneously; otherwise,
there can arise a linear term of $\delta \mathbf{\Lambda }$ in $\delta S$,
the coupled system will go instable. This shows that the gauge field $%
\mathbf{A}$ will condense together with the Higgs fields $\psi $ when it is
coupled with the latter. In such case, the transition is cooperative, one
usually says that $\varphi $ is the primary order parameter, and $\mathbf{%
\Lambda }$ the secondary one.

With the help of Eq. (\ref{CoulombLambda}), Eq. (\ref{GL2}) can be reduced as%
\begin{equation}
\nabla^{2}\mathbf{\Lambda}-2e^{2}(\varphi^{\dagger}\varphi+\overline {%
\psi^{\dagger}\psi})\mathbf{\Lambda}=-ie(\varphi^{\dagger}\nabla
\varphi-\varphi\nabla\varphi^{\dagger}).  \label{ProcaEq}
\end{equation}
Heeding that $\langle\mathbf{A}\rangle=-\mathbf{\Lambda}$, Eq. (\ref{ProcaEq}%
) is identical to the static Proca equation \cite{Jackson} except that the
prefactor of $\mathbf{\Lambda}$ may be a function of $\mathbf{r}$, the
counterpart in Proca equation just a constant. In general, the prefactor
cannot be interpreted as the mass of the gauge field. However, once there is
a constant occurring in the prefactor, a mass $\mu$ is generated to the
gauge field, as can be seen more clearly in the following discussion.

Owing to the nonlinearity, the Ginzburg-Landau equations (\ref{GL1}) and (%
\ref{GL2}) are rather complicated to handle. As usual, let us consider the
zeroth-order approximation of Eq. (\ref{GL1}) at $T=0\mathrm{K}$ \cite%
{GL,Lifshitz},
\begin{equation}
\varphi^{\dagger}(\mathbf{x})\varphi(\mathbf{x})=\nu^{2}/2,  \label{NuTZero}
\end{equation}
where the effect of $\mathbf{\Lambda}$\ and\ the zero-point fluctuations of
the fields $\psi$ and $\mathbf{A}$ are omitted. From Eqs. (\ref{ProcaEq})
and (\ref{NuTZero}), it follows that%
\begin{equation}
\nabla^{2}\langle\mathbf{A}\rangle-e^{2}\nu^{2}\langle\mathbf{A}\rangle=0.
\label{GLPicture}
\end{equation}
Since the prefactor $e^{2}\nu^{2}$ is constant now, it can be interpreted as
the mass of the gauge field at zero temperature,\
\begin{equation}
\mu^{2}(T=0)=e^{2}\nu^{2}.  \label{MuTZero}
\end{equation}
As well-know, this mass can account for the Meissner effect \cite{Meissner},
i.e., the expulsion of a magnetic field from the interior of a
superconducting material, with a London penetration length $%
\lambda_{L}=\mu^{-1}$ \cite{London}.

The above formalism is Ginzburg-Landau picture of mass producing, in this
picture, the mass is represented by the equation of motion. There is another
picture, which is due to Higgs \cite{Higgs1,Higgs2,Higgs3}. In Higgs
picture, the mass is represented directly by the system Hamiltonian itself.
Since an equation of motion is in accordance with its Hamiltonian, both
pictures are equivalent physically. To show Higgs picture, we aught to
consider the zero-temperature representation $H_{0}$ of the system
Hamiltonian,%
\begin{align}
H_{0} & =e^{iD\left[ \varphi,\pi\mathbf{;\Lambda},\mathbf{E}\right] }H\left[
\psi,\pi;\mathbf{A,E}\right] e^{-iD\left[ \varphi,\pi \mathbf{;\Lambda},%
\mathbf{E}\right] }  \notag \\
& \equiv H[\chi,\pi;\mathbf{W,E}],
\end{align}
where%
\begin{align}
\chi(\mathbf{x}) & =e^{iD\left[ \varphi,\pi\mathbf{;\Lambda},\mathbf{E}%
\right] }\psi(\mathbf{x})e^{-iD\left[ \varphi,\pi\mathbf{;\Lambda },\mathbf{E%
}\right] },  \label{ChiField} \\
\mathbf{W}(\mathbf{x}) & =e^{iD\left[ \varphi,\pi\mathbf{;\Lambda },\mathbf{E%
}\right] }\mathbf{A}(\mathbf{x})e^{-iD\left[ \varphi ,\pi\mathbf{;\Lambda},%
\mathbf{E}\right] }.
\end{align}
As is well known, within the temporal gauge, the electromagnetic field still
has gauge degrees of freedom, the Hamiltonian $H[\psi,\pi;\mathbf{A,E}]$
remains invariant under any time-independent gauge transformation. As
another representation of $H[\psi,\pi;\mathbf{A,E}]$, the Hamiltonian $H%
\left[ \chi,\pi;\mathbf{W,E}\right] $ is also gauge invariant,%
\begin{equation}
H\left[ \chi,\pi;\mathbf{W,E}\right] =e^{iG\left[ \theta,\chi ,\pi,\mathbf{E}%
\right] }H\left[ \chi,\pi;\mathbf{W,E}\right] e^{-iG\left[ \theta,\chi,\pi,%
\mathbf{E}\right] },
\end{equation}
where $\theta\in%
\mathbb{R}
$ and%
\begin{equation}
G\left[ \theta,\chi,\pi,\mathbf{E}\right] =\int\mathrm{d}\mathbf{x}\text{ }%
[-ie\theta\left( \pi\chi-\chi^{\dagger}\pi^{\dagger}\right) +\nabla
\theta\cdot\mathbf{E}].
\end{equation}

As Higgs \cite{Higgs1,Higgs2,Higgs3} did, we\ can re-gauge the fields $\chi(%
\mathbf{x})$ and $\mathbf{W}(\mathbf{x})$\ by letting $\theta (\mathbf{x})$
equal to the phase of $\chi(\mathbf{x})$, that is,%
\begin{equation}
\chi(\mathbf{x})=\rho(\mathbf{x})e^{i\theta(\mathbf{x})},\text{\ }\rho(%
\mathbf{x})\in%
\mathbb{R}
,
\end{equation}
so as to eliminate the Goldstone bosons represented by the phase of $\chi(%
\mathbf{x})$. Upon such a choice of $\theta(\mathbf{x})$, the Hamiltonian $%
H_{0}$ can be expressed as%
\begin{align}
\hspace{-0.2in}H_{0} & =\int\mathrm{d}\mathbf{x}\text{ }\Big[\widetilde{\pi }%
^{\dagger}\widetilde{\pi}+(\nabla+ie\widetilde{\mathbf{A}})\rho\cdot
(\nabla-ie\widetilde{\mathbf{A}})\rho  \notag \\
& +m^{2}\rho^{2}+g\rho^{4}+\frac{1}{2}\mathbf{E}^{2}+\frac{1}{2}(\nabla
\times\widetilde{\mathbf{A}})^{2}\Big],  \label{HCoupled}
\end{align}
where
\begin{align}
\widetilde{\pi}(\mathbf{x}) & =e^{iG\left[ \theta,\chi,\pi,\mathbf{E}\right]
}\pi(\mathbf{x})e^{-iG\left[ \theta,\chi,\pi,\mathbf{E}\right] },
\label{PiWave} \\
\widetilde{\mathbf{A}}(\mathbf{x}) & =e^{iG\left[ \theta,\chi,\pi ,\mathbf{E}%
\right] }\mathbf{W}(\mathbf{x})e^{-iG\left[ \theta,\chi ,\pi,\mathbf{E}%
\right] }.
\end{align}

From Eqs. (\ref{ChiField}) and (\ref{DPiLambda}), it follows that
\begin{equation}
\chi(\mathbf{x})=\psi(\mathbf{x})+\varphi(\mathbf{x}).
\end{equation}
Without loss of generality, we can, from Eq. (\ref{NuTZero}), choose $%
\varphi(\mathbf{x})$ as real,%
\begin{equation}
\varphi(\mathbf{x})=\nu/\sqrt{2}.
\end{equation}
That leads to%
\begin{equation}
\rho(\mathbf{x})=\left( \phi(\mathbf{x})+\nu\right) /\sqrt{2},
\label{RhoPhi}
\end{equation}
where%
\begin{equation}
\phi(\mathbf{x})=\sqrt{2}e^{iG\left[ \theta,\widetilde{\psi},\pi ,\mathbf{E}%
\right] }\psi(\mathbf{x})e^{-iG\left[ \theta,\widetilde{\psi},\pi,\mathbf{E}%
\right] }.
\end{equation}
Eq. (\ref{RhoPhi}) indicates that $\phi(\mathbf{x})$ is merely a real field.

Substitution of Eq. (\ref{RhoPhi}) into Eq. (\ref{HCoupled}) gives rise to%
\begin{align}
H_{0} & =\int\mathrm{d}\mathbf{x}\text{ }\Big[\frac{1}{2}p^{2}+\frac{1}{2}%
(\nabla+ie\widetilde{\mathbf{A}})\phi\cdot(\nabla-ie\widetilde{\mathbf{A}}%
)\phi  \notag \\
& -m^{2}\phi^{2}+g\nu\phi^{3}+\frac{g}{4}\phi^{4}+\frac{1}{2}\mathbf{E}^{2}+%
\frac{1}{2}(\nabla\times\widetilde{\mathbf{A}})^{2}  \notag \\
& +\frac{1}{2}e^{2}\nu^{2}\widetilde{\mathbf{A}}\cdot\widetilde{\mathbf{A}}%
+e^{2}\nu\phi\widetilde{\mathbf{A}}\cdot\widetilde{\mathbf{A}}  \notag \\
& -\frac{g\nu^{4}}{4}+\frac{1}{2}p_{0}^{2}\Big],  \label{HiggsRepre}
\end{align}
where%
\begin{align}
p(\mathbf{x}) & =[\widetilde{\pi}^{\dagger}(\mathbf{x})+\widetilde{\pi }(%
\mathbf{x})]/\sqrt{2}, \\
p_{0}(\mathbf{x}) & =i[\widetilde{\pi}^{\dagger}(\mathbf{x})-\widetilde{\pi }%
(\mathbf{x})]/\sqrt{2}.
\end{align}
Paying attention to%
\begin{equation}
e^{iG\left[ \theta,\chi,\pi,\mathbf{E}\right] }\left( \frac{\chi (\mathbf{x}%
)+\chi^{\dagger}(\mathbf{x})}{\sqrt{2}}\right) e^{-iG\left[ \theta,\chi,\pi,%
\mathbf{E}\right] }=\phi(\mathbf{x})+\nu,
\end{equation}
we have
\begin{gather}
\left[ p(\mathbf{x}),\phi(\mathbf{x}^{\prime})\right] =-i\delta (\mathbf{x}-%
\mathbf{x}^{\prime}), \\
\left[ p_{0}(\mathbf{x}),H_{0}\right] =0.
\end{gather}
That is to say, the $\phi(\mathbf{x})$ and $p(\mathbf{x})$ become a new pair
of canonical field operators whereas $p_{0}(\mathbf{x})$ is just a constant.
Physically, that is because $p_{0}(\mathbf{x})$ corresponds to the
eliminated degree of freedom $\phi_{0}(\mathbf{x})=-i\left( \chi(\mathbf{x}%
)-\chi^{\dagger}(\mathbf{x})\right) /\sqrt{2}$.

Eq. (\ref{HiggsRepre}) forms Higgs picture of mass producing. Comparing it
with Eq. (\ref{GLPicture}), one sees immediately that the mass of the gauge
field in Higgs picture is identical to that in Ginzburg-Landau picture, as
is expected.

The above theory for a complex field coupled with an Abelian gauge field can
be easily generalized to the case of an $N$-component field coupled with a
non-Abelian gauge field, with the same conclusion that the gauge field will
obtain a mass through spontaneous symmetry breaking. This manner of mass
producing is the well-known Higgs mechanism, which was successively
contributed by London \cite{London}, Ginzburg and Landau \cite{GL}, Anderson
\cite{Anderson1,Anderson2}, Higgs \cite{Higgs1,Higgs2,Higgs3}, and Kibble
\cite{Kibble}.

\textbf{Remark}: Spontaneous symmetry breaking can occur without any
fundamental Higgs field in superconductivity, it is thus called "dynamical
symmetry breaking", which implies that Higgs fields are "normally" needed
for symmetry breaking. From the viewpoint of the extended ensemble theory,
that is a prejudice because superconductivity, Goldstone bosons and Higgs
mechanism share the same physical ground. This is in accordance with the
standpoint of Huang expressed in his book \cite{Huang2}.

\section{Summary and Conclusions}

So far, a possible extension of Gibbs ensemble theory has been postulated,
as a personal review of the author, to enable a microscopic description to
phase transitions and spontaneous symmetry breaking. The extension is
founded on three hypotheses, which root, in physics, from Landau's ideas on
phase transitions: order parameter, variational principle, representation
transformation, and spontaneous symmetry breaking. In this sense, the
extended Gibbs ensemble theory can be viewed as a microscopic realization of
the Landau phenomenological theory of phase transitions.

Within the framework of the extended ensemble theory, a phase transition
occurs according to the principle of least entropy, which manifests itself
as an equation of motion of order parameter. This equation determines the
evolution of order parameter with temperature, and thus controls the change
in representation of system Hamiltonian. Different phases correspond to
different representations, and vice versa. A system Hamiltonian will realize
its symmetric representation in the disordered phase, and asymmetric one in
the ordered phase. The change in symmetry results from the change in
representation. That is the Landau mechanism responsible for phase
transitions and spontaneous symmetry breaking in the extended ensemble
theory, it holds in the whole range of temperature, including the critical
region.

Physically, phase transitions originate from the wave nature of matter. A
system always stays in its normal phase at sufficiently high temperatures,
which is disordered and structureless. As temperature decreases, matter
waves will interfere automatically with one another. This interference makes
the system structured and transform into its ordered phase. That is the
physical picture for phase transitions in the extended ensemble theory.

Also, the extended ensemble theory has been applied to typical quantum
many-body systems, with the conclusions as follows.

For the ideal Fermi gas, it can not produce superconductivity, its normal
phase is stable at any temperature.

Negative-temperature and laser phases arise from the same mechanism as phase
transitions, they are both instable, and will finally turn into the
positive-temperature phase. That is the microscopic interpretation for
negative temperatures and laser phases.

For the conventional weak-coupling low-$T_{c}$ superconductors, the
mean-field solution due to Bardeen, Cooper and Schrieffer is derived anew,
and proved to be stable within the extended ensemble theory.

For the ideal Bose gas, it can produce Bose-Einstein condensation only in
the thermodynamic limit, which agrees with Einstein's prediction. But that
holds in the sense of Lebesgue integration rather than Riemann integration.
In general, the order parameter for BEC can not be interpreted as the number
of condensed particles. Besides, there can not exist any supercurrent or
vortex in the ideal Bose gas, the condensation is always homogeneous.

The ideal phonon gas can not produce Bose-Einstein condensation, neither the
photon gas in a black body.

It is not admissible to quantize Dirac field using Bose-Einstein statistics.
Otherwise, the system is instable. That is a statistical rather than
mechanical reason why Dirac field has to be quantized using Fermi-Dirac
statistics. This conclusion is in accordance with the spin-statistics
theorem in quantum field theory.

A structural phase transition belongs physically to the Bose-Einstein
condensation occurring in configuration space. For a double-well anharmonic
system, it is instable at low temperatures, and will undergo a structural
phase transition at a finite temperature. Physically, that is because the
position fluctuation is a monotonically decreasing function of temperature.

For the $O(N)$-symmetric vector model, the $O(N)$ symmetry will break down
spontaneously, with the presence of Goldstone bosons. In essence, this SSB
is a Bose-Einstein condensation. If the system is coupled with a gauge
field, it can cause the latter to condense together with itself. The
cooperative condensation can be described by the Ginzburg-Landau equations.
After the condensation, the gauge field can obtain a mass, which is the
so-called Higgs mechanism.

For an interacting Bose gas, it is stable only if the interaction is
repulsive. The BEC present in this system can be described by the
generalized Ginzburg-Landau equation. If the interaction is weak, the BEC is
a \textquotedblleft$\lambda$\textquotedblright-transition, and its
transition temperature can be lowered by the repulsive interaction. As a
characteristic property of this \textquotedblleft$\lambda$%
\textquotedblright-transition, the specific heat at constant volume $C_{V}$
will vanish linearly as $T\rightarrow0$.

If liquid $^{4}\mathrm{He}$ could be regarded as a weakly interacting Bose
gas, and if the $\lambda $-transition were a Bose-Einstein condensation,
then its specific heat at constant pressure $C_{P}$ would show a $T^{3}$ law
at low temperatures, which is in agreement with the experiment. However, if
temperature goes lower further, it is predicted that $C_{P}$ will exhibit a
linear behavior at relatively lower temperatures, which is in need of
experimental verifications.

\begin{acknowledgments}
The author would like to thank Prof. Zheng-zhong Li, Prof. Jin-ming Dong,
and Prof. Wei-yi Zhang for their helpful discussions. He is also deeply
grateful to Prof. Vladimir Rittenberg for his valuable advice on how to
revise the paper.
\end{acknowledgments}

\newpage

\appendix

\section{Frequency Summation\label{FSum}}

In this appendix, we shall first show that, when $\mu>0$, the ideal Bose gas
is still well defined within the extended ensemble theory, and then discuss
the frequency summation, which is very useful when we attempt to solve
problems through Green's functions.

Upon a transformation as in Eq. (\ref{Sphi}), the statistical average
defined by Eq. (\ref{Average}) can always be transformed into a statistical
average with respect to the Hamiltonian $H(b)$ of Eq. (\ref{Hibg}). It is
thus sufficient for us to show that the statistical average with respect to $%
H(b)$ is properly defined when $\mu >0$. As is well known, every statistical
average can be calculated through a corresponding Green's function, and
every Green's function can be derived from the generating functional of the
system. Therefore, we need only to prove that the generating functional is
properly defined when $\mu >0$.

According to the path-integral formalism \cite{Negele}, the generating
functional $W[\phi ^{+},\phi ]$ for $H(b)$ can be written as follows,
\begin{widetext}%
\begin{equation}
W[\phi ^{+},\phi ]=\mathcal{N}^{-1}\int \mathcal{D}b^{+}\mathcal{D}b\text{%
\thinspace }\exp \!\!\left( \frac{1}{\hbar }\sum_{\mathbf{k},n}\left\{ b_{%
\mathbf{k,}n}^{+}\left[ i\hbar \omega _{n}-\left( \frac{\hbar ^{2}\mathbf{k}%
^{2}}{2m}-\mu \right) \right] b_{\mathbf{k},n}-\phi _{\mathbf{k},n}b_{%
\mathbf{k,}n}^{+}-\phi _{\mathbf{k,}n}^{+}b_{\mathbf{k},n}\right\} \right) ,
\label{WGenerate}
\end{equation}%
where $\omega _{n}=2n\pi /\beta \hbar $ ($n\in
\mathbb{Z}
$) denotes the Mutsubara frequency, and $\mathcal{N}$ the normalization
factor,
\begin{equation}
\mathcal{N}=\int \mathcal{D}b^{+}\mathcal{D}b\text{\thinspace }\exp
\!\!\left( \frac{1}{\hbar }\sum_{\mathbf{k},n}b_{\mathbf{k,}n}^{+}\left[
i\hbar \omega _{n}-\left( \frac{\hbar ^{2}\mathbf{k}^{2}}{2m}-\mu \right) %
\right] b_{\mathbf{k},n}\right) .  \label{NGaussian}
\end{equation}

If $\mu <0$, one gets immediately the familiar result,
\begin{equation}
W[\phi ^{+},\phi ]=\exp \!\!\left( -\frac{1}{\hbar }\sum_{\mathbf{k},n}\phi
_{\mathbf{k},n}\frac{1}{i\hbar \omega _{n}-\left( \frac{\hbar ^{2}\mathbf{k}%
^{2}}{2m}-\mu \right) }\phi _{\mathbf{k,}n}^{+}\right) .  \label{WWG}
\end{equation}%
\end{widetext}
Since $\hbar ^{2}\mathbf{k}^{2}/\left( 2m\right) -\mu >0$, the normalization
factor $\mathcal{N}$ converges.

When $\mu \geq 0$, the normalization factor $\mathcal{N}$ will diverge.
However, that does no harm because the normalization factor can be cancelled
by the numerator of Eq. (\ref{WGenerate}), as is frequently encountered in
quantum field theory \cite{Greiner}. After the cancellation, the rest is
still Eq. (\ref{WWG}) except $\mu \geq 0$. As mentioned in Sec. \ref%
{BECIdeal}, an unbounded function as integrand is permissible within
Lebesgue integration. Therefore, the sum in the exponent of Eq. (\ref{WWG})
is proper as a Lebesgue integral. The resulting $W[\phi ^{+},\phi ]$ is also
right because a function is permitted to be unbounded in Lebesgue
integration.

In a word, the generating functional for $H(b)$ is always properly defined
no matter how large the chemical potential $\mu $ is, which ends our proof.

Now, there is no mathematical limit on the chemical potential of the ideal
Bose gas, as is the case for the ideal Fermi gas. That is rational and
significant, any limit on chemical potential should be set by the physical
theory itself rather than by the mathematical theory involved. In the
extended ensemble theory, the chemical potential of a system is determined
only by the physical requirement: the conservation of particles, whether the
system obeys Bose-Einstein statistics or Fermi-Dirac statistics; there is no
limit set by the mathematical tool, i.e., Lebesgue integration.

As a consequence of Eq. (\ref{WWG}), we obtain%
\begin{align}
\mathcal{G}(\mathbf{k},i\omega_{n}) & =-\left( -\frac{1}{\hbar}\right) ^{-2}%
\frac{\delta^{2}W[\phi^{+},\phi]}{\delta\phi_{\mathbf{k,}n}^{+}\delta \phi_{%
\mathbf{k},n}}  \notag \\
& =\frac{1}{i\omega_{n}-\hbar^{-1}\left( \frac{\hbar^{2}\mathbf{k}^{2}}{2m}%
-\mu\right) },
\end{align}
where $\mathcal{G}(\mathbf{k},i\omega_{n})$ represents the temperature
Green's function which is defined as \cite{Fetter}%
\begin{equation}
\mathcal{G}(\mathbf{k},\tau-\tau^{\prime})=-\mathrm{Tr}\big(T_{\tau }\{b_{%
\mathbf{k}}(\tau)b_{\mathbf{k}}^{\dagger}(\tau^{\prime})\}\rho (H(b))\big).
\end{equation}
As a function of $\mathbf{k}$, $\mathcal{G}(\mathbf{k},i\omega_{n})$ can be
unbounded if $\mu\geq0$, but that will be all right because $\mathcal{G}(%
\mathbf{k},i\omega_{n})$ can only appear in the integrand of a Lebesgue
integral over $\mathbf{k}$.

Those discussions indicate that, in the manipulations of $+\infty $ and $%
-\infty $, one can benefit greatly from Lebesgue integration. That is
because Lebesgue integration is designed, \textit{ab initio}, on the set of
extended real numbers: $%
\mathbb{R}
^{\sharp }=%
\mathbb{R}
\cup \{+\infty \}\cup \{-\infty \}$ where $%
\mathbb{R}
=(-\infty ,+\infty )$ \cite{Hewitt}. Evidently, those benefits can not be
provided by Riemann integration. More benefits of Lebesgue integration will
be seen from the following discussions.

When one evaluates the statistical average of an observable, he will
encounter the summation over Mutsubara frequencies. As an illustration, let
us consider
\begin{align}
\int_{%
\mathbb{R}
^{3}}\mathrm{d}\mathbf{k\,}\overline{b_{\mathbf{k}}^{\dag }b_{\mathbf{k}}}&
=\int_{%
\mathbb{R}
^{3}}\mathrm{d}\mathbf{k\,}\mathrm{Tr}\big(b_{\mathbf{k}}^{\dag }b_{\mathbf{k%
}}\rho (H(b))\big)  \notag \\
& =\int_{%
\mathbb{R}
^{3}}\mathrm{d}\mathbf{k\,}\left( -\frac{1}{\beta \hbar }\right)
\sum_{n}e^{i\omega _{n}\eta }\mathcal{G}(\mathbf{k},i\omega _{n}),
\label{Summation}
\end{align}%
where $\eta =0^{+}$. We shall discuss the thermodynamic limit case. Here, it
is important to note that the frequency summation always goes ahead of the
Lebesgue integral over $\mathbf{k}$ because the thermodynamic limit must be
taken at last.

To perform the frequency summation, it is helpful to make use of the complex
function,
\begin{equation}
f(z)=\frac{1}{e^{\beta \hbar z}-1},
\end{equation}%
which has simple poles at the Mutsubara frequencies, i.e., $z=i\omega
_{n}=i2n\pi /\beta \hbar $, and transform the summation into a contour
integral \cite{Fetter},%
\begin{equation}
\left( -\frac{1}{\beta \hbar }\right) \sum_{n}e^{i\omega _{n}\eta }\mathcal{G%
}(\mathbf{k},i\omega _{n})=\int_{C}\frac{\mathrm{d}z}{2\pi i}e^{i\eta z}f(z)%
\mathcal{G}(\mathbf{k},z),
\end{equation}%
where the contour $C$ is depicted in Fig. 7. It should be emphasized that
the above procedure requires that the function $\mathcal{G}(\mathbf{k},z)$
must be analytic on the whole imaginary axis lest there be any poles other
than $i\omega _{n}$ on this axis. Sometimes, this requirement can not be met
by every $\mathbf{k}$, e.g.,%
\begin{equation}
\mathcal{G}(\mathbf{k},z)=\frac{1}{z-\hbar ^{-1}(\frac{\hbar ^{2}\mathbf{k}%
^{2}}{2m}-\mu )},\text{ }\mu \geq 0.  \label{Singular}
\end{equation}%
Each such $\mathcal{G}(\mathbf{k},z)$ that satisfies $|\mathbf{k|}=\sqrt{%
2m\mu }/\hbar $ is singular on $z=0$, and thus can not meet the requirement.
If such singular $\mathbf{k}$'s constitute just a null set $D$, i.e., $%
\mathcal{G}(\mathbf{k},z)$ satisfies the requirement almost everywhere on $%
\mathbb{R}
^{3}$, which is the usual case, then those $\mathbf{k}$'s can be left out of
consideration for they give merely a null contribution to the Lebesgue
integral over $\mathbf{k}$. Therefore, we can rewrite Eq. (\ref{Summation})
as
\begin{align}
& \int_{%
\mathbb{R}
^{3}}\mathrm{d}\mathbf{k\,}\left( -\frac{1}{\beta \hbar }\right)
\sum_{n}e^{i\omega _{n}\eta }\mathcal{G}(\mathbf{k},i\omega _{n})  \notag \\
& =\int_{E}\mathrm{d}\mathbf{k}\int_{C}\frac{\mathrm{d}z}{2\pi i}e^{i\eta
z}f(z)\mathcal{G}(\mathbf{k},z),  \label{Sum0}
\end{align}%
where $E=%
\mathbb{R}
^{3}\backslash D$. Here the set $D$ is removed form $%
\mathbb{R}
^{3}$ so that the function $\mathcal{G}(\mathbf{k},z)$ can meet the
requirement everywhere on $E$.
\begin{figure}[htbp]
\hspace*{-1.7cm}\includegraphics[scale=0.45,angle=-90]{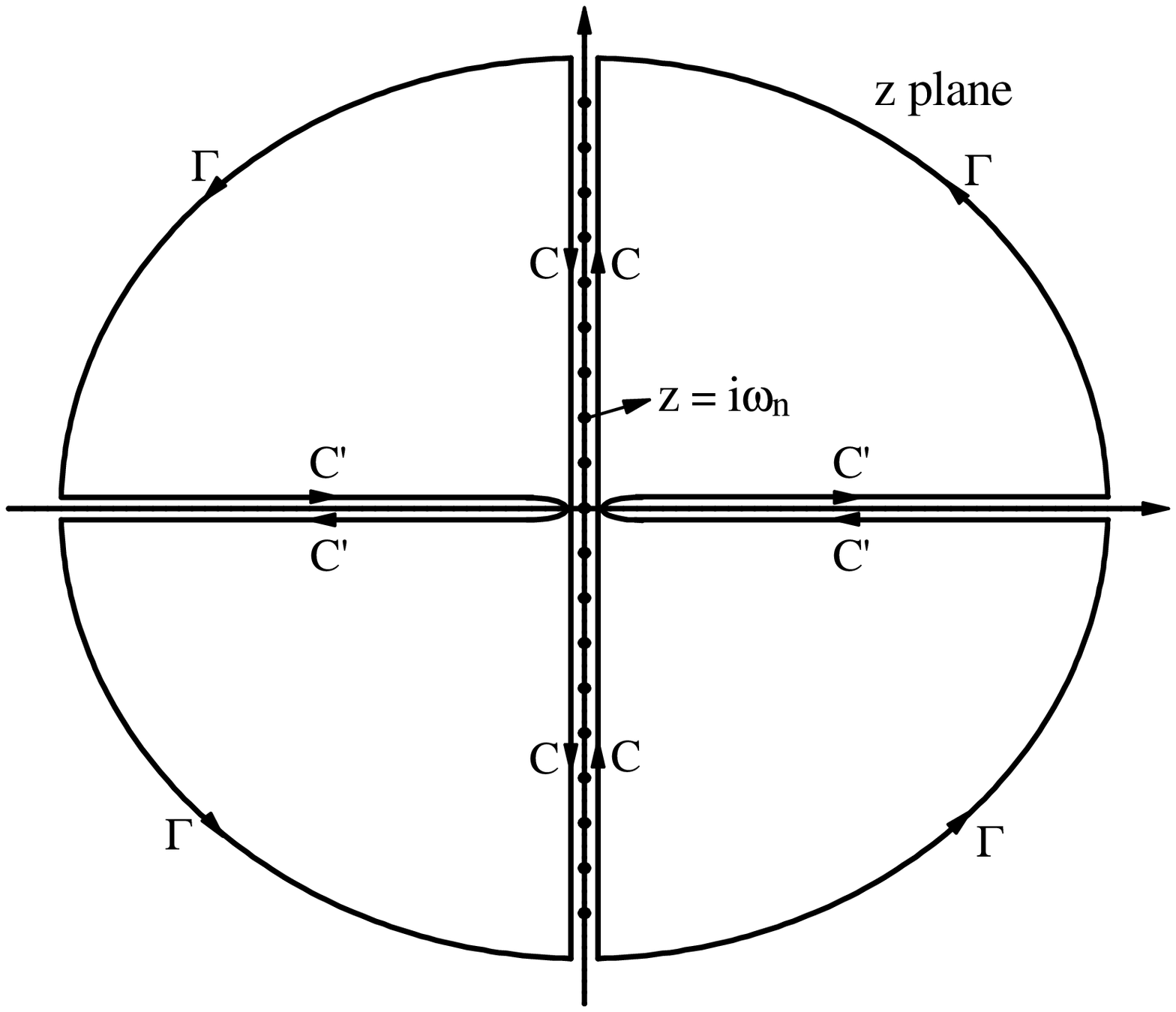}
\caption{The counters for the frequency summation of a Bose
system.}
\end{figure}%

Generally, $\mathcal{G}(\mathbf{k},z)$ are analytic on both the upper and
lower half complex-$z$ planes \cite{Zubarev,Fetter}, one can therefore
deform the contour $C$ into the contours $\Gamma $ and $C^{\prime }$, which
results in
\begin{equation}
\int_{C}\frac{\mathrm{d}z}{2\pi i}e^{i\eta z}f(z)\mathcal{G}(\mathbf{k}%
,z)=\int_{C^{\prime }}\frac{\mathrm{d}z}{2\pi i}f(z)\mathcal{G}(\mathbf{k}%
,z).  \label{Sum1}
\end{equation}%
The contribution from the contour $\Gamma $ vanishes owing to the
convergence factor $e^{i\eta z}$ \cite{Fetter}. The right-hand side can be
simplified as
\begin{equation}
\int_{C^{\prime }}\frac{\mathrm{d}z}{2\pi i}f(z)\mathcal{G}(\mathbf{k},z)=%
\mathcal{P}\int_{-\infty }^{+\infty }\mathrm{d}\omega \,f(\omega )\mathcal{A}%
(\mathbf{k},\omega ),
\end{equation}%
where $\mathcal{A}(\mathbf{k},\omega )$ is the spectral intensity,
\begin{equation}
\mathcal{A}(\mathbf{k},\omega )=-\frac{1}{2\pi i}\left[ \mathcal{G}(\mathbf{k%
},\omega +i0^{+})-\mathcal{G}(\mathbf{k},\omega -i0^{+})\right] ,
\end{equation}%
and $\mathcal{P}$ represents the principal value,
\begin{widetext}
\begin{equation}
\mathcal{P}\int_{-\infty }^{+\infty }\mathrm{d}\omega \,f(\omega )\mathcal{G}%
(\mathbf{k},\omega )=\lim_{\alpha \rightarrow 0^{+}}\left( \int_{-\infty
}^{-\alpha }\mathrm{d}\omega \,f(\omega )\mathcal{A}(\mathbf{k},\omega
)+\int_{\alpha }^{+\infty }\mathrm{d}\omega \,f(\omega )\mathcal{A}(\mathbf{k%
},\omega )\right) .  \label{Sum2}
\end{equation}

Substituting Eqs. (\ref{Sum1}--\ref{Sum2}) into Eq. (\ref{Sum0}), one has
\begin{eqnarray}
\int_{%
\mathbb{R}
^{3}}\mathrm{d}\mathbf{k}\left( -\frac{1}{\beta \hbar }\right)
\sum_{n}e^{i\omega _{n}\eta }\mathcal{G}(\mathbf{k},i\omega _{n}) &=&\int_{E}%
\mathrm{d}\mathbf{k}\lim_{\alpha \rightarrow 0^{+}}\left[ \int_{-\infty
}^{-\alpha }\mathrm{d}\omega \,f(\omega )\mathcal{A}(\mathbf{k},\omega
)+\int_{\alpha }^{+\infty }\mathrm{d}\omega \,f(\omega )\mathcal{A}(\mathbf{k%
},\omega )\right]  \notag \\
&=&\lim_{\alpha \rightarrow 0^{+}}\int_{E}\mathrm{d}\mathbf{k}\left[
\int_{-\infty }^{-\alpha }\mathrm{d}\omega \,f(\omega )\mathcal{A}(\mathbf{k}%
,\omega )+\int_{\alpha }^{+\infty }\mathrm{d}\omega \,f(\omega )\mathcal{A}(%
\mathbf{k},\omega )\right]  \notag \\
&=&\lim_{\alpha \rightarrow 0^{+}}\left[ \int_{-\infty }^{-\alpha }\mathrm{d}%
\omega \,f(\omega )\int_{E}\mathrm{d}\mathbf{k}\mathcal{A}(\mathbf{k},\omega
)+\int_{\alpha }^{+\infty }\mathrm{d}\omega \,f(\omega )\int_{E}\mathrm{d}%
\mathbf{k}\mathcal{A}(\mathbf{k},\omega )\right]  \notag \\
&=&\mathcal{P}\int_{-\infty }^{+\infty }\mathrm{d}\omega \,f(\omega )\int_{E}%
\mathrm{d}\mathbf{k}\mathcal{A}(\mathbf{k},\omega ).
\end{eqnarray}%
\end{widetext}%

Now, supplementing the set $E$ with the null set $D$, one arrives at%
\begin{align}
& \int_{%
\mathbb{R}
^{3}}\mathrm{d}\mathbf{k}\left( -\frac{1}{\beta \hbar }\right)
\sum_{n}e^{i\omega _{n}\eta }\mathcal{G}(\mathbf{k},i\omega _{n})  \notag \\
& =\mathcal{P}\int_{-\infty }^{+\infty }\mathrm{d}\omega \,f(\omega )\int_{%
\mathbb{R}
^{3}}\mathrm{d}\mathbf{k}\mathcal{A}(\mathbf{k},\omega ).  \label{BSum}
\end{align}%
As usual, by introducing the density of states $\mathcal{N}(\omega )$,
\begin{equation}
\mathcal{N}(\omega )=\frac{1}{\left( 2\pi \right) ^{3}}\int_{%
\mathbb{R}
^{3}}\mathrm{d}\mathbf{k}\mathcal{A}(\mathbf{k},\omega ),
\end{equation}%
Eq. (\ref{BSum}) can be reduced into%
\begin{align}
& \frac{1}{\left( 2\pi \right) ^{3}}\int_{%
\mathbb{R}
^{3}}\mathrm{d}\mathbf{k}\left( -\frac{1}{\beta \hbar }\right)
\sum_{n}e^{i\omega _{n}\eta }\mathcal{G}(\mathbf{k},i\omega _{n})  \notag \\
& =\mathcal{P}\int_{-\infty }^{+\infty }\mathrm{d}\omega \frac{\mathcal{N}%
(\omega )}{e^{\beta \hbar \omega }-1}.  \label{Principal}
\end{align}%
This is a very useful identity, as its applications, one can easily obtain
the final results of Eqs. (\ref{Lesb1}) and (\ref{Lesb3}). It should be
stressed that Eq. (\ref{Principal}) holds whether the chemical potential $%
\mu $ is less than, equal to, or lager than zero. If $\mu <0$, it reduces to
the common result given in Ref. \cite{Fetter}. It is a generalization of the
common result when $\mu \geq 0$. This generalization removes the
mathematical limit on the chemical potential. One can not reach the result
of Eq. (\ref{Principal}) within Riemann integration.

Obviously, the above analyses for the ideal Bose gas are also suitable for
other Bose systems, e.g., the interacting Bose gas, which will be studied in
Sec. \ref{WIBG}.

\textbf{Remark}:\textbf{\ }If the system obeys Fermi-Dirac statistics, the
principal value is unnecessary. Instead of Eq. (\ref{Principal}), one has%
\begin{align}
& \frac{1}{\left( 2\pi \right) ^{3}}\int_{%
\mathbb{R}
^{3}}\mathrm{d}\mathbf{k}\left( \frac{1}{\beta \hbar }\right)
\sum_{n}e^{i\omega _{n}\eta }\mathcal{G}(\mathbf{k},i\omega _{n})  \notag \\
& =\mathcal{P}\int_{-\infty }^{+\infty }\mathrm{d}\omega \frac{\mathcal{N}%
(\omega )}{e^{\beta \hbar \omega }+1}  \notag \\
& =\int_{-\infty }^{+\infty }\mathrm{d}\omega \frac{\mathcal{N}(\omega )}{%
e^{\beta \hbar \omega }+1}.  \label{Fermi}
\end{align}%
That is simply because the Fermi distribution function is bounded and
continuous at $\omega =0$ for any finite temperature $\beta >0$. A more
direct reason for the unnecessity of the principal value consists in the
fact that, unlike the Bose system, $z=0$ is not the pole of the complex
function,
\begin{equation}
f(z)=\frac{1}{e^{\beta \hbar z}+1},
\end{equation}%
because the Mutsubara frequencies for a Fermi system are all nonzero, i.e., $%
\omega _{n}=\left( 2n+1\right) \pi /\beta \hbar \neq 0$ ($n\in
\mathbb{Z}
$). As a consequence, the integration contour can goes as in Fig. 8, which
gives straightforwardly the final result of Eq. (\ref{Fermi}). Evidently,
that makes the Fermi system easier to handle mathematically than the Bose
system.
\begin{figure}[htbp]
\hspace*{-1.7cm}\includegraphics[scale=0.45,angle=-90]{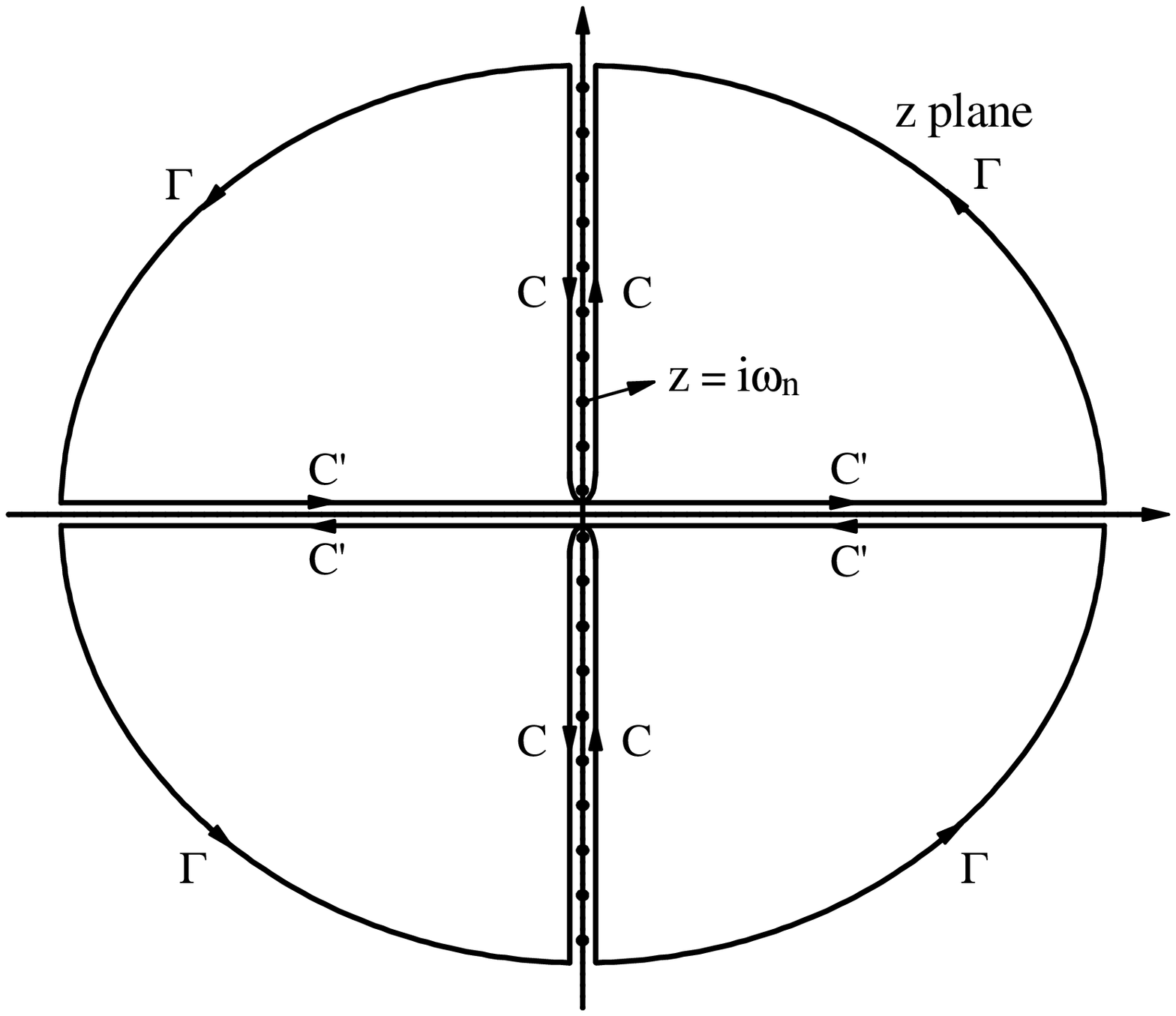}
\caption{The counters for the frequency summation of a Fermi system.}
\end{figure}%

\section{Low-temperature Thermal Activations of a Weakly Interacting Bose
System \label{TAWIBSLT}}

In Sec. \ref{WIBG}, our discussions were confined within the Hartree-Fock
approximation of Eqs. (\ref{IBECHF1}--\ref{IBECHF2}) and the short-range
approximation of Eq. (\ref{IBECSR}). Within those approximations, one sees
from Eqs. (\ref{X12LTc}) and (\ref{X32LTc}) that the thermal activations at
low temperatures can be expanded into even power series of $T$. We shall
show that, beyond those approximations, this feature still holds as long as
the interaction is weak and perturbative.

Under the condition assumed, observables can be calculated via GF. The
calculations can be transformed finally into the integrals of the following
form,
\begin{equation}
\mathcal{P}\int_{-\infty }^{+\infty }\mathrm{d}\omega \frac{g\left( \omega ,%
\widetilde{\mu }\right) }{e^{\beta \omega }-1},
\end{equation}%
where $g\left( \omega ,\widetilde{\mu }\right) $ is a real-valued function
of $\omega $ and $\widetilde{\mu }$, which comes from a summation of GF over
$\mathbf{k}$. As a function of $\omega $, $g\left( \omega ,\widetilde{\mu }%
\right) $ should be bounded at bottom or decrease fast with $\omega $ so
that the integral can converge. In both cases, we can cut off the integral
from below at a certain frequency $-\omega _{c}$ ($\omega _{c}>0$), that is,
\begin{equation}
\mathcal{P}\int_{-\infty }^{+\infty }\mathrm{d}\omega \frac{g\left( \omega ,%
\widetilde{\mu }\right) }{e^{\beta \omega }-1}=\mathcal{P}\int_{-\omega
_{c}}^{+\infty }\mathrm{d}\omega \frac{g\left( \omega ,\widetilde{\mu }%
\right) }{e^{\beta \omega }-1}.
\end{equation}%
According to the definition of principal value, the right-hand side can be
written as
\begin{widetext}%
\begin{equation}
\mathcal{P}\int_{-\omega _{c}}^{+\infty }\mathrm{d}\omega \frac{g\left(
\omega ,\widetilde{\mu }\right) }{e^{\beta \omega }-1}=\int_{-\omega
_{c}}^{-0^{+}}\mathrm{d}\omega \frac{g\left( \omega ,\widetilde{\mu }\right)
}{e^{\beta \omega }-1}+\int_{0^{+}}^{+\infty }\mathrm{d}\omega \frac{g\left(
\omega ,\widetilde{\mu }\right) }{e^{\beta \omega }-1}.
\end{equation}%
Observe that
\begin{equation}
\frac{1}{e^{\beta \omega }-1}+\frac{1}{e^{-\beta \omega }-1}=-1,
\end{equation}%
we have%
\begin{equation}
\mathcal{P}\int_{-\infty }^{+\infty }\mathrm{d}\omega \frac{g\left( \omega ,%
\widetilde{\mu }\right) }{e^{\beta \omega }-1}=-\int_{-\omega _{c}}^{0}%
\mathrm{d}\omega \,g\left( \omega ,\widetilde{\mu }\right) -\int_{-\omega
_{c}}^{-0^{+}}\mathrm{d}\omega \frac{g\left( \omega ,\widetilde{\mu }\right)
}{e^{-\beta \omega }-1}+\int_{0^{+}}^{+\infty }\mathrm{d}\omega \frac{%
g\left( \omega ,\widetilde{\mu }\right) }{e^{\beta \omega }-1}.
\end{equation}%
Integrated by substitution, it turns into%
\begin{align}
\mathcal{P}\int_{-\infty }^{+\infty }\mathrm{d}\omega \frac{g\left( \omega ,%
\widetilde{\mu }\right) }{e^{\beta \omega }-1}& =-\int_{0}^{\omega _{c}}%
\mathrm{d}\omega \,g\left( \omega -\omega _{c},\widetilde{\mu }\right)
+k_{B}T\int_{0}^{+\infty }\mathrm{d}x\frac{g\left( \omega _{c}+k_{B}Tx,%
\widetilde{\mu }\right) }{e^{x+\beta \omega _{c}}-1}  \notag \\
& +k_{B}T\int_{0}^{\beta \omega _{c}}\mathrm{d}x\frac{g\left( k_{B}Tx,%
\widetilde{\mu }\right) -g\left( -k_{B}Tx,\widetilde{\mu }\right) }{e^{x}-1}.
\end{align}%
The second integral on the right-hand side vanishes exponentially as $%
T\rightarrow 0$, it can thus be neglected. The upper limit of the third
integral can be set equal to $+\infty $ at low temperatures for the
integrand decreases as $e^{-x}$ when $x$ becomes very large,
\begin{equation}
\mathcal{P}\int_{-\infty }^{+\infty }d\omega \frac{g\left( \omega ,%
\widetilde{\mu }\right) }{e^{\beta \omega }-1}=-\int_{0}^{\omega
_{c}}d\omega \,g\left( \omega -\omega _{c},\widetilde{\mu }\right)
+k_{B}T\int_{0}^{+\infty }dx\frac{g\left( k_{B}Tx,\widetilde{\mu }\right)
-g\left( -k_{B}Tx,\widetilde{\mu }\right) }{e^{x}-1}.  \label{POmega}
\end{equation}%
The second integral on the right-hand side can expanded as%
\begin{equation}
\int_{0}^{+\infty }\mathrm{d}x\frac{g\left( k_{B}Tx,\widetilde{\mu }\right)
-g\left( -k_{B}Tx,\widetilde{\mu }\right) }{e^{x}-1}=2\sum_{n=1}^{+\infty
}\zeta (2n)g^{(2n-1)}\left( 0,\widetilde{\mu }\right) \left( k_{B}T\right)
^{2n-1}.
\end{equation}%
Substituting it into Eq. (\ref{POmega}) results in%
\begin{equation}
\mathcal{P}\int_{-\infty }^{+\infty }\mathrm{d}\omega \frac{g\left( \omega ,%
\widetilde{\mu }\right) }{e^{\beta \omega }-1}=-\int_{0}^{\omega _{c}}%
\mathrm{d}\omega \,g\left( \omega -\omega _{c},\widetilde{\mu }\right)
+2\sum_{n=1}^{+\infty }\zeta (2n)g^{(2n-1)}\left( 0,\widetilde{\mu }\right)
\left( k_{B}T\right) ^{2n},
\end{equation}%
the sum is an even power series of $T$. To the fourth order, it becomes
\begin{equation}
\mathcal{P}\int_{-\infty }^{+\infty }\mathrm{d}\omega \frac{g\left( \omega ,%
\widetilde{\mu }\right) }{e^{\beta \omega }-1}=-\int_{0}^{\omega _{c}}%
\mathrm{d}\omega \,g\left( \omega -\omega _{c},\widetilde{\mu }\right)
+2\zeta (2)g^{\prime }\left( 0,\widetilde{\mu }\right) \left( k_{B}T\right)
^{2}+2\zeta (4)g^{(3)}\left( 0,\widetilde{\mu }\right) \left( k_{B}T\right)
^{4},
\end{equation}%
\end{widetext}
it is sufficient for the discussion of the low-temperature properties of the
system. This result implies that the renormalized chemical potential $%
\widetilde{\mu }$ and the internal energy $E$ are even functions of $T$ at
low temperatures, which leads to the conclusion of Eq. (\ref{CvTV}): the
specific heat $C_{V}$ of a weakly interacting Bose gas will vanish linearly
as $T\rightarrow 0$.

The above result reminds us of the specific heat of the electron gas in a
normal metal, which also vanishes linearly as $T\rightarrow 0$. This linear
behavior of the electron gas can be explained analogously. There, what is
concerned is the integral,%
\begin{equation}
\int_{-\infty }^{+\infty }\mathrm{d}\omega \frac{g\left( \omega ,\mu \right)
}{e^{\beta \omega }+1},
\end{equation}%
where $\mu $ is the chemical potential. Along the same way as for the Bose
gas, it can be expressed as
\begin{widetext}%
\begin{equation}
\int_{-\infty }^{+\infty }\mathrm{d}\omega \frac{g\left( \omega ,\mu \right)
}{e^{\beta \omega }+1}=\int_{0}^{\omega _{c}}\mathrm{d}\omega \,g\left(
\omega -\omega _{c},\mu \right) +2\sum_{n=1}^{+\infty }\left( 1-\frac{1}{%
2^{2n-1}}\right) \zeta (2n)g^{(2n-1)}\left( 0,\mu \right) \left(
k_{B}T\right) ^{2n}.
\end{equation}%
To the second order, it is
\begin{equation}
\int_{-\infty }^{+\infty }\mathrm{d}\omega \frac{g\left( \omega ,\mu \right)
}{e^{\beta \omega }+1}=\int_{0}^{\omega _{c}}\mathrm{d}\omega \,g\left(
\omega -\omega _{c},\mu \right) +\zeta (2)g^{\prime }\left( 0,\mu \right)
\left( k_{B}T\right) ^{2},  \label{LowFermi}
\end{equation}%
\end{widetext}
which is sufficient for the low-temperature properties of the electron gas
in a normal metal because its Fermi temperature is very high, in comparison
to the critical temperature of the BEC happening in a weakly interacting
Bose gas. Eq. (\ref{LowFermi}) also implies that the chemical potential $\mu
$ and the internal energy $E$ are even functions of $T$ at low temperatures,
and that the specific heat at constant volume $C_{V}$ will vanish linearly
as $T\rightarrow 0$. For example, one can easily verify, with the help of
Eq. (\ref{LowFermi}), that the $\mu $, $E$ and $C_{V}$ for the ideal Fermi
gas have the low-temperature forms,%
\begin{eqnarray}
&&\mu (T)=\mu (0)\left[ 1-\frac{\pi ^{2}}{12}\left( \frac{k_{B}T}{\mu (0)}%
\right) ^{2}\right] , \\
&&\frac{E}{N}=\frac{3}{5}\mu (0)\left[ 1+\frac{5\pi ^{2}}{12}\left( \frac{%
k_{B}T}{\mu (0)}\right) ^{2}\right] , \\
&&\frac{C_{V}}{Nk_{B}}=\frac{\pi ^{2}}{2}\frac{k_{B}T}{\mu (0)},
\end{eqnarray}%
which are familiar results. Those discussions show that there exists
similarity even between so different systems!

\end{document}